\documentclass{svjour2}                    
\smartqed  

\usepackage{graphicx}
\usepackage{mathtools}
\usepackage{amssymb}
\usepackage{bbm}

\usepackage{hyperref}

\begin{document}

\title{Universality classes of interaction structures for NK fitness landscapes}


\titlerunning{Interaction structures of NK landscapes}        

\author{Sungmin Hwang \and Benjamin Schmiegelt \and Luca Ferretti \and
  Joachim Krug}


\institute{S. Hwang \and B. Schmiegelt \and J. Krug \at
              Institute for Theoretical Physics, University of Cologne \\
              Tel.: +49 221 2818\\
	      Fax: +49 221 5159\\
              \email{hwang@thp.uni-koeln.de, schmiegb@thp.uni-koeln.de, krug@thp.uni-koeln.de}        \\
              \emph{Present address of S. Hwang:} LPTMS, Universit\'{e} Paris-Sud 11, Orsay, France
          \and
           L. Ferretti \at
Integrative Biology group, The Pirbright Institute, United Kingdom \\
\email{luca.ferretti@pirbright.ac.uk}   
}

\date{Received: date / Accepted: date}

\maketitle

\begin{abstract}
Kauffman's NK-model is a paradigmatic example of a class of stochastic models of genotypic fitness landscapes that aim to
capture generic features of epistatic interactions in multilocus systems.
Genotypes are represented as sequences of $L$ binary loci. The fitness assigned to a genotype is a sum of contributions, each of which is a random function defined on a subset of $k \leq L$ loci. These subsets
or neighborhoods determine the genetic interactions of the model. Whereas earlier work on the NK model suggested that most of its
properties are robust with regard to the choice of neighborhoods,
recent work has revealed an important and sometimes counter-intuitive
influence of the interaction structure  on the properties of NK fitness landscapes. Here we review 
these developments and present new results concerning the number of local fitness maxima and the statistics of 
selectively accessible (that is, fitness-monotonic) mutational pathways. In particular, we develop a unified framework for computing
the exponential growth rate of the expected number of local fitness maxima as a function of $L$, and identify two different
universality classes of interaction structures that display different asymptotics of this quantity for large $k$. Moreover, we
show that the probability that the fitness landscape can be traversed along an accessible path decreases exponentially in $L$ for 
a large class of interaction structures that we characterize as
locally bounded. Finally, we discuss the impact of the NK interaction
structures on the dynamics of evolution using adaptive walk models.   

\end{abstract}


\newcommand{\prob}[1]{{\mathbb{P}\left[{#1}\right]}}
\newcommand{\mean}[1]{{\mathbb{E}\left[{#1}\right]}}
\newcommand{\var}[1]{{\text{var}\left[{#1}\right]}}
\newcommand{\cov}[2]{{\text{cov}\left[{#1},{#2}\right]}}
\newcommand{\cumul}[2][]{{\kappa_{#2}\left[#1\right]}}

\newcommand{\sgn}{\operatorname{sgn}}

\newcommand{\funcsig}[3]{{{#1}:{#2}\rightarrow{#3}}}

\newcommand{\suml}{\sum\limits}
\newcommand{\prodl}{\prod\limits}
\newcommand{\liml}{\lim\limits}
\newcommand{\intl}{\int\limits}

\newcommand{\integral}[3]{{\int_{#2}^{#3}\mathrm{d}{#1}}}
\newcommand{\integrall}[3]{{\intl_{#2}^{#3}\mathrm{d}{#1}}}

\newcommand{\nats}{{\mathbb{N}}}
\newcommand{\reals}{{\mathbb{R}}}
\newcommand{\ints}{{\mathbb{Z}}}
\newcommand{\compl}{{\mathbb{C}}}


\newcommand{\Mod}{ \, \textrm{mod} \,}

\newcommand{\nmax}{{\#}_{\text{max}}}
\newcommand{\nmaxModel}[1]{{\#}_{\text{max}}^{\mathrm{#1}}}

\newcommand{\growthFactor}[2]{\lambda_#1^\mathrm{#2}}
\newcommand{\ProbMax}{\pi_\mathrm{max}}
\newcommand{\ProbMaxModel}[1]{\pi_\mathrm{max}^{\mathrm{#1}}}
\newcommand{\Vecbf}[1]{\mathbf{#1}}
\newcommand{\Measure}[1]{\mathcal{D} \Vecbf{#1}}
\newcommand{\MeasureUQ}{\frac{\Measure{u} \Measure{q} }{(2\pi)^L}\, e^{-i \Vecbf{u}\cdot \Vecbf{q} } \, \Theta(\Vecbf{u} > 0) }
\newcommand{\MeasureY}{\Measure{y} \mathcal{P}(\Vecbf{y})}

\newcommand{\CharFuncBase}[1]{\phi_f\left(#1\right)}
\newcommand{\CharFunc}[2]{\Phi^{\mathrm{#2}}(\Vecbf{#1})}

\newcommand{\Avr}[1]{\left \langle #1 \right \rangle}
\newcommand{\ProbS}{\mathcal{P}(s)}

\newcommand{\Order}[1]{O \left( #1 \right)}
\newcommand{\Action}{\mathcal{F}}

\newcommand{\defemph}[1]{\textbf{#1}}

\newcommand{\secref}[1]{Sec. \ref{#1}}
\newcommand{\figref}[1]{Fig. \ref{#1}}
\renewcommand{\eqref}[1]{Eq. (\ref{#1})}


\newcommand{\lset}[1][L]{{\mathcal{#1}}}
\newcommand{\gspace}[1][\lset]{{\mathbb{H}_{#1}}}
\newcommand{\hamming}[2]{{d_{h}\left({#1},{#2}\right)}}
\newcommand{\mut}[1]{{\Delta_{#1}}}
\newcommand{\rankscape}[1][F]{{\mathcal{R}\left[{#1}\right]}}
\newcommand{\npaths}[2][]{{\#}^{{#1}}_{\text{p}}\left[#2\right]}
\newcommand{\nspaths}[2][]{{\#}^{{#1}}_{\text{dp}}\left[#2\right]}
\newcommand{\npathsg}[1][]{{\#}^{{#1}}_{\text{p}\Omega}}
\newcommand{\nspathsg}[1][]{{\#}^{{#1}}_{\text{dp}\Omega}}
\newcommand{\ppaths}[1]{{\pi}_{\text{p}}\left[#1\right]}
\newcommand{\pspaths}[1]{{\pi}_{\text{dp}}\left[#1\right]}
\newcommand{\ptrav}{{\pi}_{\text{t}}}
\newcommand{\pstrav}{{\pi}_{\text{st}}}
\newcommand{\nopt}{{\#}_{\text{opt}}}
\newcommand{\flsspace}[1][\lset]{{\mathbb{F}_{#1}}}
\newcommand{\proj}[1][{\lset[M]}]{{\downarrow_{#1}}}
\newcommand{\nkstruct}[1][B]{{\mathcal{#1}}}

\newcommand{\Comment}[1]{\par\noindent\colorbox{blue!30}
	{\parbox{\dimexpr\textwidth-2\fboxsep\relax}{Hwang: #1}}}

\section{Introduction}

\subsection{Probabilistic models of fitness landscapes}

Biological evolution can be conceptualized as a search
process in the space of gene sequences guided by the fitness
landscape, a mapping that assigns a measure of reproductive value to
each genotype \cite{Kondrashov2015,Svensson2012,deVisser2014}.
The relationship between genotype and fitness is exceedingly complex, as it is mediated in a highly nonlinear way 
by the multidimensional organismic phenotype that interacts with the environment and thereby determines reproductive
success. A common strategy to deal with this complexity is to shortcut the intermediate phenotypic level by assigning fitness directly to genotypes. This leads to probabilistic
models that define fitness landscapes in terms of ensembles of random
functions on a suitably chosen discrete space \cite{Stadler1999}. 
The idea that unmanageable complexity can be replaced by randomness is familiar from the statistical physics of disordered systems, and there are strong
links between the two fields \cite{Stein1992}.

The prime example of a
genotype space is the Hamming graph $\mathbb{H}_A^L$, the set of all sequences of length $L$ with
symbols taken from an alphabet of size $A$ and equipped with the
Hamming metric which counts the number of symbols in which two
sequences differ. The alphabet size is $A=4$
for nucleotide sequences and $A=20$ for proteins. In the context of
classical genetics $A$ denotes the number of alleles that can
be present at a certain genetic locus. Many studies
including the present one restrict their scope to binary sequences
with $A=2$, where the corresponding binary sequence space $\mathbb{H}_2^L$
is an $L$-dimensional hypercube.

The probabilistic approach was pioneered by Kauffman and Levin
\cite{Kauffman1987}, who considered the conceptually simplest case
where fitness values of different genotypes are drawn independently
from a common probability distribution. With reference to an
earlier publication by Kingman where a similar scheme was introduced in a
setting with an infinite number of alleles \cite{Kingman1978}, the
uncorrelated model is known as the House-of-Cards landscape
(HoC). In the words of Kingman, the rationale behind this term is the
idea that any mutation completely destroys ``\textit{the biochemical `house of
cards' built up by evolution}''. 
The assumption that a single mutation leads to a fitness value for the offspring
that is uncorrelated with the parent is clearly unrealistic, and indeed recent empirical studies have shown that the HoC model
overestimates the ruggedness of real fitness landscapes \cite{Hartl2014,Neidhart2013,Szendro2013,deVisser2014,Weinreich2013}. 
In subsequent work, Kauffman and collaborators therefore
devised a class of fitness landscape models known as NK models in which the correlation between fitness values can be 
tuned \cite{Kauffman1993,Kauffman1989}. The construction of these models was clearly influenced by the concurrent (though somewhat earlier) developments
in the theory of disordered systems \cite{Stein1992}, as evidenced by the frequent references to spin glasses in the original paper \cite{Kauffman1989}.

\subsection{NK models and ruggedness}

In NK fitness landscapes, the fitness is written as a sum of contributions, each of which depends in a HoC-like fashion
on a subset of loci. As a consequence, a mutation at a particular locus changes only the contributions of those subsets
that contain this locus, whereas all other contributions remain unchanged. In this way, the level of fitness correlations
can be controlled through the size and composition of the interacting subsets. In the original formulation of the model, 
the number of subsets is taken to be equal to the number of loci, and each subset is associated to a specific 
locus which it contains together with $K$ others. In later work some of these constraints have been relaxed 
\cite{Altenberg1997,Neidhart2013}, and below in \secref{Sec:Representations} 
we provide a formal definition of the model that 
allows to incorporate various generalizations in a unified way. 
NK fitness landscapes constructed according to the original version of
the model will be referred to as classical. Even within the
set of classical NK landscapes there are obviously many distinct,
deterministic or stochastic schemes by which
loci can be assigned to interacting subsets. This assignement is the
key structural degree of freedom of the NK model, and can be viewed as
a crude representation of genetic architecture. 
For convenience,
our nomenclature differs in two respects from that of the original definitions of Kauffman and coworkers: First, we 
denote the number of loci by $L$ rather than $N$; second, the size of
interacting subsets is denoted by $k = K+1$ throughout this article. 


Since its introduction three decades ago the NK-model has been widely
applied in investigations of fundamental questions of evolutionary
theory \cite{Ohta1997,Oestman2012,Welch2005} as well as for the analysis of empirical fitness
landscapes \cite{Rowe2010}. But also beyond the original context of evolutionary biology, the model provides a remarkably versatile framework for 
exploring how structural constraints give rise to diversity and
complexity in the solution spaces of various optimization
problems. Correspondingly, NK fitness landscapes appear in fields 
ranging from evolutionary computation to management science and
economics \cite{Buzas2014,Levinthal1997,Manukyan2016,Richter2014,Valente2014,Wright2000}.  

Much of the extensive, if somewhat scattered literature has investigated features of NK fitness landscapes that are relevant
to the efficiency of mutational searches, 
particularly the statistics of fitness maxima \cite{Buzas2014,Durrett2003,Evans2002,Limic2004,Weinberger1991}. 
At least under conditions of low mutation
supply where populations explore the landscape through single mutational steps, local fitness maxima present obstacles to the 
search process, and their role in slowing down evolutionary progress has been a concern in evolutionary theory ever since
the fitness landscape concept was first introduced in the 1930's \cite{Haldane1931,Wright1932}. The existence of multiple
fitness peaks is therefore the criterion that is most commonly used to specify what it means for a fitness landscape to 
be rugged \cite{Crona2013,Poelwijk2011,deVisser2014,Whitlock1995}. In the related context of spin glasses, the fitness peaks correspond to 
metastable states \cite{Haldane1931} that govern the low-temperature behavior of these systems \cite{Dean2000,deOliviera1999}. 
  
Recent theoretical and empirical studies have identified alternative measures of fitness landscape ruggedness that focus 
on the mutational pathways along which local or global fitness peaks can be reached
\cite{Carneiro2010,Poelwijk2007,Szendro2013,deVisser2014,Weinreich2005}. Under conditions of low mutation supply and large fitness differences,
mutational pathways are accessible to the evolving population only if fitness increases monotonically along the path, a condition
that often strongly reduces the combinatorial abundance of possible evolutionary trajectories implied by the high connectivity
of genotype space \cite{Berestycki2016,Franke2011,Franke2012,Hartl2014,Hegarty2014,Nowak2013,Weinreich2006}. 
The basic evolutionary
dynamics in this regime is captured by adaptive walk models,
in which a genetically homogeneous (monomorphic) population moves
towards higher fitness along the network of accessible
pathways in single mutational steps 
\cite{Gillespie1984,Kauffman1987,Macken1989,Nowak2015,Orr2002}. Adaptive walks terminate
at local fitness maxima, and the number of steps required to
reach a maximum from a random starting point is a convenient measure
of landscape ruggedness. At least in order of magnitude, the
length of adaptive walks is expected to be comparable to the typical
distance between maxima and also to the correlation length of the
fitness correlation function \cite{Nowak2015,Reidys2002,Stadler1996,Stadler1999,Weinberger1991}. 

\subsection{Aims and scope}
\label{Outline}

In this article we review our current understanding of how the
ruggedness of NK fitness landscapes, as quantified by the number of
fitness peaks, the number of accessible paths and the length of
adaptive walks, depends on 
the parameters of the landscape. These parameters comprise the following elements:

\begin{itemize}

\item the number of loci $L$ and the size $k$ of interacting groups of loci;

\item the scheme according to which loci are assigned to groups,
  henceforth referred to as the NK structure of the
  model; and 

\item the probability distribution from which the fitness values
  assigned to the configurations of the interacting groups are drawn.

\end{itemize}
Early numerical work on NK landscapes suggested 
that the number of fitness peaks and the
length of adaptive walks is determined primarily by the parameters $k$
and $L$, with little or no dependence on the NK structure. The two specific
structures considered were the adjacent neighborhood model
  (AN), where the loci belonging to the same interacting subset are adjacent along the
sequence, and the random neighborhood model (RN) where the
members of each group are chosen at random among all loci. Based on
simulations of these two models Kauffman writes that \textit{``whether the $K$ epistatic inputs
  to a gene are its neighbors or random among the $N$ has almost no
  bearing on the lengths of walks to optima''} \cite{Kauffman1993}.
Weinberger concluded from an approximate analytic investigation that \textit{``the topography of
  $N-k$ landscapes seems to be independent of how the neighborhoods
  are chosen''} \cite{Weinberger1991}, and similar statements can still be
found in the current literature \cite{Tomassini2008}. Some support for this hypothesis derives from the fact that the fitness
correlation function of classical NK-landscapes has a
universal form that is completely specified by $k$ and $L$
\cite{Campos2002,Campos2003,Neidhart2013}. 

On the other hand, recent numerical simulations of
accessible pathways and adaptive walks revealed significant differences between different
NK structures \cite{Nowak2015,Schmiegelt2014}, and a survey of earlier work
suggested that similar (if less pronounced) differences exist also
with regard to the statistics of fitness peaks. The
handful of available exact results for the asymptotic growth rate of the number
of maxima with $L$ display a distinct dependence on the fitness
distribution which gives way to universal behavior only when $k$ is
large \cite{Durrett2003,Evans2002,Limic2004,Nowak2015}. Since these rigorous
analyses were restricted to the AN model, no 
conclusions could be drawn with regard to the dependence on the
interaction structure. The latter was addressed numerically by Buzas and
Dinitz, who found a correlation between the number of fitness peaks and the
rank of the structure \cite{Buzas2014,Nowak2015}. The rank is a measure of the
connectivity of the genetic architecture that will be formally defined
below in \secref{sec:NKModel}.

In the next section we introduce the mathematical framework needed to
define the quantities and models of interest. We then embark on a
detailed investigation of the mean number of local fitness maxima in
NK landscapes, focusing specifically on the exponential growth rate
$\lambda_k$ of this quantity for large $L$. Starting from two exactly solvable cases,
the block model (BN) where the interacting subsets are disjoint
\cite{Orr2006,Perelson1995,Schmiegelt2014} and a novel mean
  field model (MF)  where all possible subsets
contribute to the fitness landscape with equal weights, we identify
two classes of NK structures characterized by distinct asymptotic
behaviors of $\lambda_k$ for large $k$. We systematize and expand
the range of exact expressions that have been reported for $\lambda_k$
for the AN model, which is known from previous work to share the
asymptotic behavior of the BN model. We then demonstrate that the RN
model falls into the class of the MF model, thus establishing that AN
and RN structures are \textit{not} equivalent even for large $k$. 

In Sections 4
and 5 we review the state of the art regarding accessible pathways and
adaptive walks in NK fitness landscapes, and sketch a proof 
of the asymptotic absence of globally accessible pathways for a large class of NK structures.
Finally, Section 6 summarizes the
paper and provides an outlook on open problems. 
Some derivations and proofs and a description of the numerical
algorithm used to count the number of fitness maxima are relegated to
Appendices. 

\section{Mathematical background and definitions}
\label{Sec:Representations}

\subsection{Genotype space}

We assume that the genome of an individual consists of a fixed number $L$ of independently mutable loci labeled by an index set $\lset = \{l_1,\ldots ,l_L\}$, called the \defemph{locus set}.
Generally each locus could be found in many different states, or \defemph{alleles}.
For simplicity it is usually assumed that each locus can be found in the same number $A$ of states labeled $\{a_1,\ldots,a_A\}$.
Here however, as mentioned in the introduction, we focus on the case $A = 2$, choosing $a_1 = 1$ and $a_2 = -1$  as the only possible alleles at each locus.
These may e.g.\ be interpreted as the wild type and a mutated type.
A genotype corresponds to an assignment of alleles to each locus, or equivalently (assuming an ordering of $\lset$) a sequence of alleles, i.e.\ for $A=2$ a sequence of $L$ binary values $\sigma = \{\sigma_{l_1}, \ldots \sigma_{l_L}\} \in \{\pm 1\}^L$.
The space of all genotypes will be denoted $\mathbb{H}_\mathcal{L}$.
Taking the genotype space as a vertex set for a simple undirected graph and drawing edges between any two genotypes differing at exactly one locus, we arrive at the 
\defemph{Hamming graph} $\mathbb{H}^L_2$, the $L$-fold Cartesian graph product of the complete graph on two vertices.
For higher values of $A$, the resulting graph would have been the Hamming graph $\mathbb{H}^L_A$, the $L$-fold Cartesian graph product of the complete graph on $A$ vertices.
This mutation graph defines all possible changes in genotypes due to single point mutations.
While it is possible for an offspring to accumulate multiple point mutations relative to its parent, if the mutation rate is small in comparison to the inverse of the product of 
$L$ and the population size, then double mutants are unlikely to appear and an asexual population may only explore the genotype space by single steps along the Hamming graph.
In this regime, the Hamming graph is indeed the graph of all possible mutational transitions.

The graph metric of the Hamming graph is the \defemph{Hamming metric} 
\begin{equation}
d_h(\sigma,\theta) = \sum_{i=1}^L
\left(1-\delta_{\sigma_i\theta_i}\right) 
\end{equation}
measuring the number of loci at which two genotypes differ and thereby the minimal number of mutational steps needed to be taken to reach one from the other.
We define the operators $\Delta_l:\gspace\rightarrow\gspace$ for all
$l \in \lset$ such that 
\begin{equation}
\label{def:delta}
(\Delta_l \sigma)_m = \left(1-2\delta_{lm}\right) \sigma_m.
\end{equation}
This (single-locus) mutation operator switches the allele at the $l$-th locus of a genotype, corresponding to one edge attached to $\sigma$ in the Hamming graph.
These operators are then extended to (multi-locus) mutation operators $\mut{\lset[M]}$ for all $\lset[M] \subseteq \lset$, such that $\mut{\lset[M]} = \prod_{l\in\lset[M]} \mut{l}$.
Because loci are mutationally independent the order of operations in the product does not matter and all mutation operators commute.
Furthermore mutation operators are self-inverse and form a group that leaves the metric invariant,
\begin{equation}
    d_h(\sigma, \theta) = d_h(\mut{\lset[M]}\sigma, \mut{\lset[M]}\theta),
\end{equation}
and
\begin{equation}
    d_h(\sigma,\mut{\lset[M]}\sigma) = |\lset[M]|.
\end{equation}
The maximal distance between two genotypes on the Hamming graph is $L$.
For each genotype $\sigma$, there is exactly one genotype at this distance, the \defemph{antipode} $\mut{\lset}\sigma$.
If two genotypes share an edge in the mutation graph, or equivalently lie at Hamming distance $1$, then we say they are adjacent.
A sequence of adjacent genotypes $(\sigma^{(0)},\ldots,\sigma^{(n)})$ is called a \defemph{(mutational) path(-way)}.
Here $\sigma^{(0)}$ is the initial genotype and $\sigma^{(n)}$ the final genotype, and 
$n$ is the path length.
Each path may also be expressed as an initial genotype together with a sequence of $n$ loci $(m_1,\ldots,m_n) \in \lset^n$, so that $\sigma^{(i)} = \mut{m_i}\sigma^{(i-1)}$.
Here we require, if not mentioned otherwise, paths to be simple.
This means paths may not visit any genotype more than once.
We apply this constraint because accessible pathways, which will be discussed in more detail in \secref{Sec:AccessiblePathways}, are strictly fitness increasing and thus can never loop back to a previous genotype.


\subsection{Fitness landscapes}

A \defemph{fitness landscape} is a mapping $\funcsig{F}{\gspace}{\reals}$ assigning each genotype a real-valued fitness.
Starting from an initial genotype $\sigma$, a mutation $\mut{\lset[M]}$ induces a fitness change which we will write in the shorthand notation 
\begin{equation}
\label{def:deltaf}
\mut{\lset[M]}F(\sigma) = F(\mut{\lset[M]}\sigma) - F(\sigma).
\end{equation}
The operator $\mut{\lset[M]}$ may be understood here as a difference
operator mapping a fitness landscape to a function which assigns to each genotype the selection coefficient associated with application of the set of mutations $\lset[M]$.

Asexual populations may be viewed as distributions on the genotype space.
Due to selection these distributions typically tend to move towards higher fitness and stagnate at local fitness maxima of the fitness landscape.
Mutation and genetic drift introduce noise resulting in distributions of finite width (in terms of genotype distance).
If selection is significantly outweighing the mutational input, then this width will be very small and populations are effectively localized at exactly one majority genotype.
Over time mutations will occur, which, due to strong selection, will fixate to become the new majority genotype if and only if they increase fitness.
The resulting dynamics is that of an \defemph{adaptive walk}, a time-
and space-discrete Markov process over the genotype space, where the
population moves stepwise in the direction of strictly increasing fitness (see \secref{Sec:AdaptiveWalks}). 

To describe not the actual probabilities, but rather only the possibility of such a walk taking certain mutational paths, it is useful to introduce the reduced notion of a \defemph{fitness graph}.
The fitness graph of a fitness landscape is the orientation of the mutation graph $\gspace$, such that arrows point towards higher fitness \cite{Crona2013,Franke2011,deVisser2009}.
For convenience we will assume that no two genotypes have exactly the same fitness, i.e.\ $\mut{\lset[M]}F(\sigma) \neq 0$ for all $\sigma$ and $\mut{\lset[M]}$.
Then the fitness graph is well-defined and acyclic (see \figref{FitnessGraphs} for some simple examples).

The fitness graph contains only information about signs of local mutation effects and as such may not convey enough information about the original fitness landscape.
For example local maxima can be identified from the fitness graph, but the global one cannot be determined.
As an intermediate reduction one may consider only ranks of fitness values:
The \defemph{ranked fitness landscape} $\rankscape$ of a fitness landscape $F$ is again a fitness landscape, such that $\rankscape(\sigma)$ is the rank of $F(\sigma)$ if all $2^L$ fitness values are ordered in ascending order \cite{Crona2017}.
The ranked fitness landscape's fitness graph is the same as that of the original landscape.

Despite recent progress in the large-scale analysis of empirical fitness landscapes \cite{Bank2016,Kouyos2012,Pokusaeva2017},
most available data sets are restricted to small numbers
of loci \cite{Hartl2014,Szendro2013,deVisser2014,Weinreich2013}, and measuring fitness landscapes
on a genome-wide level remains an insurmountable challenge.
We also cannot hope to describe specific landscapes exactly from their underlying biological and chemical structure.
Thus the approach taken is to consider probabilistic models of fitness landscapes, based on theoretical or empirical principles, to describe typical properties of such landscapes.
Let $\flsspace = \reals^{\gspace}$ be the space of all fitness landscapes over the locus set $\lset$.
Then a \defemph{fitness landscape model} is a probability measure over $\flsspace$.

Several such models have been studied.
From a mathematical viewpoint, the simplest non-trivial model is probably the \defemph{House-of-Cards (HoC)}  \defemph{model} \cite{Kauffman1987}.
In this model all fitness values $\{F(\sigma)\}_{\sigma\in\gspace}$ are chosen i.i.d.\ from some continuous real-valued \defemph{base fitness distribution} $p_f$.
Continuity guarantees that almost surely no two fitness values are equal.
The HoC model's ranked fitness landscape is independent of the actual choice of $p_f$, reducing the calculation of ranked properties, such as the number of local maxima, to combinatorial problems.

The HoC model, however, does not allow for correlations between mutational effects on the same locus and thus lacks a structure on loci.
One possible (though most extreme way) of associating fitness benefits with certain alleles at specific loci is to assign fitness values $f_l(\sigma_l)$ to each allele of each locus and define the total fitness as
\begin{equation}
\label{linear_fitness}
    F(\sigma) = \sum_{l\in\lset} f_l(\sigma_l).
\end{equation}
If the values of $f_l$ are chosen i.i.d.\ from a continuous probability distribution, then $f_l$ is effectively a HoC landscape over one locus.
This \defemph{linear model} is the opposite extreme of the HoC landscape.
Given the fitness difference between two alleles on one background, the fitness effect on every other background is identical, i.e.\ $\Delta_l F(\sigma) = \Delta_l f_l(\sigma_l)$ depends only on $\sigma_l$.

A canonical way of quantifying the degree of correlation in a fitness landscape model is through 
the \defemph{distance correlation function} $\rho(d)$ defined as \cite{Stadler1996,Stadler1999}
\begin{equation}
    \label{DistanceCorrelationFunction}
    \rho(d) = \frac{\mathbb{E}_{d_h(\sigma,\theta)=d}\left[F(\sigma)F(\theta)\right] - \mathbb{E}_\sigma\left[F(\sigma)\right]^2}{\mathbb{E}_\sigma\left[F(\sigma)^2\right] - \mathbb{E}_\sigma\left[F(\sigma)\right]^2},
\end{equation}
where the one-point expectations are taken over all $\sigma\in\gspace$ and the two-point expectations 
over all combinations of $\sigma\in\gspace$ and $\theta\in\gspace$ such that their Hamming distance is exactly $d$. For the HoC model $\rho(d) = \delta_{d,0}$, whereas for the linear
model $\rho(d) = 1 - d/L$.

\subsection{Epistasis}

The linear model is \defemph{non-epistatic}, meaning that each mutation has a fixed effect on overall fitness, independent of the states of other loci.
In contrast \defemph{epistasis} refers to the dependence of mutational effects on the state of other loci
\cite{Ferretti2016,Philipps2008,deVisser2011}.
Formally we say that two loci $l$ and $m$ are epistatic (for a genotype $\sigma$), if
\begin{equation}
    \mut{l}F(\sigma) \neq \mut{l}F(\mut{m}\sigma).
\end{equation}
It is useful to further differentiate \defemph{magnitude} and \defemph{sign} epistasis \cite{Weinreich2005}.
Sign epistasis is present if the equation above also hold after application of the sign function on both sides, i.e.\ if
\begin{equation}
    \sgn\mut{l}F(\sigma) \neq \sgn\mut{l}F(\mut{m}\sigma).
\end{equation}
In this case mutations on $m$ can affect whether mutations on $l$ are beneficial or not.
If sign epistasis is not present, then there is only magnitude epistasis, in which $m$ can affect the quantitative benefit of a mutation on $l$, but cannot change it from beneficial to deleterious.
In this case it is easy to show that the fitness landscape has a unique maximum \cite{Weinreich2005}.
Note that $l$ is epistatic with $m$ if $m$ is epistatic with $l$, but the same is not true for sign epistasis.
If however $l$ is sign epistatically dependent on $m$, as well as the other way around, then one speaks of \defemph{reciprocal sign epistasis} \cite{Poelwijk2007,Poelwijk2011}, see 
\secref{Sec:AccessiblePathways} for further discussion.

An alternative description of epistasis as function of distance on the
hypercube is provided by the $\gamma$ statistic introduced in \cite{Ferretti2016}.
For a given focal mutation $l$ and a set of mutations $\lset[M]$, it is defined as the correlation between fitness effects of parallel transported arrows in the fitness graph,
\begin{equation}
   \gamma_{l,\mathcal{M}} = \frac{\text{Cov}\left[\Delta_l F(\sigma),\Delta_l F\left(\Delta_{\mathcal{M}}\sigma\right)\right]}{\mathbb{E}\left[\left(\Delta_lF(\sigma)\right)^2\right]}
\end{equation}
where the mean and covariance are taken over all (or a subset of) genotypes $\sigma$.
For the case when $\mathcal{M}$ consists of a single locus 
$\mathcal{M} = \{m\}$, $\gamma_{l,m}$ quantifies the average strength of epistasis on mutation $\Delta_l$ due to prior application of mutation $\Delta_m$.
Different values of $\gamma_{l,m}$ indicate the prevalence of no, magnitude-only, sign- or reciprocal epistasis for $\gamma_{l,m} = 1$, $1 > \gamma_{l,m} > 0$, $1 > \gamma_{l,m} > -\frac{1}{3}$ and
$\gamma_{l,m} < 0$, respectively.

\subsection{Fourier-Walsh decomposition}

Being functions over a finite commutative group, fitness landscapes admit a Fourier decomposition of the form \cite{Reidys2002,Stadler1999,Weinberger1991a}
\begin{equation}
\label{Fourier}
F(\sigma) = \sum_{g \in \wp(\lset)} \hat{F} (g)\prod_{l \in g} \sigma_l,
\end{equation}
where $\wp$ denotes the power set and the $\hat{F}(g)$ are 
\defemph{Fourier coefficients}. As there are $2^L$ subsets of $\lset$, the mapping between the fitness values $F(\sigma)$ and the 
Fourier coefficients $\hat{F}(g)$ is one-to-one and invertible. The decomposition \eqref{Fourier} is an expansion in eigenfunctions of the graph Laplacian of the hypercube, which is also known
as a Walsh transform in computer science \cite{Weinreich2013}.

The linear fitness landscape \eqref{linear_fitness} is a special case of \eqref{Fourier} where the $\hat{F}(g)$ are  
nonzero only when $g$ is the empty set  or a single
locus. Correspondingly, terms containing products of $p \geq 2$ locus
contributions encode epistatic interactions of order $p$. Specifically, $\hat{F}(g)$ is proportional to the $\vert g \vert$-way
epistasis among the loci in the subset $g$ averaged over all genetic backgrounds \cite{Poelwijk2016}. 
The  \defemph{Fourier spectrum} of a fitness landscape is obtained by summing the squares of the Fourier coefficients for each order $p$, which provides a measure for the strength of 
epistasis of different orders \cite{Neidhart2013,Weinreich2013}. Note, however, that the presence or absence of sign epistasis depends on the specific values of the coefficients $\hat{F}(g)$ and 
cannot be read off from the Fourier spectrum. The Fourier spectrum is related to the distance correlation function \eqref{DistanceCorrelationFunction} 
through a one-dimensional linear mapping involving discrete orthogonal polynomials \cite{Stadler1999}.

\subsection{Local maxima}
\label{Sec:RepsLocalMaxima}

A \defemph{local fitness maximum} is a genotype $\sigma$, such that all single-locus mutations have lower fitness than $\sigma$, i.e.\ such that $\mut{l}F(\sigma) < 0$ for all $l\in\lset$.
Thus a local maximum is a sink in the fitness graph.
Different concepts of local maxima may be used, e.g. one could require $\mut{l}F(\sigma) < -\epsilon$ for some $\epsilon > 0$ as to limit the definition to more selectively robust maxima.
Since one can have double mutants for sufficiently large mutation rate, it may also be of interest to consider maxima which are robust up to higher distance, i.e. $\sigma$ with $\mut{\lset[M]}F(\sigma) < 0$ for all $\lset[M]\subset\lset$ such that $|\lset[M]| \leq D$, where $D$ is the number of simultaneous mutations considered.
Here we will only consider the simple first definition.

We will denote the expected number of local maxima as $\nmax$, possibly with an index describing the model.
There are $2^L$ genotypes and thus the fraction of genotypes expected to be local maxima can be written $\pi_{\text{max}} = 2^{-L}\nmax$.
Provided the fitness landscape model of interest is \defemph{homogeneous}, in the sense that all genotypes
are statistically equivalent, 
$\pi_{\text{max}}$ is also the probability that a randomly chosen genotype is a local maximum.
We will use this in \secref{Sec:LocalFitnessMaxima} to study the expected number of local maxima.
Two examples of fitness landscape models that are not homogeneous can be found in 
\cite{Hwang2017,Neidhart2014}. 

\subsection{NK model}
\label{sec:NKModel}

Both the HoC and the linear model are extreme cases.
Realistically we expect some intermediate structure with some ruggedness but still correlated mutation effects.
The idea of Kauffman's NK model \cite{Kauffman1993,Kauffman1989} is to introduce a parameter $k$ to the system, which is able to interpolate between the HoC and the linear model.
The model is constructed starting from the linear model \eqref{linear_fitness}. 
However each fitness contribution $f_l$ is now not only dependent on
$\sigma_l$, but also on the states of an additional set of $k-1$ other loci.
The concrete choice of these additional loci may vary and will be discussed later.
The fitness values of the fitness contributions $f_l$, now functions of $k$ alleles, are then assumed to be randomly distributed in accordance with the HoC model.
In this way $f_l$ can still be interpreted as the fitness contribution of locus $l$, but now being dependent on a few other locus states.
At $k=1$, there are no additional locus dependencies and the linear model is retrieved.
For $k=L$, each $f_l$ must necessarily be a HoC landscape over all of $\lset$ and thus $F(\sigma)$ is itself a HoC landscape.
Intermediate values of $k$ are able to interpolate between these cases or between different amounts of ruggedness.

\begin{figure}
   \centering
   \includegraphics[width=0.4\textwidth]{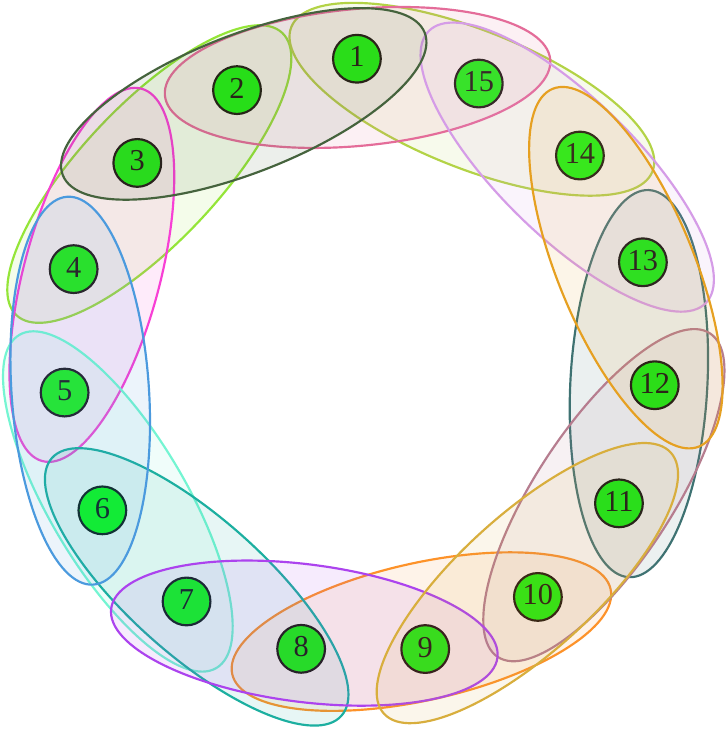}
   \hfill\includegraphics[width=0.4\textwidth]{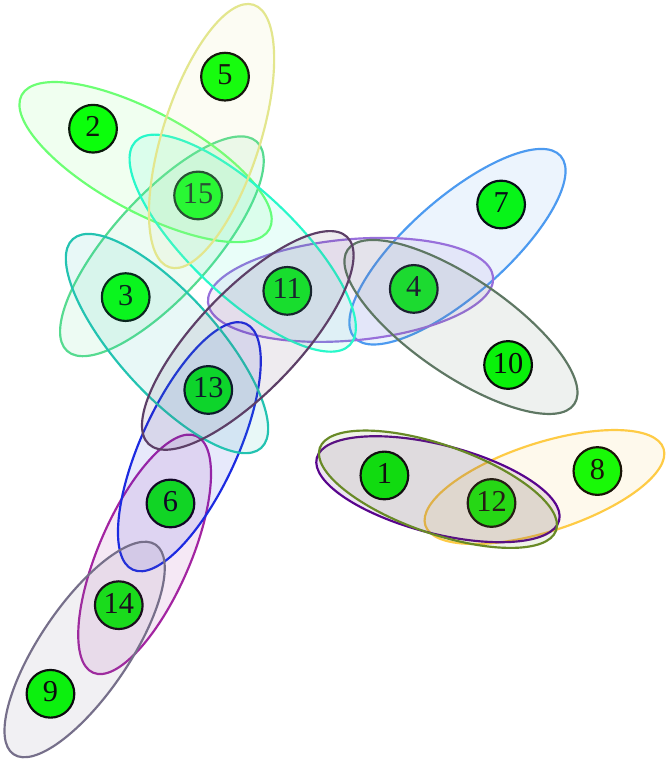}\\
   \includegraphics[width=0.4\textwidth]{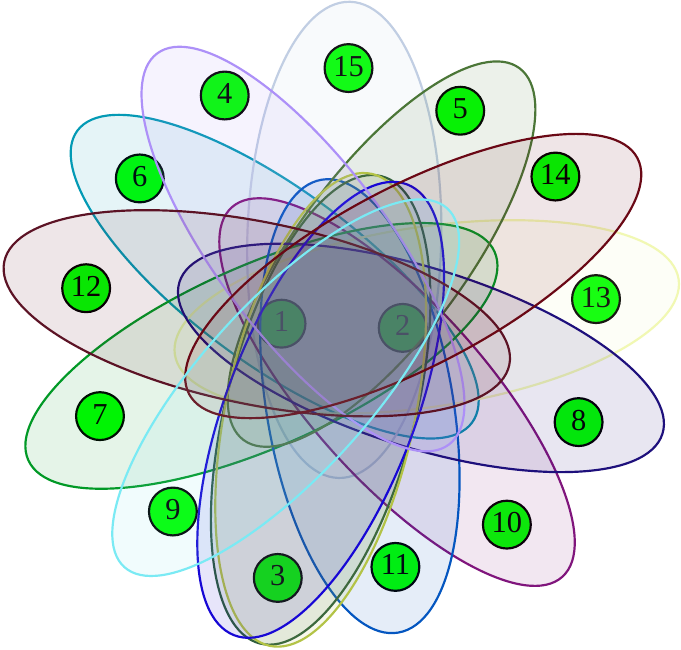}
   \hfill\includegraphics[width=0.35\textwidth]{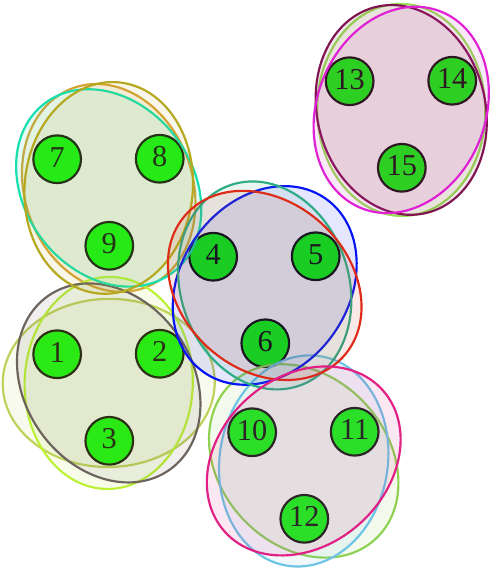}
   \caption{\label{fig:NKhypergraphs}Examples of NK structure hypergraphs. Each ellipse represents one NK edge
and nodes represent different loci. Top left: AN structure
   with $k=3$ and $L=15$. Top right: uRN structure with $k=2$ and
   $L=15$. Bottom left: SN structure with $k=3$ and $L=15$. Bottom right: BN structure
with $k=3$ and $L=15$. We choose $k=2$ for the uRN model for the sake of readability. For higher $k$ uRN structures
are usually not planar anymore.}
\end{figure}

We will however define a generalization of the NK model first.
The \defemph{(generalized) NK model} over a locus set $\lset$ is parametrized by a multiset $\nkstruct$ containing subsets of $\lset$.
This multiset can be interpreted as the edge set of a (multi-)hypergraph over the set of loci $\lset$ (\figref{fig:NKhypergraphs}).
We call this hypergraph the \defemph{NK structure (hypergraph)} and its edge sets (the elements of $\nkstruct$) \defemph{NK edges}, \defemph{NK blocks} or \defemph{NK neighborhoods}.
By $|\nkstruct|$ we denote the total number of elements (multiplicities included) of $\nkstruct$ and we index the NK edges (in some fixed manner) by natural numbers $\{1,\ldots,|\nkstruct|\}$, i.e.\ $B_1\ldots B_{|\nkstruct|}$.
Then we assign to edge $B_i$ a HoC landscape $f_i$ over $\gspace[B_i]$, i.e.\ a completely random landscape over a subset of loci.
Finally the total fitness is defined as
\begin{equation}
\label{FitnessNK}
    F(\sigma) = \suml_{i=1}^{|\nkstruct|} f_i\left(\proj[B_i]\sigma\right).
\end{equation}
Here $\proj[B_i]\sigma$ is the projection of $\sigma$ onto the subset of loci $B_i$, i.e.\ $\proj[{\lset[M]}]:\gspace\rightarrow\gspace[{\lset[M]}]$ such that $\left(\proj[{\lset[M]}]\sigma\right)_m = \sigma_m$ for all $m\in\lset[M]$.
The projection of a genotype onto $\lset[M]$ retains all alleles at loci in $\lset[M]$, but discards all other loci in $\lset\setminus\lset[M]$.
The orthogonal projection $\proj[{\lset[L]\setminus\lset[M]}]$ yields those alleles that have been discarded by $\proj[{\lset[M]}]$, and 
$\proj[{\lset\setminus\lset[M]}]\sigma$ is called the background genotype of $\sigma$ relative to the projection onto $\lset[M]$.
The union (in the sense of relations) of the two orthogonal projections returns the original genotype. 
Consider for example a locus set $\lset = \{l_1,l_2,l_3,l_4\}$ and an NK edge $\{l_2, l_3\} \subseteq \lset$.
The projection of genotype $(-1,-1,1,1)$ onto the edge is then $(-1,1)$ (assuming ordering as above).

Partially in order to avoid certain inconvenient edge cases we make the following restrictions on the NK structure:
\begin{enumerate}
 \item
 For every $l\in\lset$ there exists a $B_i$ with $l\in B_i$.
 This assures that there are no neutral mutations and that no two fitness values are equal, almost surely.
 \item
 $\frac{1}{|\nkstruct|}\suml_{i=1}^{|\nkstruct|} |B_i| = k$, where $k$ is a constant generalizing the parameter $k = K+1$ in the original NK model.
\end{enumerate}
Together they imply that $|\nkstruct|k \geq L$.

There are obviously many possible choices of the interactions, however some specific further conditions are of interest.
First note that the partial landscape $f_i$ only contributes to a mutation effect $\Delta_lF(\sigma)$ if $l \in B_i$.
The effects on those partial landscapes are all identical and independent and thus we have
\begin{equation}
    \mean{\Delta_l F(\sigma)} = 0
\end{equation}
and
\begin{equation}
    \var{\Delta_l F(\sigma)} = 2\sigma_f \cdot |\{B_i\in\nkstruct | l\in B_i\}|
\end{equation}
    where $\sigma_f$ is the variance of the base fitness distribution and the second term counts the number of NK edges containing $l$.
The distribution of this variance over loci is important to the behavior of the model.
In the most extreme case the variance of few a loci may be on the order of $L$, while other loci are contained only in one NK edge each.
Then the high-variance loci will mostly determine the fitness of a genotype, while the other loci only introduce slight variations.
Such a high-variance locus would be largely independent of the state of other loci. An example for this kind of structure will be 
introduced below in \secref{SpecificStructures}. 
In contrast, if each locus appears in an equal number of NK edges, all loci have equal-variance effects and none is special.
We call such a structure \defemph{regular}.
Due to the definition of $k$, the common number of NK edges containing a specific locus is then $k|\nkstruct|/L$.

We say an NK structure is \defemph{uniform} if $|B_i| = k$ for all $B_i$.
This is equivalent to the hypergraph being $k$-uniform. For uniform structures the Fourier decomposition \eqref{Fourier} contains products
of locus variables up to order $k$ only.

\begin{figure}
   \centering
   \includegraphics[width=0.4\textwidth]{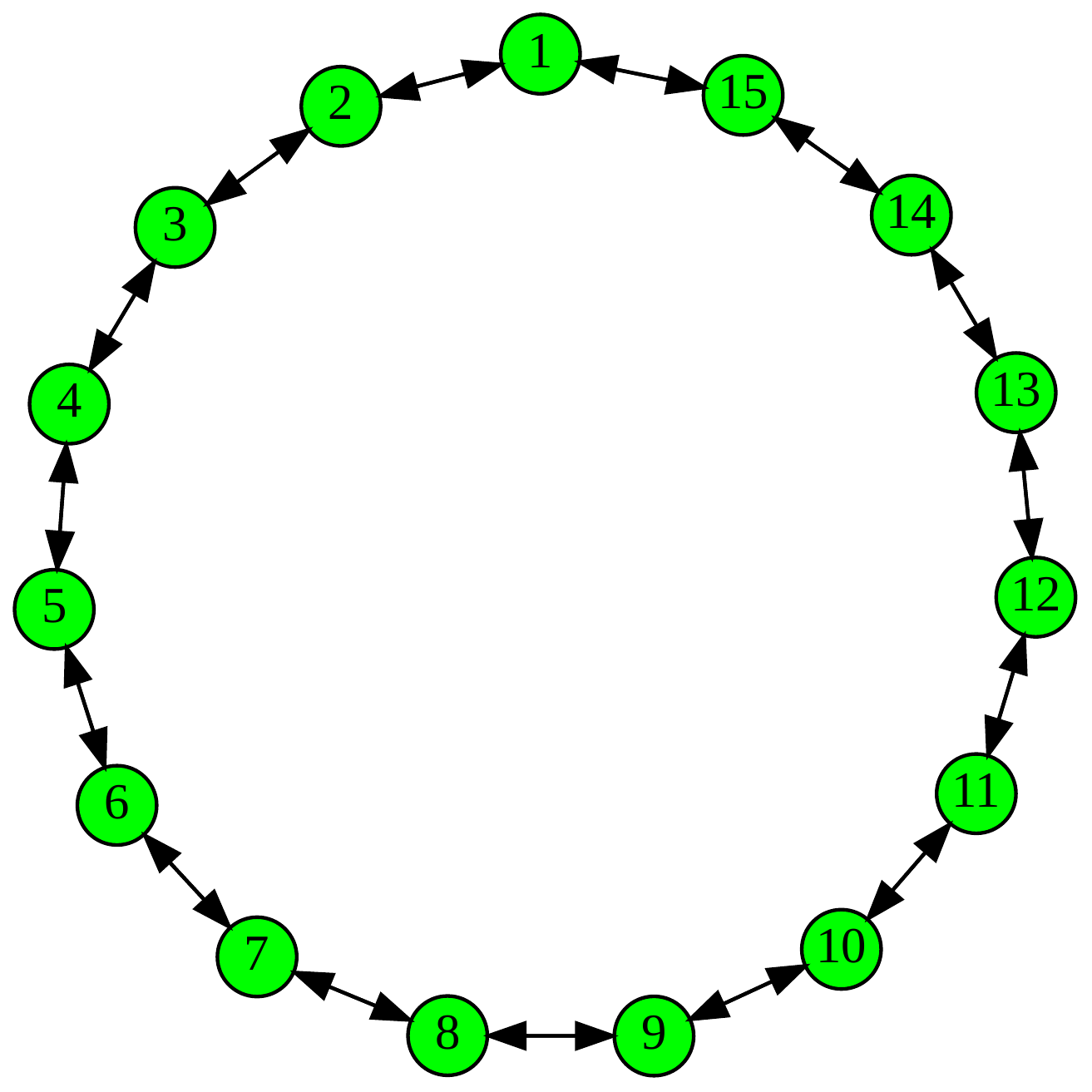}
   \hfill\includegraphics[width=0.4\textwidth]{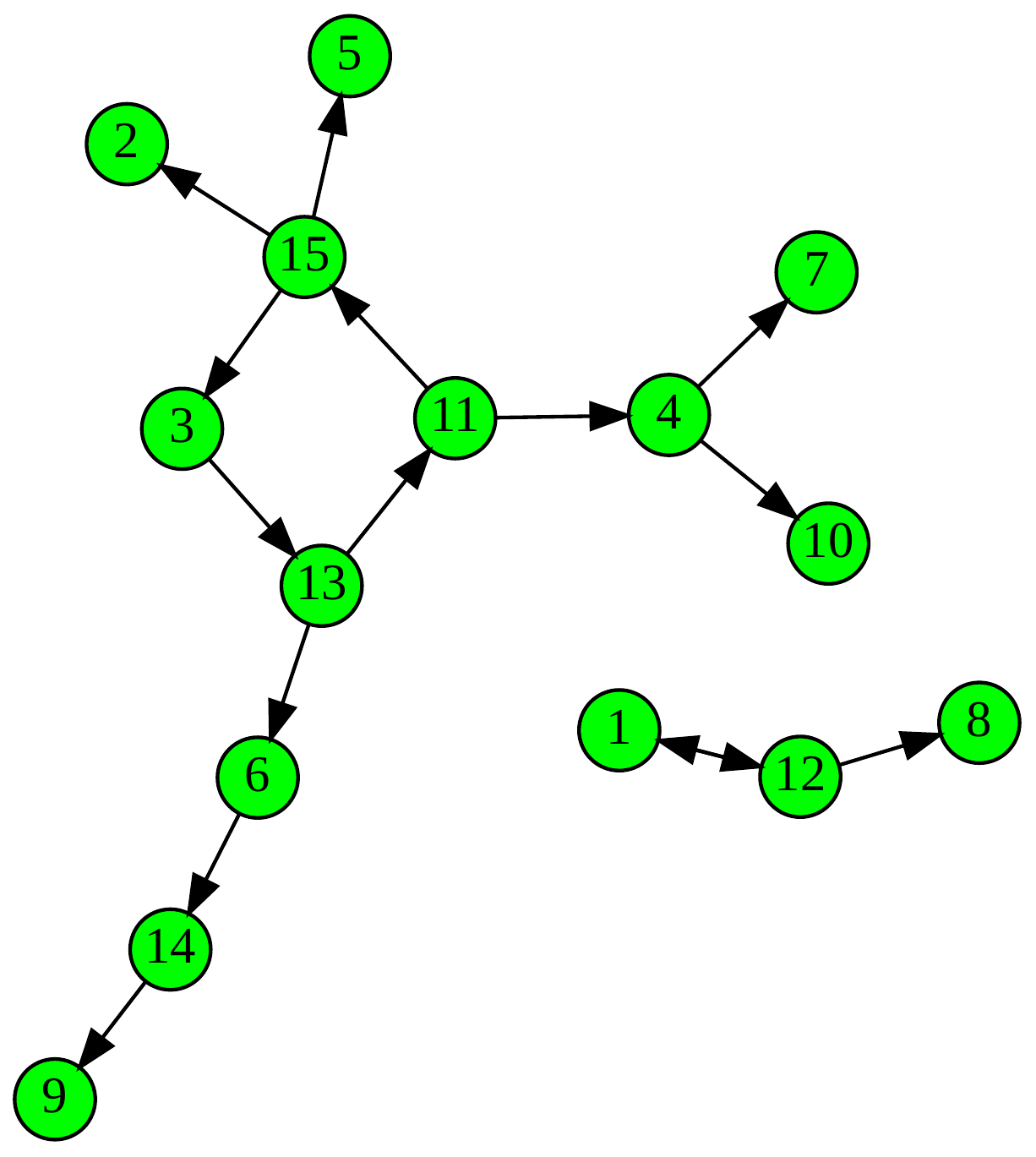}
   \\ \includegraphics[width=0.4\textwidth]{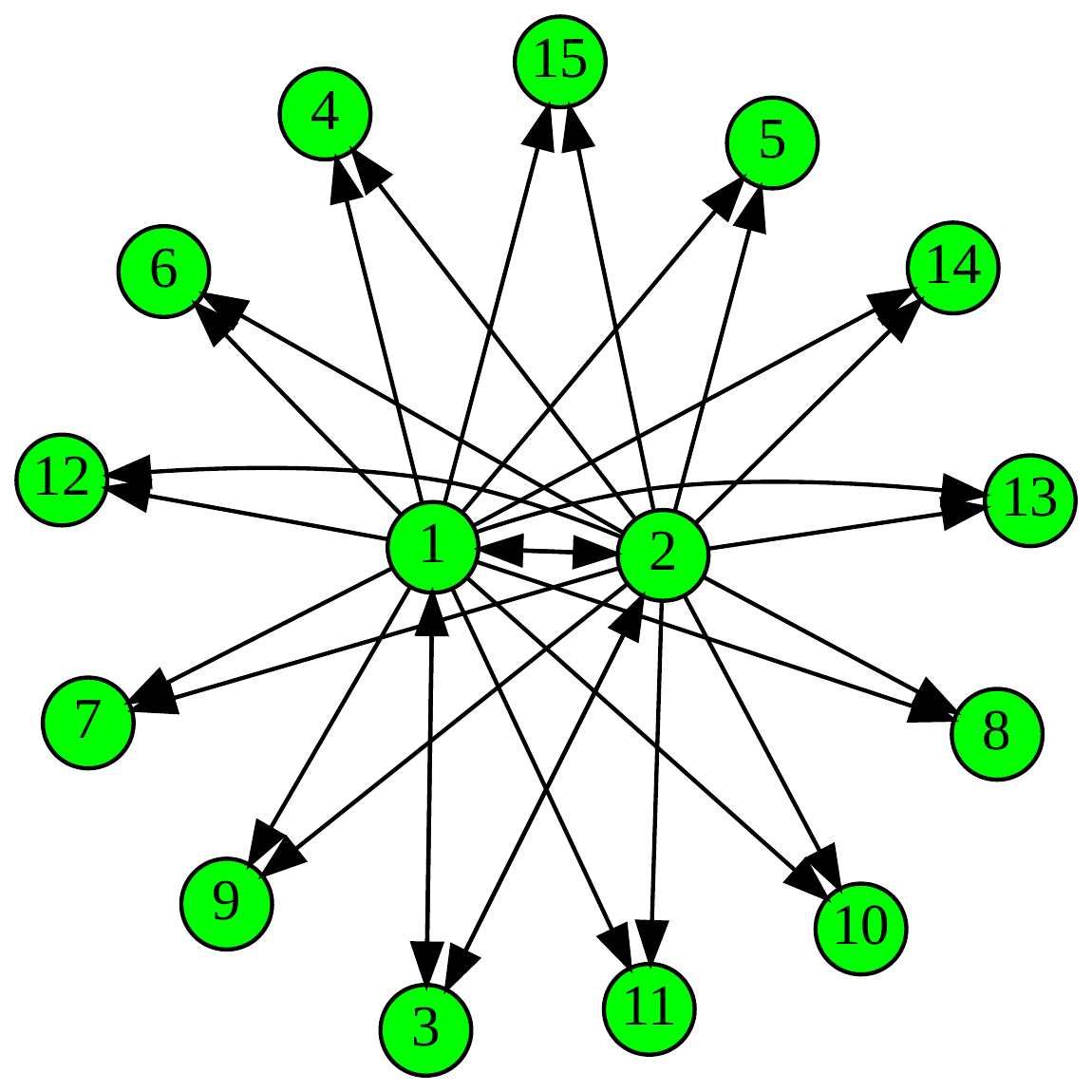}
   \hfill\includegraphics[width=0.35\textwidth]{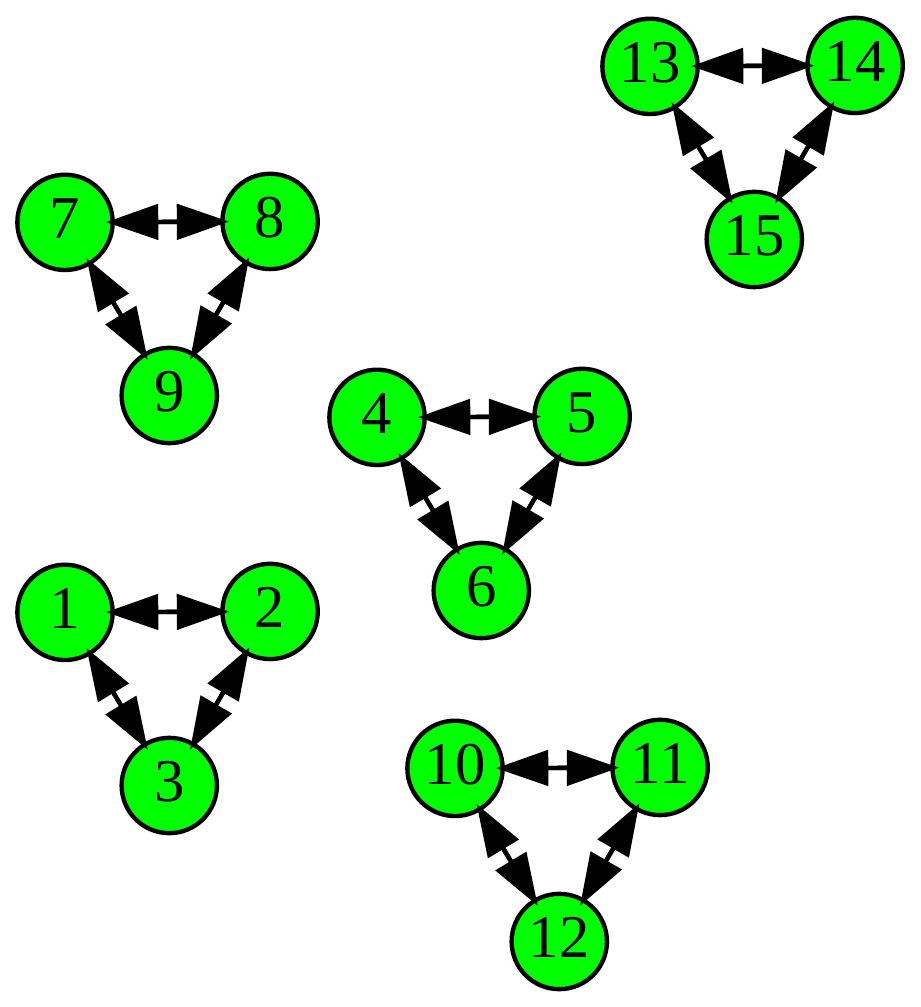}
   \caption{\label{fig:NKgraphs}Simplified NK structures for the examples shown in Fig.
\ref{fig:NKhypergraphs}. Nodes represent loci and an arrow from one locus to another implies that the fitness
contribution of the destination is dependent on the
   state of the source. In addition all locus contributions are dependent on their own state, but the resulting
mandatory loops are not depicted. With $k=2$ the uRN structure forms components consisting of a single cycle
with tree appendages.} 
\end{figure}

We say a uniform NK structure is \defemph{classical} if $|\nkstruct| = L$ and $l\in B_l$ for all $l\in\lset$.
This definition encompasses the class of NK models considered in the original articles by Kauffman et al. \cite{Kauffman1993,Kauffman1989}.
This subclass has nice properties which allow for a simpler graphical representation of the NK structure:
the \defemph{simplified NK structure (graph)} is the directed simple graph over $\lset$ with an arrow from $l$ to $m$ if $l\in B_m$ (\figref{fig:NKgraphs}).
Campos et. al. \cite{Campos2002,Campos2003} show that the distance correlation function is independent of the concrete structure choice for classical structures, and takes the universal form\footnote{Note that incorrect expressions for $\rho(d)$ appear in some of the literature preceding \cite{Campos2003}.}
\begin{equation}
    \label{NKDistanceCorrelation}
    \rho(d) = \frac{(L-k)!(L-d)!}{L!(L-k-d)!}.
\end{equation}
The corresponding Fourier spectrum was computed in \cite{Neidhart2013}.

It is sometimes useful to consider the  \defemph{incidence matrix} of the structure hypergraph, i.e.\ the matrix with elements $b_{l,r} \in \{0,1\}$, where $l\in\lset$ and $r \in\{1\ldots|\nkstruct|\}$ and $b_{l,r} = 1$ if and only if $l\in B_r$.  As a measure of the structuredness of an NK structure the \defemph{rank} defined as 
\begin{equation}
\label{rankdef}
r(\nkstruct) \equiv   \left|\bigcup_{i=1}^{|\nkstruct|} \wp(B_i)\right|
\end{equation}
has been introduced \cite{Buzas2014,Nowak2015}. It is equal to the
number of nonzero coefficients in the Fourier expansion
\eqref{Fourier}.
The ranks for some of the specific NK structures that will be discussed in the next subsection are listed in Table \ref{NKRanks}.

\subsection{Specific structure choices}
\label{SpecificStructures}

\begin{table}
    \centering
    \begin{tabular}{l|l|l}
        \rule{0pt}{1em}NK structure & Condition & Rank \\\hline
        \rule{0pt}{1em}BN  & exact & $\frac{L}{k}\left(2^k-1\right)+1$ \\
        \rule{0pt}{1em}uRN & $L\gg k$ & $L\left(2^k-k\right)+1$ \\
        \rule{0pt}{1em}AN  & $L \geq 2k-1$ & $L 2^{k-1}+1$ \\
        \rule{0pt}{1em}SN  & exact & $(L-k+2)2^{k-1}$
    \end{tabular}
    \caption{\label{NKRanks}Ranks for some classical NK structures. Results are taken from \cite{Nowak2015} except for SN. The values 
    for uRN and BN are the largest and smallest possible ones for classical structures \cite{Buzas2014}.}
\end{table}

So far no specific NK structure choice was made.
In this subsection we introduce a few common, for the most part classical, NK structure choices.

\begin{itemize}

\item
In the \defemph{block neighborhood (BN)} (with $L$ being an integer multiple of $k$) $\lset$ is divided into $\frac{L}{k}$ disjoint $k$-subsets and the simplified structure graph is the union of complete symmetric graphs on each of these subsets \cite{Perelson1995}. 
Each block effectively behaves as an independent HoC landscape.
In contrast to the general case, analytical calculations are thus relatively simple, provided that the properties of the HoC model are already known, e.g. for the number of local maxima and the number of accessible pathways \cite{Orr2006,Perelson1995,Schmiegelt2014}.
The BN is uniform, regular and classical.
\item
In the \defemph{adjacent neighborhood (AN)}, loci are put on a circle and NK edges are given by the $k-1$ nearest neighbors of each locus on this ring.
This is one of Kauffman's original choices.
Similar to the BN, the AN is uniform, classical and regular.
In contrast to the BN there is however no independence between subsets of loci.

\item
In the \defemph{random neighborhood (RN)}, each classical NK structure is chosen with uniform probability.
This structure is generally neither uniform, nor regular.
In the \defemph{uniform random neighborhood (uRN)}, each uniform classical NK structure is chosen with uniform probability.
In the \defemph{regular random neighborhood (rRN)}, each regular classical NK structure is chosen with uniform probability.
In the \defemph{uniform, regular random neighborhood (urRN)}, each uniform and regular classical NK structure is chosen with uniform probability.
The last three modifications of the RN structure limit the space of possible structures to choose from.
The random variant as used by Kauffman et al. \cite{Kauffman1989,Weinberger1991}, is actually our uRN\@.
We expect all four variants to behave similarly, at least for large $k$, as the variation in regularity and uniformity will naturally shrink with increasing $k$.

\item
In the \defemph{star neighborhood (SN)} $k-1$ loci are chosen as \defemph{center loci} and they are contained in every block $B_l$.
The other $L-k+1$ loci are called \defemph{ray loci}. A block $B_l$ associated with a ray locus contains the locus itself along with the $k-1$ center loci. When $l$ is a center locus, 
the remaining ($k$'th) element in $B_l$ is set to one of the ray loci (but the same for each center locus).
We introduce this structure as a stark contrast to the other models described above \cite{Schmiegelt2016}.
While it too is classical and uniform, it is strongly non-regular.
The center loci are present in $L$ NK edges giving them correspondingly large variances in mutational effects, while all other loci are only contained in a single NK edge.
Furthermore distances in this structure are very small.
Each pair of loci is in at most distance $2$ along the structure hypergraph, while for all other models described above, the average distance between loci scales with $L$ at constant $k$.
These differences will result in qualitatively different behavior of properties discussed later on.
Note however that the distance autocorrelation function \eqref{NKDistanceCorrelation} is the same for the SN structure as for all other classical structures at equal $k$ and $L$.

\item
The \defemph{mean field structure (MF)} is not classical, containing each possible uniform edge exactly once.
It is thus uniform and also regular.
We use this mean field model as a slight variation from the original structures but with nice mathematical properties.
Effectively we are distributing the average interaction strength of NK edges over all possible choices of these edges.

\end{itemize}

\section{Local fitness maxima}
\label{Sec:LocalFitnessMaxima}

In this section, we begin by introducing a general formalism for calculating the number of local maxima $\nmax$  that can be applied to any of the (generalized) NK structures considered in this review.
The primary goal of this formalism is to estimate the exponential growth rate $\growthFactor{k}{model}$ defined by the relation $\nmax \sim (2 \growthFactor{k}{model})^L$.
The factor $2$ is conventionally introduced in the literature simply
to express the fact that the number of genotypes in the hypercube 
$\mathbb{H}_\mathcal{L}$ increases as $2^L$. 
Since the NK model is homogeneous, $(\growthFactor{k}{model})^L$ may thus be interpreted as the probability that a randomly chosen genotype is a local maximum. As the number of fitness maxima cannot be smaller than 1, the bounds $1/2 \leq \growthFactor{k}{model} \leq 1$ apply. 

In order to minimize the notational burden unavoidable for the large degree of generalization to be pursued, we shall take a heuristic approach by starting with the HoC model as the simplest example and then extend our analysis to the NK model with arbitrary interaction structure.
On this journey, we first encounter two exactly solvable cases, the
block neighborhood (BN) and mean field (MF) models. Whereas the BN
model was originally studied by Perelson and Macken
\cite{Perelson1995}, the MF model is introduced for the first time in
the present work. 
In contrast to the strong universality hypothesis proposed by
Weinberger \cite{Weinberger1991} and cited above in \secref{Outline}, the
distinct asymptotic behaviors exhibited by these two models exemplify our main finding that two different universal behaviors are realized depending on the choice of the NK structure.

To further investigate the range of possible behaviors, we then move our attention to two classical examples, the adjacent neighborhood (AN) and random neighborhood (RN) structures. 
From our analysis of the AN model we recover most of the known exact
results for $\growthFactor{k}{AN}$ that were obtained previously
\cite{Durrett2003,Evans2002,Limic2004} and 
subsequently extend these to a larger class of base distributions $ p_f $.
At the same time we strive to make the mathematical structure behind the formalism transparent to readers with a physics background,
such as to enable them to more easily address future challenges in this field.
Finally, we move on to a variant of the RN model where an exact solution for $\growthFactor{k}{RN}$ can be obtained in the limit $k\to\infty$. 
Asymptotically we will find that $\growthFactor{k}{RN}$ follows the
same behavior as $\growthFactor{k}{MF}$. Since the AN and BN models
are known to display the same asymptotics, this implies that the AN and RN
models are asymptotically distinct. For readers who want to get a
quick overview of the results presented in this section a summary is
provided in \secref{Minima:Summary}.

\subsection{Number of local maxima for HoC fitness landscapes}

As explained above in Sec.~\ref{Sec:RepsLocalMaxima}, if we limit our interest to the mean number of local maxima, it is sufficient to pick an arbitrary reference genotype $\sigma$ and 
focus on the problem of finding the probability $\ProbMax$ for $\sigma$ being a local maximum.
Once this is established, the total number of local maxima is trivially recovered by multiplying $\ProbMax$ by the number of genotypes $2^L$.

For the HoC model, following this procedure is quite straightforward:
Because the fitness values of $\sigma$ and its neighbors are statistically independent and $F(\sigma)$ should be the largest among $L+1$ random variables, it is obvious that the probability $\ProbMax$ is $ (L+1)^{-1} $ \cite{Kauffman1987}. 
More detailed  statistical properties of $\nmaxModel{HoC}$ can be found in \cite{Macken1989,Schmiegelt2014}.

However, for later purposes, let us forget this result for a moment and introduce a more general formalism for computing $\ProbMax$.
Let $h_0$ and $h_l$ denote the fitness values of genotypes $\sigma$ and $\Delta_l \sigma$, respectively, i.e., $h_0 = F(\sigma)$ and $h_l =  F(\Delta_l \sigma)$.
Then, $\sigma$ is a local maximum if  $h_0 > h_l$ or $u_l \equiv h_0 - h_l > 0$ for all $1\le l \le L$.
Using the vector notation $\Vecbf{u} \equiv (u_1,u_2, \cdots , u_L)$, 
the joint probability density of the $u_l$ is given by
\begin{align}
	\mathcal{P}(\Vecbf{u}) = \int \prod_{l=0}^{L} dh_l \, p_f(h_l) \prod_{l=1}^{L} \delta(u_l - (h_0 - h_l) ),
	\label{PDFHoC}
\end{align}
or alternatively, the characteristic function reads
\begin{align}
\Phi(\Vecbf{q}) &= \int \prod_{l=1}^{L} du_l e^{
	i \sum_{l=1}^{L} q_l u_l
} \mathcal{P}(\Vecbf{u}) 
	= \int \prod_{l=0}^{L} dh_l \, p_f(h_l) e^{
		i \sum_{l=1}^{L} q_l (h_0 - h_l) 
} \nonumber\\
	&= \CharFuncBase{
		\sum_{l=1}^L q_l
	}   \prod_{l=1}^{L} \CharFuncBase{-q_l} 
	=
	\int dy \, p_f{
		(y)
	}   \prod_{l=1}^{L} \CharFuncBase{-q_l} \exp\left ( i y \sum_{l=1}^L q_l \right)
	\label{CFHoc}
\end{align}
where $\CharFuncBase{q}$ is the characteristic function of $p_f(h)$.
By performing the inverse Fourier transform of $\Phi(\Vecbf{q})$ 
and then integrating over only positive values of $u_l$, we obtain
\begin{align}
	\ProbMax &=   \int_{0}^{\infty}  \prod_{l=1}^{L} du_l \,   \mathcal{P}(\Vecbf{u})
	= \int \MeasureUQ    \Phi(\Vecbf{q}), 
	\label{ProbMax}
\end{align}
where we have introduced a symbol $\Measure{}$ to denote the integration over $L$-dimensional real space (i.e., $\Measure{v} = \prod_{l =1}^{L} dv_l$).
Moreover, to encode the positivity condition for $\Vecbf{u}$, we define the theta function $ \Theta(\Vecbf{u} > 0) $ such that it is one if all the elements of $\Vecbf{u}$ are positive and zero otherwise.

Now, we are ready to calculate $\ProbMax$. 
Inserting \eqref{CFHoc} into \eqref{ProbMax} and making use of the integral representation of the delta function
\begin{align}
	\delta(q) =  \int \frac{dy}{2\pi} e^{i y q}
\end{align}
leads us to write 
\begin{align}
	\ProbMax &=  \int dy \, p_f(y) \int \MeasureUQ    \prod_{l=1}^{L} \CharFuncBase{-q_l}   e^{i q_l y} \nonumber\\
	&= \int dy \, p_f(y) \left[
		\int_{0}^{\infty} du  \, p_f(y- u)
	\right]^L.
\end{align}
Finally, by realizing that $G(y) = \int_{0}^{\infty} dm\, p_f(y- m) =  \int_{-\infty}^{y} dy\,  p_f(y)$ is the cumulative base distribution, the substitution $ x = G(y)$ is evaluated to
\begin{align}
	\ProbMaxModel{HoC} = \int_{0}^{1} dx \, x^L = \frac{1 }{L+1},
	\label{ProbMaxHoC}
\end{align}
which is the desired result for the HoC model. 
The fact that $ \ProbMaxModel{HoC} $ decays algebraically in $L$ implies $\growthFactor{k}{HoC} = 1$.

\subsection{Number of local maxima for NK fitness landscapes}
By the construction of the NK model as described in \eqref{FitnessNK}, the fitness $F(\sigma)$ of a sequence $ \sigma $ is the sum of 
HoC fitness values defined on the subspaces $ \gspace[B_r] $ spanned by the edge sets or NK blocks $B_r$.
Since a characteristic function is a natural object when dealing with a random quantity constructed from the sum of independent random variables, we will build our approach 
upon the characteristic functions of the NK blocks.
Specifically, we expect the characteristic function of $\Vecbf{u}$ to be of the form 
\begin{align}
\label{CFNKProduct}
	\Phi(\Vecbf{q}) = \prod_{r  = 1}^{\vert \nkstruct \vert } \Phi_r(\Vecbf{q}),
\end{align}
where $ \Phi_r(\Vecbf{q}) $ denotes the characteristic function of $\Vecbf{u}$ within the NK block $B_r$.
Because each HoC model is defined only on a subset of $\lset$, it is convenient to employ the incidence matrix notation $b_{l,r}$ that indicates the presence (absence) of a locus $l$ in a neighborhood set $r$, i,e, $b_{l,r} = 1\, (0)$ if $l \in B_r$ ($l \notin B_r$).
In terms of these variables, the characteristic function $\Phi_r$ 
can be rewritten in the following form:
\begin{align}
	\Phi_r(\Vecbf{q}) 
	&= \CharFuncBase{
	\sum_{l=1}^L q_l b_{l,r}
	}   \prod_{l=1}^{L} \CharFuncBase{-q_l}^{b_{l,r}} \nonumber\\
	&= \int dy_r \, p_f{ (y_r) } \prod_{l=1}^{L} \left[
		\CharFuncBase{-q_l} e^{i y_r q_l }
	\right]^{b_{l,r}}.
	\label{CFNKGeneral}
\end{align}
Once the full characteristic function \eqref{CFNKProduct} has been derived, $\ProbMax$ is readily calculated by inverse Fourier transform along the lines of 
\eqref{ProbMax}, i.e., 
\begin{align}
\ProbMax = \int \MeasureUQ  \int \MeasureY   \prod_{r=1}^{|\nkstruct|} \prod_{l=1}^{L} \left[
\CharFuncBase{-q_l} e^{i y_r q_l }
\right]^{b_{l,r}},
\label{ProbMaxFormalism}
\end{align}
where $\mathcal{P}(\Vecbf{y}) = \prod_r  p_f{ (y_r) }$, the $|\nkstruct|$-dimensional base fitness distribution.

Below we will follow these steps to compute $\ProbMax$ for several known NK structures as well as for the MF structure introduced in Sec.~\ref{SpecificStructures}.
By doing so, we will recover earlier results and obtain new insights into how 
the universal and non-universal behavior of $\ProbMax$ is shaped by the interaction structure and the base fitness distribution.

\subsubsection{Block neighborhood}
In the BN model, the NK structure $\nkstruct$ comprises mutually non-overlapping sets of size $k$. 
Each $B_r$ thus defines an independent module in which the loci are correlated among each other but not with the loci outside of the module.
This non-overlapping property facilitates the analysis dramatically since it allows us to write $\ProbMax$ in a factorized form, 
$\ProbMaxModel{BN} = \prod_{r} \ProbMax^r$ where $ \ProbMax^r $ is simply $\ProbMaxModel{HoC}$ for $k$ loci, as given by \eqref{ProbMaxHoC}.
Putting everything together, we find
\begin{align}
	\ProbMaxModel{BN} \equiv (\growthFactor{k}{BN})^L =   \prod_{r=1}^{|\nkstruct|} \frac{1}{k+1} = \left(
		\frac{1}{k+1}
	\right)^{L/k},
	\label{ProbMaxBN}
\end{align}
where we have used the fact that the number of blocks is $L/k$.
Equivalently, the mean number of local maxima is
\begin{align}
\mean{\nmaxModel{BN}} = 2^L \left(
\frac{1}{k+1}
\right)^{L/k}.
\end{align}
As consistency checks, one can immediately show that inserting $k =1$ and $k=L$ recovers $ \mean{\nmaxModel{BN}} = 1$ for additive landscapes and $ \mean{\nmaxModel{BN}} = (L+1)^{-1}  2^L$ for HoC landscapes, respectively.

This closed form solution allows us to study the asymptotic behaviors in various limits. The most interesting scaling limits include i) $L\to\infty$ for $k$ fixed and ii) 
the joint limit $L, k\to\infty$ with fixed $\alpha = k /L$.
In the first limit, it is clear that $ \ProbMaxModel{BN} $ increases exponentially with $L$ with an exponential growth rate
\begin{align}
\ln \growthFactor{k}{BN} \equiv \lim_{L\to\infty} \frac{\ln \ProbMaxModel{BN}}{L} =  \ln \left(
\frac{1}{k+1}
\right)^{1/k} = -\frac{\ln k}{k} + \Order{\frac{1}{k^2}}
\label{NumMaxBN}
\end{align}
as $L \to \infty$.
As $k\to\infty$, $ \ln \growthFactor{k}{BN} $ converges to the theoretical upper bound, namely zero. 
Thus, for larger $k$, we expect more rugged fitness landscapes.

In such a large $k$ limit, the second scaling limit, where $\alpha$ is kept fixed, provides a better understanding of the behavior of $\ProbMaxModel{BN}$.
In this limit, it is evident that the leading exponential behavior of $ \mean{\nmaxModel{BN}} $ should be $2^L$.
The correction to this exponential behavior should be at most algebraic as already seen in the HoC model.
In the case of BN, this correction may be easily evaluated to 
\begin{align}
\label{BNjointlimit}
	\ProbMaxModel{BN}  = \left (\frac{1}{L\alpha}  + \Order{L^{-2}} \right )^{1/\alpha} \sim L^{-1/\alpha}.
\end{align}
A more detailed analysis of the BN model has been conducted in the literature~\cite{Perelson1995,Schmiegelt2014}, and in particular, the second moment of $ \nmaxModel{BN} $ is given by
\begin{align}
	\mean{(\nmaxModel{BN})^2} = \left(
		\mean{\nmaxModel{BN}}
	\right)^2 \left( 
		1 + \frac{k-1}{2^{k+1}}
	\right)^{L/k}.
\end{align}

\subsubsection{MF neighborhood}
\label{sec:LocalMaximaMF}
The mean-field NK structure is another extreme type of NK model. In this case 
the neighborhood set $\nkstruct[B]$ contains all possible subsets of size $k$, which effectively makes the fitness landscape unstructured in contrast to the block model which has a well-defined modular structure. 
By construction, the size of $\nkstruct[B]$ is given by $|\nkstruct[B]| = \binom{L}{k}$ unlike classical NK structures that satisfy $|\nkstruct[B]| = L$.
Because of this huge combinatorial factor, one might wonder if an additional normalization that rescales the overall fitness to a reasonable level should be introduced.
While this might be necessary for other applications, we do not bother with it here since the number of local maxima only depends on the fitness ordering between neighboring genotypes and not on the overall fitness scale.

Additionally, we assume that the base fitness distribution $p_f(h)$ is a standard Gaussian distribution. 
This assumption is made for two reasons.
First of all, the choice of Gaussian distribution greatly simplifies the analysis of $\nmaxModel{MF}$.
Secondly and more importantly, the number of local maxima $ \nmaxModel{RN} $ for the RN NK structure complemented by a large class of base fitness distributions 
will be shown to follow the same limiting behavior as $\nmaxModel{MF}$.
Verifying this claim for universal behavior will be the main topic of \secref{sec:LocalMaximaRN}.

Recalling the fact that $\CharFuncBase{q} = e^{-q^2/2}$ for a standard Gaussian distribution and using the first identity in \eqref{CFNKGeneral}, the characteristic function is readily obtained as 
\begin{align}
\CharFunc{q}{MF} = \exp\left(
	- \sum_l \binom{L-1}{k-1} q_l^2 - \sum_{l>m} \binom{L-2}{k-2} q_l q_m
\right).
\end{align}
The two binomial numbers correspond to the number of neighborhood sets that contain the locus $l$, and both the loci $l$ and $m$, respectively. 

Now, we are left to calculate $\ProbMaxModel{MF}$ using \eqref{ProbMax}. 
As mentioned before, any rescaling of fitness values should leave the quantity of interest unchanged.
Exploiting this invariance, the fitness rescaling $F(\sigma) \to F(\sigma) / \sqrt{\binom{L-2}{k-2}}$ allows the subsequent transformations 
$u_l \to u_l / \sqrt{\binom{L-2}{k-2}}$ and $q_l \to \sqrt{\binom{L-2}{k-2}} q_l$, which effectively reduces the number of free parameters to one. 
Defining 
\begin{align}
\eta = \sqrt{\frac{2 (2 L-k-1)}{L(k-1)}}, 
\label{MFSingleParameter}
\end{align}
the probability reads 
\begin{align}
\ProbMaxModel{MF} = \int \MeasureUQ \exp\left[
	-\sum_{l=1}^{L} \frac{L\eta^2}{4} q_l^2  - \frac{1}{2} \left(
		\sum_{l=1}^{L} q_l
	\right)^2
\right].
\end{align}
Finally, employing the Hubbard-Stratonovich transform, the quadratic coupling term in the square bracket is linearized and the integrals for different indices 
$l$ are completely decoupled:
\begin{align}
\ProbMaxModel{MF} =& \sqrt{\frac{L}{2\pi}} \int dy \,  e^{-Ly^2/2} \left[
\frac{1}{2\pi} \int_0^\infty du \int dq e^{
- \frac{L\eta^2}{4} q^2 - i q u +i \sqrt{L} y q
}
\right]^L\nonumber\\
=& \sqrt{\frac{L}{2\pi}} \int dy  \, e^{-Ly^2/2} \left[
\frac{1}{2} \left(\text{erf}\left(\frac{y}{\eta}\right)+1\right)
\right]^L = \sqrt{\frac{L}{2\pi}} \int dy \,  e^{L \Action_\eta(y) },
\label{ProbMaxMF}
\end{align}
where 
\begin{align}
\Action_\eta(y) = -y^2 /2 + \ln \left[
\frac{1}{2} \left(\text{erf}\left(\frac{y}{\eta}\right)+1\right)
\right].
\label{MFAction}
\end{align}
The integral in \eqref{ProbMaxMF} does not allow for a closed form solution for general $\eta$.
However, a straightforward calculation shows that the expected results can be recovered in the two limiting cases $k=1$ (for linear landscapes) and $k=L$ (for HoC landscapes).

To proceed, a reasonable scaling limit should be taken to draw some practical conclusions.
Let us first consider the large $L$ limit with fixed $k$, which was discussed above for the BN model. 
In this limit, the parameter in \eqref{MFSingleParameter} is expanded as $\eta^2 = \frac{4}{k-1} + O(1/L)$. 
The fact that $\eta$ is independent of $L$ up to leading order in $L$ suggests that the integral may be evaluated using the saddle point method up to a correction of
 $\Order{1/L}$. 
We point out that once the value $y^*$ that maximizes the ``action'' \eqref{MFAction} has been found, the value of $\growthFactor{k}{MF}$ readily follows from $\ln \growthFactor{k}{MF} = \Action_\eta(y^*)$.
Specifically, assuming $y^*$ is known, the saddle point approximation yields a rather formidable formula:
\begin{align}
\ProbMaxModel{MF} =\frac{ \left(\frac{e^{-\frac{(y^*)^2}{2}}}{2} \left(\text{erf}\left(\frac{y^* \sqrt{k-1}}{2 }\right)+1\right)\right)^L \exp \left(\frac{\sqrt{k-1} (k+1) y^* e^{-\frac{1}{4} (k-1) (y^*)^2}}{4 \sqrt{\pi } \left(\text{erf}\left(\frac{y^*}{2}  \sqrt{k-1} \right)+1\right)}\right)}{\sqrt{\frac{(k-1)^{3/2} y^* e^{-\frac{1}{4} (k-1) (y^*)^2}}{2 \sqrt{\pi } \left(\text{erf}\left(\frac{y^*}{2} \sqrt{k-1} \right)+1\right)}+\frac{(k-1) e^{-\frac{1}{2} (k-1) (y^*)^2}}{\pi  \left(\text{erf}\left(\frac{y^*}{2}  \sqrt{k-1} \right)+1\right)^2}+1}}
\end{align}
with an error of the order $\Order{L^{-1}}$.

\begin{figure}
	\includegraphics[width=1.0\textwidth]{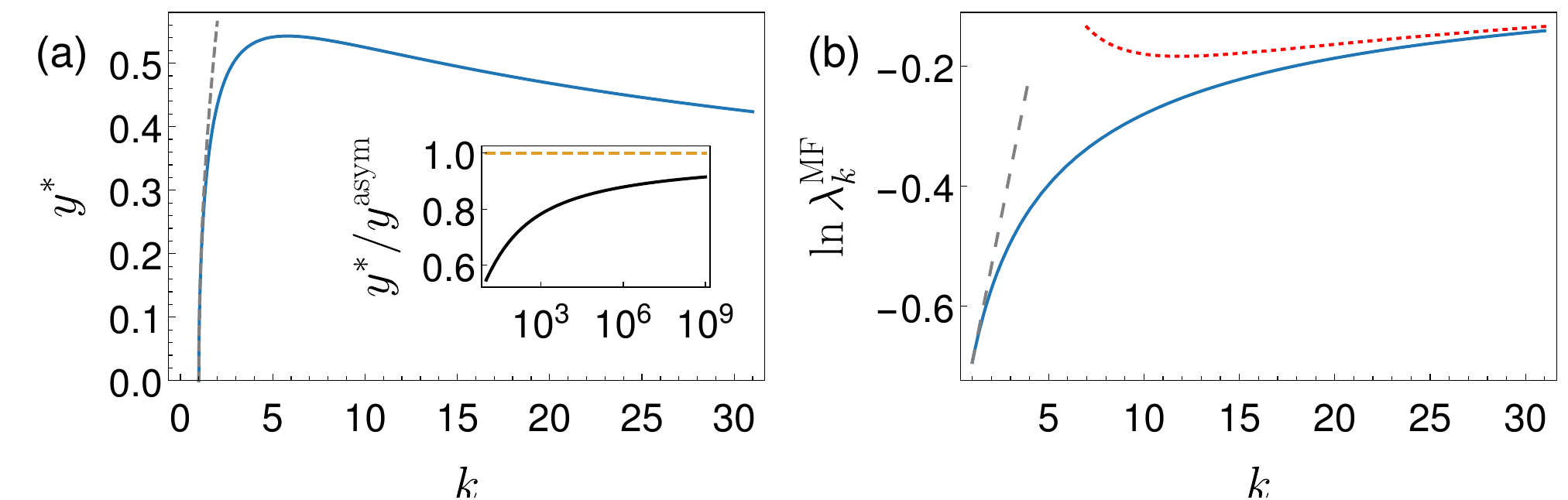}
	\caption{Plots of $y^*$ and $\ln \growthFactor{k}{MF} $ as a function of $k$. (a) 
		The solid curve indicates the numerical solution for $y^*$. 
		The dashed curve shows the first order approximation 
$\sqrt{\frac{k-1}{\pi}}$ in \eqref{LocalMaximaMFSolutionY} as $k\to 1$. 		
		(inset) The black curve represents the ratio of $y^*$
                to the first order approximation $y^\mathrm{asymp} = 2
                \sqrt{ \frac{  \log (k)}{k}}$, which is valid for $k \to \infty$. 
		It illustrates the slow convergence of $y^*$ to the first order approximation even for relatively large $k$, i.e., $ k \sim 10^9 $.
		(b) The solid line describes the actual value of $\ln \growthFactor{k}{MF}$ found numerically from the variational problem. 		
		Two dashed curves are obtained by the first order approximations for both limits, i.e., $ \frac{k-1}{2 \pi }-\ln (2) $ for $ k \to 1 $ and 
		\eqref{lambdaMFasymptotic} for $k \to \infty$. 
}
	\label{fig:MeanFieldNK}       
\end{figure}

Even though the variational problem has no closed form solution in general, one can analyze the asymptotic series expansion for small or large $\eta$ as
\begin{align}
y^* = \begin{cases} 
\frac{2}{\sqrt{\pi } \eta } + \Order{\eta^{-2}}  & \text{for} \quad  \eta \to \infty \\
 \sqrt{-2 \eta^2 \ln \eta + O( \ln (| \ln \eta |))}       & \text{for} \quad  \eta \to 0.
 \label{LocalMaximaMFSolutionY}
\end{cases}
\end{align}
Subsequently, this expansion allows us to obtain
\begin{align}
\ln \growthFactor{k}{MF}= \mathcal{F}_\eta(y^*) =\begin{cases} 
-\ln (2)+\frac{2}{\pi  \eta ^2} & \text{for} \quad  \eta \to \infty \\
\eta^2 \ln \eta + O(\eta^2 \ln (|\ln \eta|))       & \text{for} \quad  \eta \to 0.
\end{cases}
\label{ProbMaxMFAsymptotic}
\end{align}
With the results written in terms of $\eta$, the functional dependence on $k$ can be easily recovered by the relation $\eta^{-2} = \frac{k-1}{4}$ as defined in \eqref{MFSingleParameter}.
The saddle point $y^*$ that maximizes $ \Action_\eta(y) $ and the corresponding exponential factor $\ln \growthFactor{k}{MF} $ are illustrated in \figref{fig:MeanFieldNK} as a function of $k$.

The small $\eta$ expansion in \eqref{ProbMaxMFAsymptotic} translates into the expression 
\begin{equation}
\label{lambdaMFasymptotic}
\ln \growthFactor{k}{MF} \approx \frac{\ln(16) - 2 \ln(k-1)}{k-1} = - \frac{2 \ln(k)}{k} + \Order{\frac{1}{k}}
\end{equation}
which is noteworthy for two reasons. First, and most importantly, 
the leading order behavior $\ln \growthFactor{k}{MF}  \sim -\frac{2 \ln (k)}{k} $ differs from that obtained for the block model, 
$\ln \growthFactor{k}{BN}  \sim -\frac{\ln (k)}{k} $, which contradicts the claim of universality originally stated by Weinberger \cite{Weinberger1991}.
Second, the leading term in \eqref{lambdaMFasymptotic} is only logarithmically larger than the next-to-leading term.   
Thus, in the range of $k$ that is accessible to the explicit numerical evaluation of $\nmax$ for arbitrary NK structures,  
i.e., at most $ k \sim O(10^2) $, the next-to-leading correction remains substantial.
Nevertheless, the full expression in \eqref{lambdaMFasymptotic} provides an accurate approximation to the true behavior already for $k = 30$
[see \figref{fig:MeanFieldNK} (b)]. 

Although the calculation as described relies on taking the limit $L \to \infty$ before the limit of large $k$, extending the result to the joint limit $L, k \to \infty$ at fixed $\alpha = k/L$
is straightforward at least on a formal level. For this it suffices to note that \eqref{MFSingleParameter} now implies the relation $\eta \approx \frac{2}{\sqrt{k}}\sqrt{1-\frac{\alpha}{2}}$, which combined with 
the $\eta \to 0$ limit in \eqref{ProbMaxMFAsymptotic} yields 
\begin{equation}
\label{MFuniversality}
\ln \growthFactor{k}{MF} \approx -\frac{(2 - \alpha)\ln k}{k} \;\; \textrm{and} \;\; \ProbMaxModel{MF} \sim L^{-(\frac{2}{\alpha}-1)}.
\end{equation}   
The exponent of the algebraic decay of $\ProbMaxModel{}$ is different from that obtained in \eqref{BNjointlimit} for the BN structure,
but reduces to the $1/L$-behavior expected for the HoC model when
$\alpha \to 1$. A rigorous analysis based on extreme value theory
confirms this simple argument up to logarithmic corrections
(Appendix \ref{App:JointLimit}). 

As we will see in the following, the different asymptotics obtained for the BN and MF models are not just arbitrary examples created by unusual choices of NK structures, but 
in fact they appear to be robust across large classes of structures. They exemplify a somewhat surprising trend, which is that NK models with more structured interaction
schemes such as the BN model result in more rugged fitness landscapes. 
In the next two subsections we will explore two other NK structures, each of which follows the asymptotic behavior found for the BN and MF models, respectively.
  
\subsubsection{Adjacent neighborhood}
The regularity of the AN NK structure allows us to view our analysis from a different angle. 
This point is best described by \eqref{ProbMaxFormalism} with a slight modification given as
\begin{align}
\ProbMaxModel{AN} &= \int \MeasureY \MeasureUQ \prod_{r=1}^{L} \prod_{l=1}^{L} \left[
\CharFuncBase{-q_l} e^{i y_r q_l }
\right]^{b_{l,r}} \nonumber \\
&= \int \MeasureY \MeasureUQ  \prod_{l=1}^{L} \CharFuncBase{- q_l}^k e^{ i q_l  \sum_{r=0}^{k-1} y_{(l + r)\Mod L} },
\end{align}
where the operator $A \Mod B$ is used to denote the remainder of $A$ when divided by $B$.
Also, it is worth pointing out that the characteristic function for each locus $l$ always appears $k$ times due to the translational invariance.
Since the $k$-th power of a characteristic function is Fourier-transformed back to the $k$-th convolution of the corresponding probability density, the integrals for $\Vecbf{u}$ and $\Vecbf{q}$ may be written in terms of $\tilde{F}^{(k)}(z) = \int_{-\infty}^{z} dy \, p_f^{(k)}(y)$, where $ p_f^{(k)}(y) $ is the $k$-fold convolution of $ p_f(y) $:
\begin{align}
\ProbMaxModel{AN} &= \int \MeasureY \prod_{l=1}^{L} \tilde{F}^{(k)}  	\left(
	\sum_{r=0}^{k-1} y_{(l + r) \Mod L}
\right).
\label{ProbMaxAN}
\end{align} 
This elegant equation was first derived by Weinberger \cite{Weinberger1991}.
To understand this expression better, it is convenient to expand the product for the simplest case $k=2$. 
Then, one may identify a simple pattern of the following form
\begin{align}
	\ProbMaxModel{AN} &= \int \prod_{r=1}^{L} dy_r K_w(y_1; y_2) K_w(y_2; y_3) \cdots K_w(y_L; y_1),
\end{align}
where 
\begin{align}
 K_w(x; y) \equiv p_f(x)^w \tilde{F}^{(2)}  	\left(
x+y
\right) p_f(y)^{1-w}, 
\label{IntegralOperator}
\end{align}
with an arbitrary choice of $w \in [0,1]$.
Thus, $ \ProbMaxModel{AN} $ may be regarded as the trace of the $L$-th power of an integral operator defined by the integral kernel $ K_w(x; y) $.
One can show that the eigenvalue spectrum of the kernel does not depend on the choice of $w$ by checking that the trace of an arbitrary power of $ K_w(x; y) $ is independent of $w$. Moreover, the fact that $ K_w(x; y) $ becomes symmetric when $w=1/2$ guarantees that all the eigenvalues of this operator are real.

This construction recasts the problem of finding $\ProbMaxModel{AN}$ into an eigenvalue problem for the integral kernel $K_w(x;y)$.
In particular, the largest eigenvalue will correspond to $\growthFactor{2}{AN}$ in the limit $L\to\infty$.
A similar but not identical transfer matrix technique for $ \ProbMaxModel{AN} $ was originally introduced by Evans and Steinsaltz \cite{Evans2002}.

Finding eigenvalues of arbitrary integral operators is in general a non-trivial problem
\cite{Kanwal2012}.
However, if $ K_w(x; y) $ is separable, i.e.,  if $ K_w(x; y) $ can be cast into a sum of factorized terms of the form 
\begin{align}
	K_w(x; y) = \sum_{p=1}^{n} u_p(x) v_p(y),
	\label{IntegralOperatorSeparable}
\end{align}
the problem can be mapped to finding the eigenvalues of an $n \times n$ matrix with matrix elements given by 
\begin{align}
	T_{p q} = \int dx \, u_p(x) v_q(x).
	\label{TransferMatrix}
\end{align}
In the following, we will provide two classes of base distributions that allow for an exact solution through this technique.

As the simplest example, let us consider a random variable $Z$ with the property that $\tilde{F}^{(2)}(z) = 1-e^{-z}$ for $0 \le z \le \infty$.
In other words, the two-fold convolution of the base probability density is exponential.
From the definition of the integral kernel \eqref{IntegralOperatorSeparable} with the choice of $w=1$, one finds that 
$ K_1(x; y) = p_f(x)  - p_f(x) e^{-x} e^{-y}  $.
The corresponding matrix is readily obtained as 
\begin{align}
	T = \left( \begin{array}{cc} 
		1    &  \mean{e^{ikX}}  |_{k = i}  \\
	-\mean{e^{ikX}}  |_{k = i}  &- \mean{e^{ikX}}  |_{k = 2i}  \\
	\end{array}\right) = 
	\left( \begin{array}{cc} 
	1    & \frac{1}{\sqrt{2}} \\
	-\frac{1}{\sqrt{2}}  & -\frac{1}{\sqrt{3}}  \\
	\end{array}\right),
	\label{TransferMatrixExample}
\end{align}
where we have calculated the characteristic function of $X$ to be $\sqrt{\frac{i}{k+i}}$ by taking the square root of the characteristic function of the exponential distribution.
Finally, 
we can easily calculate the largest eigenvalue as $ \growthFactor{2}{AN} =  \frac{1}{6} \left(3-\sqrt{3}+\sqrt{6 \sqrt{3}-6}\right) \simeq 0.560622$, 
a result originally derived in \cite{Nowak2015}.

The base fitness distribution corresponding to the previous example is a gamma distribution with shape parameter $s=1/2$, and in fact
gamma-distributed fitness values appear in several earlier studies where exact results for $\growthFactor{2}{AN}$ were obtained  \cite{Durrett2003,Evans2002}. 
With the current framework at hand, it turns out that the association of solvable instances of the AN model with certain gamma distributions is not a coincidence.
Below we will show that the integral kernels $K_w(x;y)$ generated by gamma distributions with shape parameter $s$ being either a half-integer or an integer are separable 
and thus all the previously known results can be calculated in a uniform manner.

For an arbitrary shape parameter $s$, the two quantities defining $K_w(x;y)$ in \eqref{IntegralOperator} are given by
\begin{align}
p_f(z) = g_s(z) \equiv  \frac{e^{-z}z^{s-1}}{\Gamma(s)}
\end{align}
and 
\begin{align}
\tilde{F}_s^{(2)}(z) = 1-\frac{\Gamma(2s,z)}{\Gamma(2s)}.
\end{align}
Furthermore, the incomplete gamma function permits a series expansion of length $2s$,
\begin{align}
\Gamma(2s,z) = \Gamma(2s)\, e^{-z} \sum_{m=0}^{2s-1} \frac{z^m}{m!},
\end{align}
provided $2s$ is an integer.
Inserting this into \eqref{IntegralOperator}, we arrive at 
\begin{align}
K_1(x;y)
&=- \sum_{p=0}^{2s-1} \frac{e^{-2x}x^{p+s-1}}{\Gamma(s)} \times \frac{\Gamma(2s-p,y)}{p!\Gamma(2s-p)} + g_s(x). 
\end{align}
When cast into the form of \eqref{IntegralOperatorSeparable}, this shows that the integral operator $K_1(x;y)$ is mapped onto a $ (2s +1) \times (2s +1) $ matrix with entries given by
\begin{align}
	T_{pq} =\left\{ \begin{array}{cc}
	 - M_{pq} & \mathrm{if}\quad 0 \le p, q \le (2s-1)   \\ 
	L_p& \mathrm{if}\quad q=2s,\, 0 \le q \le (2s-1) \\  
	-R_q& \mathrm{if}\quad p=2s,\, 0 \le p \le (2s-1) \\  
	1 & \mathrm{if}\quad p = 2s,\, q = 2s\\  
	\end{array} 
	\right.,
	\label{TransferGamma}
\end{align}
where 
\begin{align}
M_{pq} &=\int_{0}^{\infty} dx \, \frac{e^{-2x}x^{p+s-1}}{\Gamma(s)} \times \frac{\Gamma(2s-q,x)}{q!\Gamma(2s-q)} \nonumber\\ 
&= \frac{\, _2F_1(p+s,p-q+3 s;p+s+1;-2) \Gamma (p-q+3 s)}{q! (p+s) \Gamma (s) \Gamma (2 s-q)},
\label{TransferMatrixM}
\end{align}	 
\begin{align}
L_{p} &=\int_{0}^{\infty} dx \frac{e^{-2x}x^{p+s-1}}{\Gamma(s)} = \frac{2^{-p-s}\Gamma(p+s)}{\Gamma(s)},
\end{align}	 
and
\begin{align}
R_{q} &=\int_{0}^{\infty} dx \frac{e^{-x}x^{s-1}}{\Gamma(s)} \times \frac{\Gamma(2s-q,x)}{q!\Gamma(2s-q)} \nonumber \\
&= \frac{\, _2F_1(s,3 s-q;s+1;-1) \Gamma (3 s-q)}{\Gamma (q+1) \Gamma (s+1) \Gamma (2 s-q)}.
\end{align}	 
Setting $s=1/2$ in \eqref{TransferGamma}, we immediately reproduce the transfer matrix 
obtained in the previous example \eqref{TransferMatrixExample}.
Note that since the gamma distribution converges to a Gaussian distribution as $s\to \infty$, 
	$ \growthFactor{2}{AN} $ for the Gaussian distribution can be obtained by examining the asymptotic behavior for large $s$.

To provide a larger class of exactly solvable cases, one might hope that a similar approach can be taken for random variables $Y$ that are 
transformed from a gamma distributed random variable $X$, if the transformation function $Y = f(X)$ is sufficiently simple.
One such example is $Y = -X$, and we call the corresponding distribution a \defemph{negative gamma distribution}. 
As a special case of this distribution, the value of $\growthFactor{2}{AN}$ for $s=1$ has been found in \cite{Durrett2003}.
The structural similarity possessed by the transformed distribution allows us to repeat the same procedure that we followed for the gamma distribution.
In this case, we find that the corresponding transfer matrix is of size $ (2s) \times (2s) $ with matrix elements
\begin{align}
	T_{pq} = M_{pq},
		\label{TransferNegativeGamma}
\end{align}
where $M_{pq}$ was defined in \eqref{TransferMatrixM}.
\begin{figure}
	\includegraphics[width=1.0\textwidth]{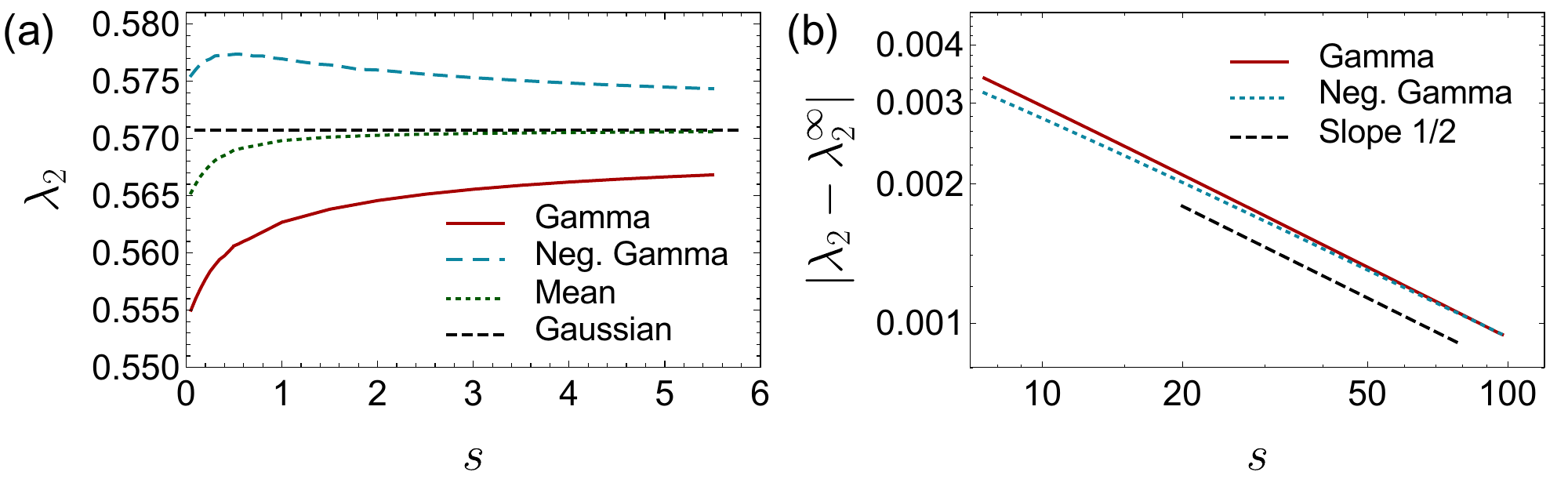}
	\caption{(a) A plot of $\growthFactor{2}{AN}$ as a function of the shape parameter $s$ for the gamma distribution (red solid line) and the negative gamma distribution (blue dashed line). 
		For integer or half-integer values of $s$, the largest eigenvalue of 
		the transfer matrix $T$ given by \eqref{TransferGamma} and \eqref{TransferNegativeGamma} is computed numerically while for the other values of $s$, estimates from numerical simulations are used. Since the convergence to the asymptotic exponential behavior is very rapid in this case, it was sufficient 
		to use sequence lengths between $L=4$ and $7$ for which the number of maxima could be determined by explicit enumeration.
		The average of the red solid curve and the blue dashed curve is represented by the green dotted curve.
		Within the resolution of the image, the average appears to quickly converge to $ \growthFactor{2}{}  = 0.5707$, the value numerically obtained for a 
		Gaussian distribution as indicated by the black dashed line.
		(b) A double-logarithmic plot showing the speed of convergence for the gamma and the negative gamma distribution. 
		The value of $\growthFactor{2}{\infty}$ is estimated from the average of the two curves using the largest shape parameter numerically available. 
		The black dashed line is provided as a visual guide to show that the curves decay algebraically as $s^{-1/2}$.
	}
	\label{fig:AdjacencyNKGamma}       
\end{figure}

Once the matrix has been set up according to \eqref{TransferGamma} or \eqref{TransferNegativeGamma}, the largest eigenvalue is computed numerically through a standard algorithm.
The behaviors of $\growthFactor{2}{AN}$ for the gamma distribution and the negative gamma distribution are illustrated in \figref{fig:AdjacencyNKGamma}. 
In particular, this shows that both curves converge algebraically as $s^{-1/2}$ to the value of the Gaussian distribution. 
Since the curve for the gamma distribution converges from below whereas the curve for the negative gamma distribution converges from above, the average of the 
two curves should provide an accurate estimate for the Gaussian distribution. 
In fact, we found that the sub-leading corrections for the two curves seem to perfectly cancel each other. Thus, one can see from the comparison with simulation results obtained for the sequence length $L=128$ (black dashed line of \figref{fig:AdjacencyNKGamma} (a)), that a very precise estimate ($\growthFactor{2}{AN}=0.5707$) can be obtained even for relatively small $s$.

At this point, it is worth noticing that $ \growthFactor{2}{AN} $ for the negative gamma distribution is maximized at $s=1/2$. 
Since our analytical framework is only applicable for $s$ being either an integer or a half-integer, extensive simulations in the vicinity of $s= 1/2$ had to be performed to create a smooth curve around 
$s=1/2$ in \figref{fig:AdjacencyNKGamma} (a).
What is more interesting about this point is the fact that the eigenvalue problem becomes trivial, as $T$ becomes a $1 \times 1$ matrix with the single element $1/\sqrt{3}$. 
Surprisingly, this number coincides with exact value of $\growthFactor{2}{BN}$, as given by \eqref{NumMaxBN}, for the case $k=2$.
We show in Appendix \ref{appendix:VariationalAnalysis} that the correspondence between the AN model with negative gamma distribution and the BN model can be extended to 
arbitrary $k$ by setting the shape parameter to $s = 1/k$. In particular, 
\begin{equation}
\label{AN=BN}
\growthFactor{k}{AN} = \growthFactor{k}{BN} = \left(\frac{1}{k+1} \right)^{1/k} \; \textrm{for} \; p_f(x) = g_{1/k}(-x). 
\end{equation}
Moreover, a variational analysis   
around the negative gamma distribution with shape parameter $1/k$, viewed as a point in the probability distribution space, proves that $ \growthFactor{k}{AN} $ is not only maximized along the $s$-axis but also extremized in the whole space of distributions with support limited to the negative real axis. 
This observation corroborates the conjecture \cite{Schmiegelt2014} that the BN model growth rate $\growthFactor{k}{BN}$
is an upper bound on $\growthFactor{k}{}$ among all possible NK structures. 

Next we discuss how our method can be generalized to larger values of $k$.
In order to avoid notational clutter, it is best to consider $k=3$.
In this particular case, we can construct a transfer matrix having state space $\mathbb{R}^2$:
\begin{align}
	K_w(x_1, x_2 ; y_1, y_2) =&  p_f(x_1)^w  p_f(x_2)^w \tilde{F}^{(3)}  	\left(x_1+x_2+y_1	\right) \nonumber \\
	&	\times \tilde{F}^{(3)}  	\left(x_2+y_1+y_2	\right)  p_f(y_1)^{1-w}  p_f(y_2)^{1-w},
\end{align}
for $0\le w \le 1$.
By expanding the state space to $\mathbb{R}^{k-1}$, a similar construction can be made for higher values of $k$.
Furthermore, once the kernel is constructed, all the procedures described for $k=2$ may be applied for arbitrary $k$ as long as the kernel is separable.
However, we found that the direct application of this approach for the gamma distribution becomes quickly unmanageable,
because the dimension of the transfer matrix increases combinatorially fast. The only result known from the literature for $k > 2$ is the value 
$\growthFactor{3}{AN} \simeq 0.61140$ for the exponential distribution \cite{Evans2002}.

Despite this limitation, one may still perform an asymptotic analysis for $\growthFactor{k}{AN}$.
In particular,  for the Gaussian distribution, it is rigorously known that \cite{Limic2004}
\begin{align}
	\ln \growthFactor{k}{AN} = -\frac{1}{k} \left( \ln k + R_{L,k}\right),
\end{align}
where $ -c \sqrt{\ln k} \le R_{L,k} \le c \ln \ln k  $ for some $c>0$.
For arbitrary distributions the same authors establish the inequality
\begin{equation}
-\frac{3}{k} \left( \ln k + o(1)\right) \le \ln  \growthFactor{k}{AN}  \le -\frac{1}{k} \left( \ln k + o(1)\right)
\end{equation}
up to discreteness effects in $L/k$. They conjecture that the coefficient 3 in the lower bound can be replaced by 1, and corroborate this claim by improved bounds for two classes of heavy-tailed
base distributions. Taken together with the identity (\ref{AN=BN}), these results lend strong support to the idea that the AN and BN models belong to the same 
universality class of NK structures, in the sense that $\ln \growthFactor{k}{AN, BN}  =  - \frac{\ln k}{k}$ to leading order in $k$ and $\ProbMaxModel{AN,BN} \sim L^{-1/\alpha}$ in the joint limit.     

\subsubsection{Random neighborhood}
\label{sec:LocalMaximaRN}
Although the random NK structure has been one of the most commonly studied neighborhood structures in the literature, little is known about the analytic behavior of $\growthFactor{k}{RN}$; in fact the 
existence of a well-defined exponential growth rate for $\nmax$ has been rigorously established only for the AN model \cite{Durrett2003}. 
In contrast to the BN or AN interaction structures which are defined in a deterministic manner, the RN model is marginally structured, in the sense that the neighborhood sets are 
realizations drawn from a random ensemble. 
Thus, it is of interest to ask how this marginal structure influences the behavior of $\growthFactor{k}{RN}$, now that we have seen that the maximally unstructured MF model 
belongs to a different universality class than the AN and BN models.
In order to answer this question, we choose to study the regular
random NK structure (rRN) as defined in \secref{SpecificStructures}.
The regular structure is chosen because it turns out to be analytically tractable. 
However, we claim that whether we assume regularity or uniformity on the NK structures should not matter  for sufficiently large $k$, 
since the fluctuations in the locus degrees or the size of NK blocks decay as $k^{-1/2}$.
We later numerically confirm that this is indeed true.


To proceed, let us first examine \eqref{CFNKGeneral}.
In contrast to the previously studied models with deterministic NK structures, the elements of the incidence matrix $b_{l,r}$ in the 
RN models may be considered as binary random variables constrained by the conditions i) $b_{l,l} = 1$ and ii)  $\sum_r b_{l,r} = k$ for all $l$ \cite{Dean2000}. 
The second condition ensures that the underlying NK structures are regular while the first condition represents the self-link condition imposed on classical NK structures. 
In our analysis, we found that the first condition does not play any significant role while introducing unnecessary complication.
Because the variable influenced by this condition is only one out of $k$ variables for each locus, the effect due to this condition should be at most $\Order{k^{-1}}$. 
Thus, as long as we focus on the leading asymptotic behavior, the condition i) can be dropped in the following analysis. 

The average over different realizations of the rRN NK structure can
now be emulated by promoting the $b_{l,r} $ to Bernoulli random variables.
These variables are assumed to be i.i.d with the Bernoulli success probability $p/L$ where $p$ is an arbitrary fixed constant in the limit $L\to\infty$.
Then, the average of a random quantity $Q$ over the rRN NK structure is given by
\begin{align}
\Avr{Q} \equiv \mathcal{N}^{-1}
	\Avr{
		Q \prod_l \delta_{\sum_{r} b_{l,r}, k} }_{\{b_{l,r}\}
	},
\label{EnsembleAverage}
\end{align}
where the angular bracket with subscript $\Avr{\cdots}_{\{b_{l,r}\}}$ indicates the average over the Bernoulli variables $\{b_{l,r}\}$, and
we have introduced a normalization constant 
\begin{align}
 \mathcal{N} = \Avr{
	 \prod_l \delta_{\sum_{r} b_{l,r}, k} }_{\{b_{l,r}\}
}.
\end{align}
Our goal is to evaluate \eqref{EnsembleAverage} for the quantity of interest, i.e., $Q = \ProbMaxModel{rRN}$. 


The normalization constant $\mathcal{N}$ is relatively simple to calculate.
Since the $\{b_{l,r}\}$ are independent Bernoulli variables, the total weight is given by the 
binomial distribution,
\begin{align}
\mathcal{N} &= \left[
\binom{L}{k} \left(1- \frac{p}{L}\right)^{L-k} \left(\frac{p}{L}\right)^{k}
\right]^L \simeq \left[
\frac{e^{-p} p^k}{k!} + O(1/L)
\right]^L.
\label{RNNormalization}
\end{align}
For the average of $Q$, it is convenient to use an integral representation for the Kronecker delta symbol,
\begin{align}
\delta_{x,n} = \frac{1}{2\pi} \int_{0}^{2\pi} e^{i(x-n)t} dt.
\end{align}
By combining this equation with \eqref{ProbMaxFormalism},
we may set up our starting equation for $ \Avr{\ProbMaxModel{rRN}} $ as
\begin{align}
&\Avr{
	\ProbMaxModel{rRN} \prod_l \delta_{\sum_{r} b_{l,r}, k} }_{\{b_{l,r}\}} = \int \Measure{t} \MeasureY \MeasureUQ \nonumber\\
& \times \Theta( 0 < \Vecbf{t} < 2\pi)  \Avr{ e^{ \sum_l i t_l (\sum_{r} b_{l,r}-k)}   \prod_{r, l} \left[
\CharFuncBase{-q_l} e^{i y_r q_l }
\right]^{b_{l,r}} 
}_{\{b_{l,r}\}}.
\end{align}
Here, we introduced another theta function $ \Theta( 0 < \Vecbf{t} < 2\pi) $ enforcing the condition $t_l \in (0,2\pi)$ for all $l$.
After averaging $\{b_{l,r}\}$ and neglecting terms of $O(1)$ in the exponential, one finds
\begin{align}
&\Avr{ e^{ \sum_l i t_l (\sum_{r} b_{l,r}-k)}   \prod_{r, l} \left[
	\CharFuncBase{-q_l} e^{i y_r q_l }
	\right]^{b_{l,r}} 
}_{\{b_{l,r}\}}	  \nonumber\\ 
&= \left( \prod_l e^{ - i t_l k}  \right)\exp\left[
- L p +  
\frac{p}{L}\sum_{l, r} \CharFuncBase{-q_l} e^{i y_r q_l + i t_l }
\right],
\end{align}
where we used the fact that $ \Avr{ e^{b_{l,r} x}}_{\{b_{l,r}\}} = 1 - \frac{p}{L} \left(1 -e^{ x} \right) = e^{- \frac{p}{L} \left(1 -e^{ x} \right) + \Order{L^{-2}}}$.
By defining a quantity $ \psi(q_l) = \frac{1}{L} \sum_{r} e^{ i q_l y_r} $,
the last term in the square bracket is succinctly written as
\begin{align}
\frac{p}{L}\sum_{l, r} \CharFuncBase{-q_l} e^{i y_r q_l + i t_l } &= p \sum_l e^{it_l}  \CharFuncBase{-q_l}  \psi(q_l).
\end{align}
After taking a short glance at the definition of $\psi(q_l)$, it is tempting to claim that this is simply the characteristic function of the base density function $p_f(y)$, because $\Vecbf{y} = \{y_r\}$ is drawn from the probability measure $\MeasureY = \prod_{r} dy_r p_f(y_r)$.
As long as $\Vecbf{y}$ is a typical realization, this claim must be true.
However, we cannot make this assumption, because the values of the $y_r$ conditioned on being a local maximum may not be typical.
Instead, we will call $\psi(q_l)$ the \defemph{sample characteristic function} realized by $\Vecbf{y}$.
Due to the structural similarity, this allows a (cumulant) expansion of the form 
\begin{align}
\ln \psi(q_l) &=  i q_l Y_1 - \frac{q_l^2}{2} ( Y_2 -  Y_1^2) + O(q_l^3),
\label{Cumulant}
\end{align}
where $Y_m = \frac{1}{L} \sum_r y_r^m$, the $m$-th \defemph{sample moment}.
However, one should keep in mind that the $Y_m$ are random variables which depend on the random vector $\Vecbf{y}$.

Now we are ready to evaluate the integrals over $t_l$. 
After applying the identity
\begin{align}
\frac{1}{2\pi} \int_{0}^{2\pi} e^{-i Q t + x e^{it}} dt = \frac{x^Q}{Q!}
\end{align}
to each of the integrals with respect to $t_l$,
we may factor out the equation as
\begin{align}
&\Avr{
	\ProbMaxModel{rRN} \prod_l \delta_{\sum_{r} b_{l,r}, k} }_{\{b_{l,r}\}
}	\nonumber\\
&=\int \MeasureY  \left[ 
\int_{0}^{\infty} \frac{d u}{2\pi} \int_{-\infty}^{\infty} dq  \, e^{-iu q - p} 
\frac{
	\left[
	p\, \psi(q ) \phi_f(-q) 
	\right]^k}{k!}
\right]^L.
\end{align}
Finally, dividing by the normalization constant \eqref{RNNormalization} yields
\begin{align}
\Avr{\ProbMaxModel{rRN}} = &\int \MeasureY   \left[ 
\int_{0}^{\infty} \frac{du}{2\pi} \int_{-\infty}^{\infty} d q \, e^{-i u q}  \left[
\psi(q ) \phi_f(-q) 
\right]^k
\right]^L.
\label{NKRNProbFullForm}
\end{align}
As expected from the fact that $p$ was introduced as an arbitrary parameter, the dependence on $p$ completely vanishes in the final equation. 

Next the $u$ and $q$ integrals in the square bracket may be evaluated by means of the steepest descent method assuming $k$ is sufficiently large. Reflecting the fact that $u$ can be arbitrarily large, $u$ is rescaled to $k u$ to have the same order in $k$ in the exponential.
Moreover, since the result of the integral should be real-valued, it is convenient to perform a complex rotation $q \to i q$.
Rewriting \eqref{NKRNProbFullForm} and denoting the integral in the square bracket by $I$, we have
\begin{align}
I = i k \int_{0}^{\infty} \frac{du}{2\pi} \int_{- \infty  i}^{ \infty i} d q \,  
\exp\left[ k \left(
u q - f(q)
\right)	
\right],
\end{align}
where 
\begin{equation}
f(q) = - \ln  \left[\psi(i q) \phi_f(-i q)\right].
\end{equation}
Then, the steepest contour is determined such that it passes through the saddle point satisfying 
the equation $u - f'(q_c) = 0$.
Along this contour, one finds 
\begin{align}
I = i k \int_{0}^{\infty} \frac{du}{\sqrt{2\pi}} \,  
\exp\left[ k \left(
u q_c - f(q_c)
\right)	
\right] \times \frac{1}{\sqrt{k f''(q_c)}}.
\end{align}
Subsequently, yet another saddle point approximation to the $u$ integral gives
\begin{align}
I &= \sqrt{ \frac{ - k}{f''(q_c)}}   \int_{0}^{\infty} \frac{du}{\sqrt{2\pi}} \,  
\exp\left[ k q_c'(u_c)\left(
u- u_c
\right)^2	
\right]
= \frac{\text{erf}\left(\frac{\sqrt{k} f'(0)}{\sqrt{-2 f''(0) }}\right)+1
}{2}  
\end{align}
where $u_c$ is defined by the relation $q_c(u_c) = 0$ (or equivalently $u_c = f'(0)$) and we used the reciprocal relation $q_c'(u) \frac{1}{f''(q_c(u))} = 1$ well known in the context of the Legendre transformation.
Surprisingly, if we are only interested in the leading behavior, this integral only depends on the two quantities $f'(0)$ and $f''(0)$.
Using the cumulant expansion \eqref{Cumulant}, we may rewrite $ I $ in terms of the first two moments as
\begin{align}
I(Y_1, Y_2) =\frac{1}{2} \left(\text{erf}\left((Y_1 - m_1) \sqrt{\frac{k}{2 ( Y_2 - Y_1^2 + m_2 - m_1^2  )     }}\right)+1\right)  
\label{NKLambdaPhiIntegral}	 
\end{align}
up corrections of the order of $k^{-1}$, 
where $m_q$ denotes the $q$-th moment of the base distribution $p_f$.
We emphasize that $I$ depends only on $Y_1$ and $Y_2$ by explicitly specifying them as the arguments of $I$.
Exploiting the fact that $\ProbMax$ is not affected by translation and scaling, we may take $m_1 = 0$ and $m_2 =1$ without loss of generality.
Hence, we have 
\begin{align}
I(Y_1, Y_2) =\frac{1}{2} \left(\text{erf}\left(Y_1 \sqrt{\frac{k}{2 ( Y_2 - Y_1^2 + 1 )     }}\right)+1\right) \left(1 + \Order{k^{-1}}\right).
\label{NKLambdaPhiIntegral2}	 
\end{align}
Since $I$ depends only on $Y_1$ and $Y_2$, 
the remaining task for the integral over $\Vecbf{y}$ is to  calculate the joint probability
\begin{align}
J(Y_1, Y_2) \equiv L^2 \int \MeasureY\, \delta\left (L Y_1 -  \sum_r y_r\right ) \delta\left (L Y_2 - \sum_r y_r^2\right ).
 \label{JointDistribution}
\end{align}
As $Y_1$ and $Y_2$ are sums of a large number of random variables, the large deviation principle implies that the joint probability 
should be of the form $  J(Y_1, Y_2) \sim e^{
	L  \mathcal{J}(Y_1, Y_2),
}$ where $  \mathcal{J}(Y_1, Y_2) $ is the corresponding rate function.
Once the joint probability is obtained for the given base distribution, we are finally ready to evaluate $ \ProbMaxModel{rRN} $ by means of the saddle point method,
\begin{align}
\ProbMaxModel{rRN}
=& \int dY_1 dY_2 \,  J(Y_1, Y_2) I(Y_1,Y_2)^L 
\nonumber \\
\sim&\,  \exp \left[{ L \mathcal{F}^{\mathrm{rRN}}(Y_1^*, Y_2^*)	
} \right],
\label{ProbMaxRegular}
\end{align}
where the starred variables $ (Y_1^*, Y_2^*) $ represent the solution of the extremum conditions on the action 
\begin{equation}
\label{RNAction}
\mathcal{F}^{\mathrm{rRN}}(Y_1, Y_2) =  \mathcal{J}(Y_1, Y_2) +  \mathcal{I}(Y_1,Y_2) \; \textrm{with}  \; \mathcal{I}(Y_1,Y_2) \equiv \ln I(Y_1, Y_2).
\end{equation}

As an example, let us suppose that our base distribution is a standard normal distribution.
This particular choice makes the calculation of the joint probability relatively easy.
Using the integral representation of the delta function, the joint distribution \eqref{JointDistribution} may be written as
\begin{align}
	J(Y_1, Y_2) =& \frac{L^2}{(2\pi)^2}\int dZ_1 dZ_2 \int \prod_r \frac{dy_r}{\sqrt{2\pi}} e^{-L\frac{Y_2}{2}} \nonumber\\
	 \times & \exp\left(  i L Y_1 Z_1 + i L Y_2 Z_2 - i Z_1\left[\sum_r y_r  \right] - i Z_2\left[\sum_r y_r^2 \right]\right) \nonumber\\
	=&  \frac{L^2}{(2\pi)^2}\int dZ_1 dZ_2 e^{L\left[
			-\frac{Y_2}{2 } + i Y_1 Z_1 + i Y_2 Z_2 + \frac{i Z_1^2}{4 Z_2} - \frac{1}{2} \ln \left( 2 i Z_2 \right)
		\right]} \nonumber\\
	\sim& \exp\left[
		\frac{L}{2}\left\{ 1 -Y_2 +\ln \left(Y_2-Y_1^2\right)\right\}
	\right],
	\label{JointDistributionGaussian}
\end{align}
where we have used the fact that the solution of the extremum conditions for $Z_1$ and $Z_2$ is given by $ Z_1 = -\frac{i Y_1}{Y_1^2-Y_2}, Z_2 = \frac{i}{2 \left(Y_1^2-Y_2\right)}$.
Once $J(Y_1, Y_2)$ is obtained, $\growthFactor{k}{rRN}$ is readily calculated by combining \eqref{JointDistributionGaussian} with \eqref{ProbMaxRegular}.

Now, we are ready to uncover the universal behavior hidden in \eqref{ProbMaxRegular}.
To describe it clearly, let us consider the limit $k \to \infty$ first.  
Note that the dependence on $k$ only appears in $ \mathcal{I}(Y_1,Y_2)$.
Examining the behavior of the error function shows that this limit effectively makes $\mathcal{I}(Y_1,Y_2)$ vanish.   
Furthermore, in the absence of the term $\mathcal{I}(Y_1,Y_2)$ in \eqref{RNAction} that prefers certain non-typical realizations of $\Vecbf{y}$, it is clear that the saddle point $(Y_1^\ast, Y_2^\ast)$ is given by the typical realizations, namely $Y_1^\ast = 0$ and $Y_2^\ast = 1$.
Hence, if we introduce the variables $\epsilon_q = Y_q - m_q$ for $q \in \{1,2\}$, they are expected to vanish for sufficiently large $k$, and this allows us to perform a series expansion with respect to these variables.
Using the general property of large deviation functions that the lowest order terms are given by a covariance matrix of $Y_1$ and $Y_2$ \cite{Touchette2009}, we find that 
\begin{align}
	\mathcal{J}(\epsilon_1, \epsilon_2) = \sum_{p, q\in \{1,2\}}  -\frac{1}{2} \epsilon_p \Sigma^{-1}_{pq} \epsilon_q + O(\mathrm{cubic\,\, in \,\, } \epsilon_q),
	 \label{JointDistributionExpansion}
\end{align}
where $\Sigma$ is the covariance matrix among $\epsilon_1$ and $\epsilon_2$. 
Specifically, the values are given by $\Sigma_{11} = L \mean{\epsilon_1^2} = 1$, $\Sigma_{12} = \Sigma_{21} = L \mean{\epsilon_2 \epsilon_1} = m_3$ and $\Sigma_{22} = L \mean{\epsilon_2^2} = m_4 - m_2^2 = m_4 - 1$.

\begin{figure}
	\includegraphics[width=1.0\textwidth]{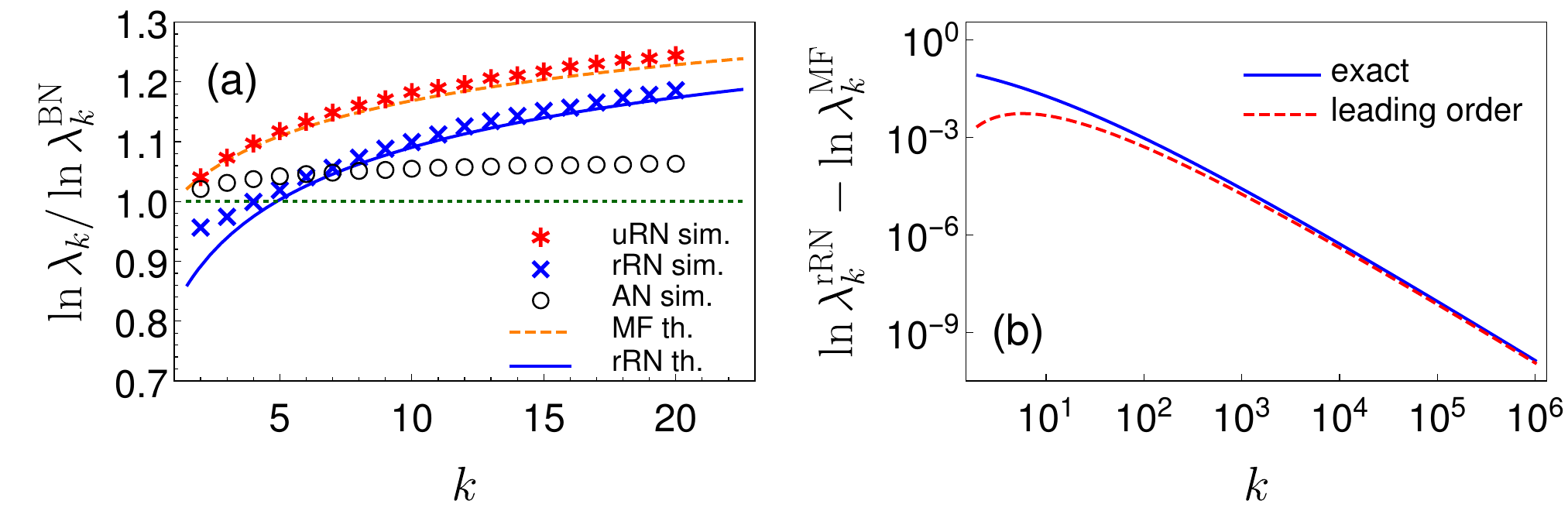}
	\caption{(a) Plots of $\ln \growthFactor{k}{ }/\ln \growthFactor{k}{BN }$ for various NK structures. The standard normal distribution is chosen as the base distribution for the simulations. The symbols denote the simulation data for each NK structure: stars for uRN model, crosses for rRN model and open circles for the AN model. In addition, analytical results for MF model (dashed) \eqref{ProbMaxMF}, rRN model (solid) \eqref{ProbMaxRegular} and BN model \eqref{ProbMaxBN} (dotted) are drawn, respectively. 
	Because the exponential growth rate is rescaled by $\growthFactor{k}{BN}$, the curve for BN is constant at $1$.
	The results for the two types of RN model converge to the asymptotic behavior of the MF model, while the curve for the AN model is expected to converge to one.
	(b) Plot of higher order corrections in the rRN model. 
	The exact value is calculated from \eqref{ProbMaxRegular} while the approximate value is obtained from \eqref{ProbMaxRegularToMF}. This shows that the leading correction behaves as $\frac{(\ln k)^2}{16 k^2}$.
	}
	\label{fig:RegularRandom}       
\end{figure}

On the other hand, the expansion for $\mathcal{I}(\epsilon_1,\epsilon_2)$ should be performed with caution due to the fact that $\sqrt{k} \epsilon_1$ cannot be assumed to be a small variable. 
Instead, the order of $\epsilon_1$ will be determined through this combined variable when $k$ is taken to $\infty$. 
Namely, the expansion takes the form 
\begin{align}
 	\mathcal{I}(\epsilon_1,\epsilon_2) = g(\sqrt{k}\epsilon_1) + g'(\sqrt{k}\epsilon_1) \frac{\epsilon_1 \sqrt{k}}{8} \epsilon_2  + O(\mathrm{quadratic\,\, in \,\, } \epsilon_q),
 	 \label{ProbExpansion}
\end{align}
with $g(\sqrt{k} \epsilon_1) =  \ln \left[ 
\frac{1}{2} \text{erf}\left(\epsilon_1 \sqrt{\frac{k}{4}}\right)+\frac{1}{2}
\right]$. 
Note that to lowest order, only $\epsilon_1$ appears.
Thus, as far as this leading order is concerned, the extremum condition for $\epsilon_2$ is readily solved by $\epsilon_2 = - \frac{\Sigma_{21}^{-1 }} {\Sigma_{22}^{-1 } }$, which then leads us to write the action 
as a one-dimensional function 
\begin{align}
\mathcal{F}^{\mathrm{MF}}(\epsilon_1) &=- \frac{1}{2} \epsilon_1^2 + \ln \left[ 
\frac{1}{2} \text{erf}\left(\epsilon_1 \sqrt{\frac{k}{4}}\right)+\frac{1}{2}
\right],
 	 \label{ProbMaxRegularToMF}
\end{align}
regardless of the specific form of the covariance matrix $\Sigma$. 
Surprisingly, if we identify $\epsilon_1$ with $y$, this is exactly the MF action \eqref{MFAction} to leading order in $k$. 
Thus, for sufficiently large $k$, the solution of \eqref{ProbMaxRegular} should converge to the MF solution.
This is confirmed by the simulation results shown in \figref{fig:RegularRandom} (a).

Next we turn to the corrections to the leading behavior.
Since these depend on the next-order terms of $ \mathcal{J}(\epsilon_1, \epsilon_2)$ which contain higher order correlations between $\epsilon_1$ and $\epsilon_2$, it is evident that this behavior is less universal than \eqref{ProbMaxRegularToMF}. 
Thus, in general not much can be said except the overall order of the corrections, which is at most $O((\epsilon_1^*)^3)$ (here the argument maximizing \eqref{ProbMaxRegularToMF} is denoted 
by $ \epsilon_1^* $).
Nevertheless, the next-order correction can be computed on a case by case basis once a distribution is given.
For the Gaussian distribution, a simple analysis shows that the next order correction of the saddle point equation gives $\epsilon_2^* = 0 \epsilon_1^* + \frac{1}{2}  (\epsilon_1^*)^2 + O((\epsilon_1^*)^3)$ and thus we arrive at 
\begin{align}
 	\mathcal{F}^{\mathrm{rRN}}(\epsilon_1^*, \epsilon_2^*) = \mathcal{F}^{\mathrm{MF}}(\epsilon_1^*) + \frac{1}{16} (\epsilon_1^*)^4 + O((\epsilon_1^*)^5 ).
\end{align}
Using the asymptotic expansion \eqref{ProbMaxMFAsymptotic}, we found that this correction is of the order of $\frac{(\ln k)^2}{k^2}$ (See \figref{fig:RegularRandom} (b)).

Now that we have established the universal behavior of $\growthFactor{k}{}$ for the case of the rRN model, it would be interesting to see if it applies also to other versions of the  RN model, 
e.g., the uniform (uRN) model.
For the sake of comparison we have performed simulations of this model with a standard normal distribution as the base distribution. 
This choice is made since it allows for an efficient numerical computation, which was first suggested in \cite{Buzas2014} (see Appendix \ref{appendix:Algorithm} for the details of the algorithm). 
Also, in order to test the effect of the self-link condition $b_{l,l} =1$ which has been ignored in the analytical calculation, the simulations were performed in the presence or the absence of this condition.
Fortunately, we found no significant difference between the results on the scale of \figref{fig:RegularRandom} and thus each NK structure is represented by a single curve without specifying 
whether the condition was implemented or not.

The simulation results turn out to be quite surprising in the sense that $ \growthFactor{k}{uRN} $ is extremely close to $\growthFactor{k}{MF}$ for all the parameter ranges we checked.
This supports our claim that a wide class of RN models is asymptotically MF-like as long as $k$ is sufficiently large.
 
Finally, we emphasize that the seemingly constant gaps shown in \figref{fig:RegularRandom} between the simulation data and the theoretical curves for the 
RN models  are artifacts that originate from the normalization by $\ln \growthFactor{k}{BN}$. 
In order to justify this statement, recall that our solution \eqref{NKLambdaPhiIntegral2} is correct up to the order of $O(k^{-1})$. 
Since the results are rescaled by $\growthFactor{k}{BN} \sim \frac{\ln k}{k}$, the gaps decay only as $(\ln k)^{-1}$, which effectively remains constant over the range of $k$ covered
in the simulations.

\subsubsection{Star neighborhood}
\label{MaximaSN}

\begin{figure}
  \begin{minipage}[c]{0.55\textwidth}
  	\includegraphics[width=\textwidth]{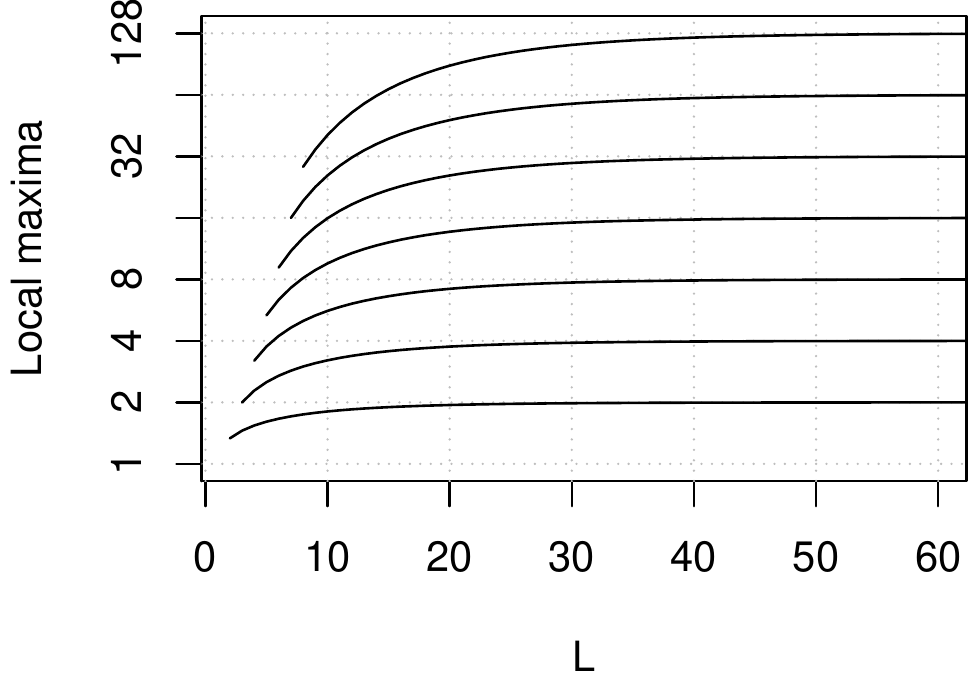}
  \end{minipage}\hfill
  \begin{minipage}[t]{0.40\textwidth}
  \vspace*{-0.5cm}
    \caption{\label{fig:MaximaSN}
		Mean number of local maxima for the SN structure versus number of loci $L$. From bottom to top the lines represent the cases of $k=2$ to $8$. 
		For large $L$ the number of maxima converges to $2^{k-1}$. Gaussian base fitness was used.
	}
  \end{minipage}
\end{figure}

As our last example of NK structures we consider the star neighborhood (SN) introduced in Sec.~\ref{SpecificStructures}. In contrast to all other structures discussed 
so far, the number of local maxima in the SN model remains finite for $L \to \infty$ and hence formally $\growthFactor{k}{SN} = 1/2$
(see \figref{fig:MaximaSN}).

The analysis of the SN model can largely be based on combinatorial arguments. 
Suppose the $k-1$ center loci are fixed in a certain configuration, 
and let us first determine the number of local maxima with respect to the remaining $L-k+1$ ray loci under this background.
Since none of the ray loci appear together in any NK edge, a mutation on one of them cannot affect the sign of the mutational effect on another, i.e. they are pairwise completely non-epistatic.
Thus each ray locus can be mutated into its state contributing higher fitness and this state is the unique (global) fitness maximum in the subspace 
of ray loci for the given background of center loci.
Since this is true for every allele combination of center loci, there can be at most $2^{k-1}$ local maxima on the star neighborhood.
This by itself already proves that $\growthFactor{k}{SN} = 1/2$.

For the scaling limit $L \to \infty$ at constant $k$, we can find bounds on the probability that the candidates for local maxima 
identified above are actually realized.
Because a mutation on a center locus affects all NK edges, the ray locus sub landscapes for each background of center loci are statistically independent.
Therefore, starting from the local maximum candidate constructed for one allele combination of the center loci and applying a mutation to one of the center loci,
the new fitness value $F'$ is a sum of $L$ i.i.d. random variables drawn from $p_f$. This is to be compared to the fitness value $F^\mathrm{cand}$ 
of the candidate configuration, which was obtained by maximizing each of the $L-k+1$ ray locus contributions between the two possible states of that locus.  
Thus $F^\mathrm{cand}$ is the sum of maxima of $L-k$ pairs of random variables drawn from $p_f$, 
plus one maximum of a pair of random variables drawn from the $k$-fold convolution of $p_f$; the last contribution originates from the special ray locus which is 
contained in all the blocks associated with the center loci.

 Except for the deterministic distribution, the expected value of the maximum of two independent draws from a probability 
distribution is always greater than the mean of the distribution itself, and therefore the mean of $F^\mathrm{cand}-F'$ grows linearly in $L$.  
At the same time the variance, as long as it exists, also grows linearly in $L$.
Thus by Chebyshev's inequality the probability for the mutation to lead us to discard the candidate local maximum is decreasing as $\frac{1}{L}$.
As the number of possible mutations of the center loci 
is also independent of $L$, it follows that the probability of each of the local maximum candidates not to be an actual local maximum is decreasing as $\frac{1}{L}$.
Therefore at constant $k$, $\mean{\nmaxModel{SN}} = 2^{k-1}\left(1-\mathcal{O}\left(\frac{1}{L}\right)\right)$.
Because we also know that $2^{k-1}$ is a strict upper bound, it follows that $\lim_{L\rightarrow\infty}\prob{\nmaxModel{SN} \neq 2^{k-1}} = 0$.

\subsection{Summary}
\label{Minima:Summary}
In this section, we have investigated the expected value of the number of local fitness maxima for various NK structures.
By developing a new analytic framework that allows us to treat different structures in a unified manner, 
we have discovered that the exponential growth rate of this quantity behaves asymptotically as 
\begin{equation}
\label{Defbeta}
\ln  \growthFactor{k}{} \sim -\beta \frac{\ln k}{k} 
\end{equation}
in the large $k$ limit, with the coefficient $\beta$ 
taking the values $\beta =1$ for  the AN and BN models and  $\beta =2$ for  the MF and RN models.
Similarly in the joint limit $k, L \to \infty$ at fixed $\alpha = k/L$, the probability $\ProbMax{}$ that a random genotype is a local maximum decays algebraically as
\begin{equation}
\label{JointGeneral}
\ProbMax{} \sim L^{-\mu}
\end{equation}
with $\mu = 1/\alpha$ for the AN and BN models, and $\mu = 2/\alpha -
1$ for the MF and RN models. The latter result has so far been
established only for the MF model, where it is modified by a
logarithmic correction (Appendix \ref{App:JointLimit}).
Although the change from $\beta = 1$ to $\beta = 2$ in
\eqref{Defbeta} may not seem very dramatic, it is important to
note that the corresponding numbers of fitness maxima $\nmax$
differ by a factor of $k^{L/k}$, which can be large already for
moderate values of $k$ and $L$. 
  
Because the AN and BN models can be considered to be more structured in a certain sense,
these results suggest that the fitness landscape is more rugged when the NK structure is more organized.
A similar conclusion was reached in \cite{Buzas2014} and \cite{Nowak2015}, where it was found that the number of maxima correlates negatively with the
rank (\ref{rankdef}) of the NK structure. Note, however, that the SN structure does not conform to this pattern, as its rank is relatively low (between the BK and AN models, see Table \ref{NKRanks}) 
whereas the number of maxima remains finite for $L \to
\infty$. 

As a next question, one might ask if other values of $\beta$ can be found or even further if other types of functional behavior can be realized for certain choices of NK structures.
Given the large variety of NK structures that is allowed by the definition of the model, the answers to both questions turn out be affirmative. 
To answer the first question, let us consider a somewhat contrived example. 
First, let us split the genotype sequence into two pieces of  size $L \rho$ and $L (1-\rho)$, respectively.
Furthermore, suppose that there is one NK block associated to each locus.
Next, let us assume that the NK blocks associated to the loci belonging to the first piece are constructed as if it were a BN model of size $L \rho$. For the second piece,
the NK blocks are created as in an RN model. 
Since there is no overlap between these two pieces by construction, the total number of local maxima is simply the product of those in each subsystem. From this, 
one may conclude that the asymptotic behavior of the exponential growth factor should be
\begin{align}
	 \ln  \growthFactor{k}{} = -\left[\rho + 2(1-\rho)\right] \frac{\ln k}{k}.
\end{align}
Thus, depending on the parameter $\rho$, the value of $\beta$ varies continuously from $2$ to $1$. 
However, this model does not allow for values that are outside of the range $1 \leq \beta \leq 2$.
In this sense, the value of $\beta$ is a measure of the amount of structure in the NK model.

In Appendix \ref{appendix:betabounds} we prove that $\beta \in [1,2]$ for all uniform
and regular structures if Gaussian fitness is assumed.
We expect this to hold for all sufficiently regular structures, although the
proof is likely to be somewhat more complicated.
Whether $\beta$ can take on other values if the base fitness
distribution is varied is open. Preliminary unpublished results for an extremely
heavy-tailed distribution suggested in \cite{Limic2004} seem to
indicate that the relation \eqref{Defbeta} may not even hold for certain uniform regular structures.
Nonetheless we expect at least distributions with finite moments to result in behavior equivalent to the Gaussian case, since for large $k$ fitness differences effectively converge to a jointly normal distribution following a kind of central limit theorem.

With regard to the second question, the example of the star neighborhood in Sec.~\ref{MaximaSN} shows that the exponential growth of the number of maxima 
with $L$ is not a general feature even among the classical NK structures. 
We attribute this inherently different behavior to the extreme non-regularity of the SN structure, where certain loci appear a macroscopic number of times.
Extending our analysis to other such non-regular structures might be an interesting future direction to further clarify the behavior of $\growthFactor{k}{}$.

\section{Accessible pathways}
\label{Sec:AccessiblePathways}

\subsection{Definitions}

There are many paths between far away genotypes.
However, some paths may be harder to take for a population, with some quasi impossible to take.
A path is called \defemph{accessible} if it increases fitness in each step \cite{Weinreich2006,Weinreich2005}.
This in particular implies that accessible paths are never circular
and that no genotypes can be visited twice on an accessible path.

We say a path from $\sigma$ to $\theta$ is \defemph{direct} if
$d_h(\sigma,\theta)$ is the number of steps taken, i.e. if the path
has minimal length, and \defemph{indirect} otherwise \cite{Wu2016}.
The number of allelic states
$A$ is largely irrelevant for the analysis of direct paths. In contrast,
indirect paths become more complex for $A > 2$ because of the
possibility of distance-neutral mutations that neither increase nor decrease the
distance to the target \cite{Zagorski2016}. 
Here we mostly restrict our analysis to the biallelic case,
where the genotype spaces are hybercubes.
Our results for the
NK-model presented in Sect.~\ref{Sec:LocallyBoundedNK} can however be
straightforwardly generalized to multiple alleles.

Direct paths on the hypercube mutate each locus at most once, i.e.\ there are no backwards mutations or mutational reversions \cite{DePristo2007}.
On the hypercube there exist exactly ${\left(d_h\left(\sigma,\theta\right)\right)!}$ direct paths between any two genotypes, in particular there are ${L!}$ direct paths between a genotype and its antipode.
The total number of (simple) paths including indirect paths is much larger, see \cite{Berestycki2014}.
In the following we denote the total number of accessible paths by $\npaths{\sigma\rightarrow\theta}$ and the number of direct accessible paths by $\nspaths{\sigma\rightarrow\theta}$.
If these numbers are non-zero we say that $\theta$ is \defemph{(direct) accessible} from $\sigma$.

Of particular interest are paths from a genotype $\sigma$ to its antipodal $\mut{\lset}\sigma$ as an approximate worst-case scenario.
Many genotypes are not accessible from their antipodal purely because their fitness is low compared to their neighbors.
As these cases are not very interesting, one may focus on high-fitness final genotypes.
Here looking at local maxima and in particular the global maximum $\Omega$ as destination seems natural \cite{Carneiro2010,Franke2011}.
We use the short-hand notation $\nspathsg = \nspaths{\mut{\lset}\Omega\rightarrow\Omega}$ and $\npathsg = \npaths{\mut{\lset}\Omega\rightarrow\Omega}$ respectively for direct and arbitrary paths to the global maximum from its antipodal.
The number of direct and indirect accessible paths to the global maximum has been studied for different fitness landscape models.
A major question of interest is the probability of existence of such paths for a large number of loci.
This problem is non-trivial.
On the one hand the number of possible paths between antipodal genotypes increases factorially (direct paths) or faster (indirect paths) with the number of loci.
On the other hand the number of fitness values needed to be found in monotonic order for a path to be accessible increases as well.
This bears similarity to certain percolation problems.
Therefore also the term \defemph{accessibility percolation} has been used to describe the probability of existence of paths to the global maximum from its antipodal
\cite{Nowak2013}.

Practically it is for some models, such as the NK model, difficult to condition on the global maximum.
Therefore it may be useful to consider a class of accessible paths larger than those discussed in the previous paragraph to describe a percolation property of the fitness landscape.
We call a landscape \defemph{(direct) traversable} if there exists a pair of genotypes at maximal distance $L$ with an accessible (direct) path between them.
This definition is more similar to traditional percolation problems, as no additional conditioning on the global maximum is required.


\subsection{House-of-Cards model}

Accessible paths to the global maximum from the antipodal point have been studied in detail in the limit of $L\rightarrow\infty$.
A simple combinatorial argument shows that $\mean{\nspathsg} = 1$ in the HoC model \cite{Franke2011}.
For this notice that any given direct path to the global maximum is accessible if all $L$ involved genotypes, excluding the global maximum itself, are ordered in ascending order.
Because all these values are i.i.d. this probability is $\frac{1}{L!}$.
As there are $L!$ such paths, the claim follows.

The distribution of $\nspathsg$ however becomes highly skewed for larger $L$, which implies that the mean is not informative of the typical behavior. 
Using the second moment method Hegarty and Martinsson showed that \cite{Hegarty2014}
\begin{equation}
    \prob{\nspathsg > 0} \sim \frac{\ln L}{L}
\end{equation}
as $L\rightarrow\infty$.
Thus the probability of finding any direct accessible path to the global maximum is decreasing in the number of loci, but slowly so.
Interestingly they also find that a slight modification of the HoC model obtained by fixing the fitness of $\mut{\lset}\Omega$ to a value corresponding to the quantile value $\alpha_L$ yields a threshold function $\alpha^\star_1(L) = \frac{\ln L}{L}$, such that $\lim_{L\rightarrow\infty} \prob{\nspathsg > 0} = 1$ for $\alpha_L = \alpha_1^\star(L) - \epsilon_L$ and $\lim_{L\rightarrow\infty} \prob{\nspathsg > 0} = 0$ for $\alpha_L = \alpha_1^\star(L) + \epsilon_L$ where $\epsilon_L > 0$ arbitrary, such that $\lim_{L\rightarrow\infty} L\epsilon_L = \infty$.
Thus the direct accessibility of the global maximum is, for large enough $L$, mainly constrained by the initial fitness and tends to $1$ in particular if the initial genotype is constrained to be the global minimum of the landscape. The limit distribution of direct accessible paths to the global maximum has been 
further studied in \cite{Berestycki2016}.

Berestycki et al. \cite{Berestycki2014} consider arbitrary length accessible paths to the global maximum from its antipodal and find a threshold behavior as well.
While $\mean{\npathsg}$ grows exponentially for $\alpha_L < \alpha^\star_2 = 1-\ln(\sqrt{2}+1) = 0.11863\ldots$, it decays exponentially to zero for $\alpha_L > \alpha^\star_2$.
Thus $\prob{\npathsg > 0}$ for the original HoC model must be asymptotically bounded from above by $\alpha^\star_2$.
Berestycki et al. conjecture that the expectation ``tells the truth'', i.e. that $\lim_{L\rightarrow\infty}\prob{\npathsg > 0} = 1$ for $\alpha_L < \alpha^\star_2$, which would also imply $\lim_{L\rightarrow\infty}\prob{\npathsg > 0} = \alpha^\star_2$ for the original HoC model.
This conjecture was proven by Martinsson \cite{Martinsson2015}.
Computational results for the HoC model with a larger number of alleles, $A > 2$, suggest that for any fixed number of alleles $\prob{\npathsg > 0}$ converges to values strictly between $0$ and $1$ as $L\rightarrow\infty$ \cite{Zagorski2016} .

\subsection{Block neighborhood}

The accessibility of the block model has been studied in \cite{Schmiegelt2014}.
Because mutational effects of loci on different blocks are completely statistically independent and fully additive, a path in the BN model is 
accessible if and only if the restriction of the path onto each block is accessible.
Additionally due to this independence of blocks, the global maximum of the full landscape will also be the global maximum on the individual blocks.
Thus, $\nspathsg$ will be a product of $\frac{L}{k}$ independent realizations of $\nspathsg$ for HoC landscapes with $k$ loci.

For the probability to find an accessible direct path to the global maximum in particular we have then
\begin{equation}
    \label{AccBN}
    \prob{\nspathsg[\text{BN}] > 0} = \prob{\nspathsg[\text{HoC}(k)] > 0}^{\frac{L}{k}}.
\end{equation}
Therefore at constant $k$, as $L$ increases, this probability decays exponentially to zero.
As explained above the direct accessibility for the HoC model goes as $\frac{\ln k}{k}$ for large $k$ and so at $\frac{k}{L} = \alpha$ fixed, asymptotically for large $L$
\begin{equation}
    \prob{\nspathsg[\text{BN}] > 0} \sim \left(\frac{\ln L}{\alpha L}\right)^{\frac{1}{\alpha}}
\end{equation}
which is still decreasing to zero, but more slowly.
In fact it is closer to the behavior of the HoC model.
The functional form is mostly the same, except for the modification by a power of $\frac{1}{\alpha}$ which implies a faster decay than in the HoC model when $\alpha < 1$.
By the same arguments \eqref{AccBN} holds for $\npathsg$ as well and again the decay at constant $k$ is exponential.
At fixed $L/k$ however, using the result for the HoC model that $\prob{\npathsg[\text{HoC}(k)] > 0}$ actually converges to a non-zero 
constant for $k \to \infty$, $\prob{\npathsg[\text{BN}] > 0}$ also converges to a non-zero constant under this scaling.


Using the same decomposition of the full path into subpaths within blocks, one can see that for the mean number of direct paths a similar equation
\begin{equation}
    \label{MeanPathsBN}
    \mean{\nspathsg[\text{BN}]} = \frac{L!}{k!^{\frac{L}{k}}}\mean{\nspathsg[\text{HoC}(k)]}^{\frac{L}{k}} = \frac{L!}{k!^{\frac{L}{k}}}
\end{equation}
holds.
The combinatorial factor describes the number of ways in which each set of direct accessible paths on blocks can be combined into a direct accessible path on the full landscape.
In fact every realization of $\nspathsg[\text{BN}]$ must be an integer multiple of this factor.
Thus $\mean{\nspathsg[\text{BN}]}$ increases super-exponentially both under constant $k$ scaling and when $k$ increases proportionally to $L$.
Due to the product structure, the actual distribution of $\nspathsg[\text{BN}]$, scaled by \eqref{MeanPathsBN} and conditioned on being larger than $0$, 
will at constant $k$ be asymptotically log-normal \cite{Schmiegelt2014}.

In summary, the BN model landscape has a large mean number of direct accessible paths, but this is actually hiding the fact 
that most landscape realizations do not contain a single such path.
However, if accessible paths exist, then the multiplicative structure guarantees that there are many. 
The decay of accessibility is much faster than in the HoC model for constant $k$ and moderately faster than in the HoC model for proportionate scaling $k \sim L$.

\subsection{Locally bounded NK structures are not traversable}
\label{Sec:LocallyBoundedNK}
\begin{figure}
    \includegraphics[width=\textwidth]{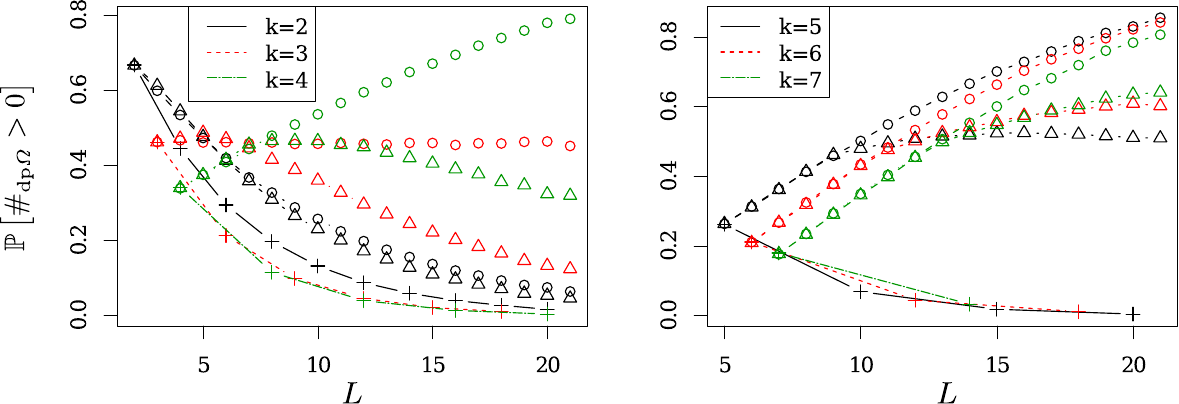}
    \caption{\label{Schmiegelt2014Fig8}
        Accessibility of the global maximum in some NK models.
        Number of loci $L$ on the x-axis and $\prob{\nspathsg > 0}$ on the y-axis for different values of $k$.
        Different NK structures are determined by symbols used:
        Triangles for AN, circles for uRN and crosses for BN.
        Adapted from \cite{Schmiegelt2014} and based on simulation results using Gaussian base fitness.
    }
\end{figure}

In the following we consider a large class of NK structures and their asymptotic traversability in the limit of large $L$ and constant $k$.

Let $\rho_l$ be the number of loci that have  graph distance $4$ or less to locus $l$ in the NK structure hypergraph.
We then say an NK structure is \defemph{(distance 4) locally bounded} if the mean of $\rho_l$ over all $l$ and with respect to realizations of randomized structures has finite limit superior.
In particular structures which are regular, uniform and have at most a linearly growing number of edges in $L$ are locally bounded (for arbitrary distances) in the limit $L\rightarrow\infty$ at constant $k$.
This holds, because the number of immediate neighbors of $l$ cannot be larger than the number of NK edges it is associated with times the number of elements in these edges.
Examples of such structures are the AN, BN and urRN models.
The boundedness property also holds for the (u)(r)RN models at constant $k$, because the degree distributions of loci become 
effectively independent and all their moments converge to $L$-independent values.
The SN and MF models are however not locally bounded because each locus can reach every other locus in two steps for the SN model and in one step for the MF model.
If $k$ is diverging as $L\rightarrow\infty$, then no uniform or regular NK structure can be locally bounded, because each locus is either 
a member of one edge with a diverging number of elements or a member of a diverging number of edges with at least one other member.

In 
\cite{Franke2011,Franke2012} direct accessibility of the global maximum has been studied for the uRN model via simulations.
Further simulation data can be found in 
\cite{Schmiegelt2014}, for the uRN model, as well as for the AN model.
As the BN model is also a representative of the class of locally bounded structures, one might have expected qualitatively similar behavior for these structures.
However the AN and uRN models seem to show a more complex behavior in the simulated parameter range, see \figref{Schmiegelt2014Fig8}. In particular,
the simulations indicate that accessibility increases with increasing $L$ for the uRN and AN models, at least for sufficiently large $k$.
Despite this apparent non-universal trend, it can be shown rigorously that the probability for the existence of traversing paths decays exponentially in $L$ for all locally bounded NK structures \cite{Schmiegelt2016}.
A short summary of the proof will be given here.

\begin{figure}
    \centering
    \includegraphics[width=0.8\textwidth]{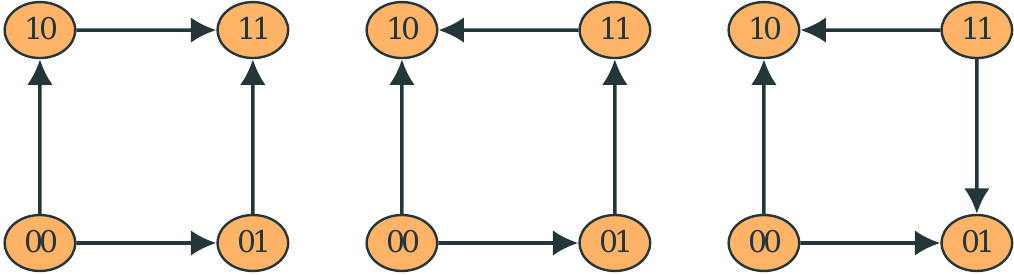}
    \\[1cm]
    \hfill\includegraphics[width=0.35\textwidth]{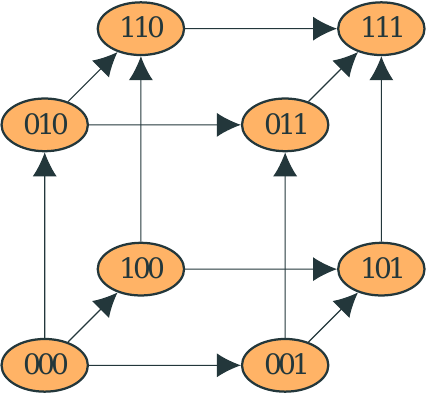}
    \hfill\includegraphics[width=0.35\textwidth]{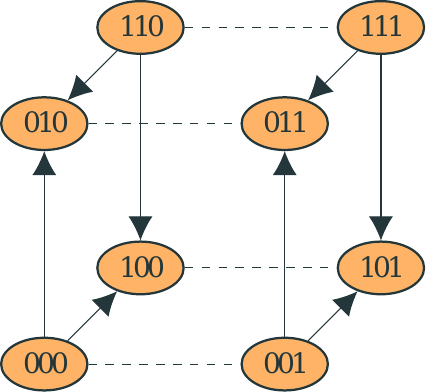}
    \hfill~
    \caption{\label{FitnessGraphs}
        Top: A single square in the fitness graph, without sign epistasis (left), with non-reciprocal sign epistasis (middle) and reciprocal sign epistasis (right).
        Bottom: Example of fitness graphs on three loci without any sign epistasis (left) and with global reciprocal sign epistasis (right). Each axis corresponds to one of the loci, and arrows point towards increasing fitness. On the right hand side loci $1$ and $2$ are globally reciprocal, as determined by the criterion that at any given point either both mutations are deleterious or both are beneficial. Additionally, as required by our definition of global reciprocal sign epistasis, the direction of arrows for loci $1$ and $2$ is identical for every state of locus $3$, i.e. the pattern in the $1-2$ plane
is simply translated along the 3-direction. 
Note that the direction of arrows for mutations on $3$ does not influence this property, and these arrows are therefore not specified. Given global reciprocal sign epistasis there can be no accessible path on the hypercube crossing the loci $1$ and $2$ simultaneously, and as a consequence one quarter of the 
nodes of the graph are always unexplorable for an adaptive walk, irrespective of its starting point.}
\end{figure}

Consider two loci $l$ and $m$.
There are four allele configurations for these two loci under any given background.
They span a $2$-dimensional hypercube, i.e. a square.
For a given background the associated fitness values can be in one of $4! = 24$ orderings.
However reducing some symmetries there are only three different types of fitness graphs, see \figref{FitnessGraphs}.
Either both sets of parallel arrows are oriented the same way, or only one of the two pairs is, or none.
The first case is the one without sign epistasis.
The second case identifies a sign epistatic dependence of one locus on the other but not the other way around.
And finally the last case shows reciprocal sign epistasis, i.e. sign epistasis between $l$ and $m$ in both directions.
This is the only case where the two-locus fitness landscape
displays two local maxima and minima, and in fact reciprocal sign epistasis is a necessary condition
for the existence of multiple maxima for any number of loci \cite{Poelwijk2011}. 
Additionally the square becomes non-traversable under reciprocal sign epistasis, 
because there is no accessible path from any corner to the antipode and the two loci cannot be 
mutated one after another on an accessible path.

In general, reciprocal sign epistasis between two loci is limited to a particular genetic background, a situation 
that we refer to as \defemph{local reciprocal sign epistasis}.
A third locus on the background may be mutated in-between $l$ and $m$ and thus allow the pathway to cross the square anyway.
Strict constraints on the traversability of the full landscape arise, however, if the reciprocal fitness ordering on 
the $l$/$m$-square is preserved for all backgrounds.
Then mutations in the background cannot influence the direction of fitness effects on $l$ and $m$, i.e. $l$ and $m$ 
are not sign epistatic with respect to any other locus under any background.
We call this \defemph{global reciprocal sign epistasis (GRSE)}.
It is identified by reciprocal sign epistasis between $l$ and $m$ on all backgrounds, as well as lack of sign epistatic dependence of $l$ and $m$ on any other locus on any background.
The existence of a single GRSE locus pair is sufficient to make the landscape non-traversable by direct or 
indirect paths, because the locus pair may never be mutated into the antipodal state together.
This then also implies $\npathsg = \nspathsg = 0$.

In locally bounded NK structures at constant $k$ as $L\rightarrow\infty$, the probability for existence of a 
GRSE pair of loci approaches unity exponentially fast in $L$.
Two loci can only be global reciprocal sign epistatic, or epistatic at all, if they share at least one NK edge.
But if they do share at least one NK edge, then there is a probability strictly between $0$ and $1$ for global reciprocal sign epistasis to occur.
For example there is a non-zero probability that $l$ and $m$ are globally reciprocal on the shared NK edge partial landscape (which is simply HoC) and that at the same time the smallest fitness difference on the NK edge containing $l$ and $m$ is larger than the largest fitness difference on all of the other NK edges containing either $l$ or $m$.
Of course the exact probability depends on the configuration of the NK structure around $l$ and $m$.
However only the NK edges containing $l$ and $m$ are of relevance.
Fitness values on other edges cannot contribute to fitness differences of mutations on $l$ and $m$.
Thus the subgraph of the NK structure around $l$ and $m$ determines the probability of $l$ and $m$ being global reciprocal sign epistatic.
Two pairs $l_1,m_1$ and $l_2,m_2$ of loci are then independent in their property of GRSE if they do not share any NK edges at all, i.e. if they lie in NK structure graph distance of at least $2$.

\begin{figure}
    \includegraphics[width=0.9\textwidth]{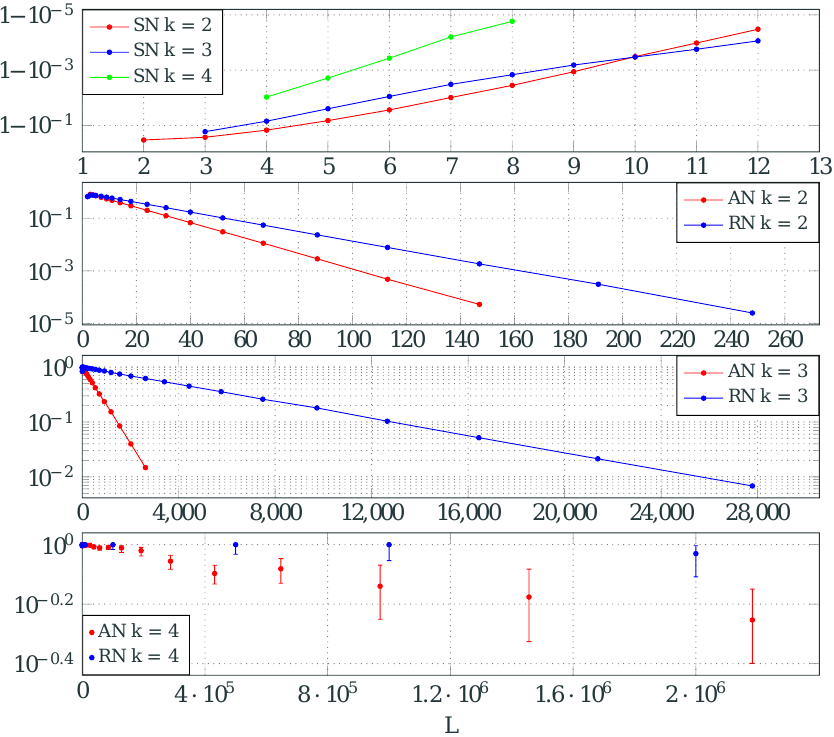}
    \caption{\label{GRSEProb} Simulation results for the 
    	probability not to find any global reciprocal sign epistatic pair of loci in the SN, AN and uRN models as a function of the number of loci and for different $k$.
    	The number of simulation runs per data point vary, but the
        error bars indicate the 95\% Clopper-Pearson confidence
        interval for the estimate. In the upper panels, error bars would be too small to be distinguishable from the data points. 
    	In the AN and uRN models, which are locally bounded structure
        choices, the probability decreases exponentially to zero for
        all $k\geq2$. This means that for large number of loci and
        fixed mean connectivity $k$, there will asymptotically almost
        surely be pairs of loci that are reciprocally sign epistatic on all backgrounds, such that adaptively accessible paths crossing the whole landscape become impossible.
    	The SN model, as a non-locally bounded alternative structure,
        behaves completely different. The probability not to find global reciprocal sign epistasis seems to approach 
unity as $L\rightarrow\infty$, resulting in no possible statement on the number of accessible paths one way or another. A Gaussian base fitness distribution was used in all cases.
	} 
\end{figure}

In locally bounded structures it is possible to find a non-zero fraction of loci for which the number of loci $\rho_l$ 
at distance 4 or less is smaller than some $L$-independent constant.
By way of elimination one can choose a linearly growing subset of these loci, such that they are additionally mutually separated by distance at least $4$.
These loci are then mutually independent in their property of GRSE.
As their degree must also be bounded by an $L$-independent constant, there are only a finite number of structures possible in their immediate neighborhood.
Thus the infimum over the individual probabilities of GRSE for all possible configurations is also strictly larger than zero.
Combining the non-zero infimum with the linearly growing number of independent realizations, the probability that there is no global reciprocal sign epistasis at all is at most a value smaller than $1$ taken to the power of a non-zero fraction of $L$.
Thus the probability to find GRSE approaches $1$ at least exponentially and the traversability decreases at least exponentially to zero, as do $\prob{\nspathsg > 0}$ and $\prob{\npathsg > 0}$.

This is consistent with the more precise result for the BN model, in which the traversability decreases exactly exponentially without any polynomial correction and with a growth rate derived from the corresponding property in the HoC model. 
The argument cannot however be applied to the SN model, as this structure is not locally bounded.

For the AN and uRN model these results seem to contradict the simulation results presented earlier, where it appeared 
that direct accessibility of the global maximum converges to $1$ instead of $0$ in the simulated range of $L$, for sufficiently large $k$.
However this turns out to be a small system size effect only.
The argument above was purely qualitative.
The actual decay rates for accessibility may scale extremely strongly with $k$, and in fact it becomes difficult to find 
GRSE in either model even at relatively large $L$ already for small $k$
(\figref{GRSEProb}).

\section{Adaptive walks}
\label{Sec:AdaptiveWalks}

Adaptive walks are a simplified class of evolutionary dynamics that arise
from a more comprehensive description, as provided, e.g., by the 
Wright-Fisher and Moran models \cite{Park2010}, 
in the limit of \defemph{strong selection and weak mutation} 
\defemph{(SSWM)} \cite{Gillespie1984,Orr2002}. 
The weak mutation condition states that the supply
of beneficial mutations is low enough to ensure that each newly arising 
mutation either fixes or goes extinct before another mutation appears.
Apart from the brief periods during which a clone of mutants
is on its way to fixation or extinction, the population is then almost always
monomorphic. The precise form of the fixation probability depends on the underlying
population dynamical model, but often the Kimura formula \cite{Kimura1962} 
\begin{equation}
\label{Kimura}
p_\mathrm{fix}^{N}(s) = \frac{1 - e^{-2s}}{1 - e^{-2Ns}}
\end{equation} 
is employed, where $s$ is the fitness difference between the mutant and the resident type and $N$ 
denotes the population size. Within the SSWM approximation, 
strong selection refers to the condition that the magnitude
of typical fitness differences $s$ scaled with the population size $N$ 
is large, $N \vert s \vert \gg 1$. According to (\ref{Kimura}) this implies that only beneficial 
mutations that increase fitness ($s > 0$) have a chance of going to fixation in the 
population. Thus in the SSWM regime, the population
can be regarded as a point in genotype space that moves
along paths of increasing fitness in single mutational steps. 
These are exactly the accessible pathways that were discussed in the preceding section, but the viewpoint here is different: Rather than just asking whether or not accessible pathways exist,
the adaptive walk models also address the likelihood that a given path is actually traversed by the evolving population. 

In the adaptive walk setting the waiting times for mutation and fixation 
events are ignored and the process is reduced to a discrete time 
Markov chain on the set of genotypes. 
It is evident from the derivation sketched above
that the transition probability $T(\sigma \to \theta)$ between two adjacent
genotypes is given by the fixation probability of the $\theta$-mutant
in the $\sigma$-background normalized by the sum over the fixation probabilities of all fitter genotypes
that are reachable from $\sigma$, 
\begin{equation}
\label{Transition}
T(\sigma \to \theta) = \frac{p_\mathrm{fix}^{\infty}(F(\theta) - F(\sigma))}{
\sum_{\tau \in {\cal{N}}_+(\sigma)} 
p_\mathrm{fix}^\infty(F(\tau) - F(\sigma))},
\end{equation} 
where ${\cal{N}}_+(\sigma) = \{\Delta_l \sigma \vert l \in \lset, \Delta_l F(\sigma) > 0 \}$
is the set of mutational neighbors of $\sigma$ that have higher fitness, and it is understood
that $p_\mathrm{fix}^{\infty}(s) = 0$ for $s < 0$.    

Three limiting cases of the dynamics \eqref{Transition} that arise from specific assumptions about the 
scale of the fitness differences are of particular interest. First, if all fitness differences are small in 
absolute terms, then the linear approximation $p_\mathrm{fix}^\infty(s) \approx 2s$ can be employed 
and the transition probabilities become proportional to the 
(positive) fitness differences. This is the setting originally
considered by Gillespie and Orr \cite{Gillespie1983,Gillespie1984,Orr2002}, 
and further studied in \cite{Jain2011,Jain2011a,Neidhart2011,Seetharaman2014a}. 
Conversely, if all (positive) fitness differences are large, then $p_\mathrm{fix}^\infty \to 1$ for all beneficial
mutants and $T(\sigma \to \theta) \to \vert {\cal{N}}_+(\sigma)\vert^{-1}$ independent of $\theta$, which implies that
any fitter neighboring genotype is chosen with equal probability. This defines
the \defemph{random adaptive walk} introduced by Kauffman and Levin \cite{Kauffman1987}. Finally, if the
fitness differences are very inhomogeneous, such that one of them is much larger than all the others, then
the Markov chain defined by (\ref{Transition}) moves deterministically to the neighboring genotype of largest
fitness. This limit of \defemph{greedy} adaptation was also addressed by Kauffman and Levin \cite{Kauffman1987} and 
studied in detail on uncorrelated fitness landscapes by Orr \cite{Orr2003}. So-called 
\defemph{reluctant} adaptive walks that move deterministically to the element of ${\cal{N}}_+(\sigma)$ that has lowest fitness have also
been considered \cite{Nowak2015}, though they seem to lack a natural interpretation in the framework of the 
general model defined by \eqref{Transition}. 

Importantly, the trajectories of random, greedy and reluctant adaptive walks are fully
specified by the rank ordering of the fitness values. This is a property that they share with the other probes of fitness landscape
ruggedness, local maxima and accessible pathways, that have been
discussed in the preceding sections. The primary measure of ruggedness is the average number of steps required for the walk to 
reach a local fitness maximum from a random starting genotype, a quantity that will be referred to as the \defemph{length} of the 
walk $\ell$. The known results for the walk length on the uncorrelated HoC landscape to leading order in $L$ are summarized
in Table \ref{tab:hoc_walk_results}. Greedy walks reach a local maximum after a finite (small) number of steps, whereas the walk
length diverges logarthmically in $L$ for random adaptive walks and linearly for reluctant walks.   

\begin{table}
\centering
\begin{tabular}{l|l|l|l}
    \rule{0pt}{1em}Walk type & Length $\ell$ & Height $1 - \kappa / L$ & References \\\hline
    \rule{0pt}{1em}Greedy & $e-1$ & $\kappa=0.4003\ldots$ & \cite{Nowak2015,Orr2003}\\
    \rule{0pt}{1em}Random & $\ln L$ & $\kappa=0.6243\ldots$ & \cite{Flyvbjerg1992,Macken1991,Nowak2015} \\
    \rule{0pt}{1em}Reluctant & $L / 2$ & $\kappa=1$ & \cite{Nowak2015a,Nowak2015} \\
\end{tabular}
\caption{\label{tab:hoc_walk_results}Properties of adaptive walks on the House-of-Cards landscape with
fitness values distributed uniformly on the interval $[0,1]$.}
\end{table}      

Analytical results for walk lengths on correlated fitness landscapes are relatively scarce, but some progress
has recently been achieved for walks on Rough Mount Fuji landscapes \cite{Aita2000,Neidhart2014}, 
a class of models defined by a weighted superposition of 
an additive fitness landscape and an uncorrelated random (HoC) landscape \cite{Park2015,Park2016a,Park2016b}. For the discussion of adaptive
walks on NK landscapes we start from the observation that the walk length is additive over blocks for the block
neighborhood \cite{Nowak2015,Perelson1995,Seetharaman2014a}, and therefore
\begin{equation}
\label{BN_length}
\ell_\mathrm{BN} = \frac{L}{k} \ell_\mathrm{HoC}(k)
\end{equation}
holds as an exact relation. The dependence on the walk type enters through the HoC walk length $\ell_\mathrm{HoC}$, the 
asymptotics of which can be read off from Table \ref{tab:hoc_walk_results}. Although \eqref{BN_length} is not quantitatively
correct for other interaction structures, it captures several important features of the walk length in the NK model. In particular
for fixed $k$ the walk length grows linearly in $L$, and the ordering among different walk types corresponds to that obtained
for the HoC landscape \cite{Nowak2015}.

An argument due to Weinberger \cite{Weinberger1991} links the linear dependence of the walk length on the number of loci
$L$ to the exponential decay of the density of fitness maxima $\pi_\mathrm{max} \sim (\lambda_k)^L$. Since the total
number of genotypes is $2^L$ and the number of maxima $\nmax \sim (2 \lambda_k)^L$, the average ``basin of attraction'' of a 
maximum contains $2^L/\nmax = \lambda_k^{-L}$ sequences. Such a basin can be visualized as a volume with a diameter $D$ given by 
\begin{equation}
\label{basin_diameter}
D = L \vert \log_2(\lambda_k) \vert,
\end{equation}
and Weinberger claims that $D/2$ provides a lower bound on the length of any adaptive walk, in particular on the length
of a greedy (or gradient) walk. Comparison with the exact relation \eqref{BN_length} shows that the latter statement is not 
quite true. Since the greedy HoC walk length has a finite limit $e-1$ for $k \to \infty$ whereas $\ln(\lambda_k^\mathrm{BN}) 
\sim -(\ln k)/k$, we see that $\ell_\mathrm{BN} \sim L/k \ll D \sim
(L/k) \log_2 k$ for large $k$. This discrepancy may be related to the
strong clustering of local maxima that has been observed in particular
for the BN neighborhood; we will return to this point below in
Section~\ref{Sec:Outlook}. 

Nevertheless the negative correlation
between the adaptive walk length and the density of local maxima suggested by \eqref{basin_diameter} is confirmed by detailed
simulations of different types of walks on NK landscapes with different interaction structures \cite{Nowak2015}. Walk lengths are always 
shortest on BN landscapes, intermediate on AN landscapes and longest on RN landscapes, and the walk length is positively correlated
with the rank of the interaction scheme. Whether the universality results obtained for $\lambda_k$ can be extended to 
adaptive walk lengths remains an open question for future work. 

Within the framework of abstract landscape theory it has been postulated that the length of adaptive walks 
should be related to the correlation length $\xi$ of the fitness landscape, which can be generally defined in terms of the distance correlation function by \cite{Reidys2002,Stadler1996,Stadler1999}
\begin{equation}
\label{correlation_length}
\xi = \sum_{d=0}^\infty \rho(d).
\end{equation} 
Inserting the expression \eqref{NKDistanceCorrelation} for the classical NK structures one finds the simple result
\begin{equation}
\label{correlation_length_NK}
\xi = \frac{L+1}{k+1}.
\end{equation}
This is of the same leading order as, but generally smaller than \eqref{BN_length}, which is expected to be a lower bound on the adaptive walk lengths
(note that $\ell_\mathrm{HoC} \geq e-1 > 1$ according to Table \ref{tab:hoc_walk_results}).

Apart from the length of an adaptive walk it is also of interest to consider
the \defemph{height} reached, i.e. the fitness value of the local maximum at which the walk terminates.
Results for the height of adaptive walks on HoC landscapes
are summarized in Table \ref{tab:hoc_walk_results}, where fitness
values are assumed for concreteness to be uniformly distributed on
the unit interval. On this scale the expected fitness value of a
randomly chosen local maximum is $1 - 1/(L+2) \approx 1 - 1/L$ for
large $L$. It can thus be seen from Table  \ref{tab:hoc_walk_results} that random and greedy adaptive walks
terminate at local maxima of atypically high fitness, and that greedy
walks are more efficient than random walks in reaching exceptionally
high peaks. Whether or not the fitness peaks located by an adaptive
walk are typical is of interest in situations where walks are used to
explore empirical fitness landscapes that are too large for local
maxima to be enumerated exhaustively \cite{Bank2016,Kouyos2012}.    

A numerical study of walk heights on NK fitness landscapes revealed a
surprisingly complex dependence on the interaction
structure and the type of the walks \cite{Nowak2015}. For the BN and
AN structures the greedy (reluctant) walks are most (least) efficient
in locating high fitness peaks, as might be expected from the results
for the HoC landscapes, but for the RN structure this order can be
reversed in a range of $k$. At fixed $k$ the walk height generally
increases with the rank of the interaction structure.

\section{Discussion and conclusion}
\label{Sec:Summary}

\subsection{Biological implications}

Conflicting intuitions about the topography of fitness landscapes have been the cause of debate in evolutionary theory ever since the concept first appeared \cite{Gavrilets2004,Svensson2012}. 
Whereas Sewall Wright argued that these landscapes are likely to possess ``\textit{innumerable peaks...which are separated by valleys}'' and stressed the need to understand how evolution is able to find its
way ``\textit{from lower to higher peaks}'' \cite{Wright1932}, his opponent Ronald Fisher thought that the problem would not present itself because of the high dimensionality of genotype space \cite{Provine1986}. 
At its mathematical core, Fisher's argument is a statement about the overwhelming likelihood of extrema of high-dimensional differentiable functions to be saddle points rather than maxima or minima and as such, it ignores the specific, discrete structure of the space of genotypes. 

An important role of the probabilistic fitness landscape models considered in this review is that they allow us to phrase and answer questions
about the generic structure of genotypic fitness landscapes in precise mathematical terms \cite{deVisser2014}. 
In a certain sense, they show that Wright and Fisher were both right: Although it is true that the fraction of fitness
peaks among all genotypes, $\ProbMax$, generally decreases with increasing genotype dimensionality, this is more than offset by the exponential growth of the total number genotypes in such a way
that the number of peaks $\nmax$ also grows exponentially. We have seen that, in the NK models, $\ProbMax$ decays exponentially or algebraically in $L$ depending on whether the epistasis parameter
$k$ is kept constant or scaled to infinity, and the SN structure exemplifies the kind of epistatic interactions that are required for $\nmax$ 
\textit{not} to diverge when $L \to \infty$.  

The general biological message of our work is that the ruggedness of a
fitness landscape depends not only on the amount of genetic
interactions, but also on how these interactions are
organized. Whereas the fact that epistastic interactions are ubiquitous
and often lead to complex fitness landscapes is now widely
appreciated, researchers are only beginning to pose more refined
questions regarding the structure of the interactions. For example,
several recent articles have addressed the prevalence and
evolutionary role of higher-order interactions that cannot be reduced 
to contributions from pairs of loci \cite{Crona2017,Sailer2017,Weinreich2013}. 

Within the class of
NK-models, the parameter $k$ specifies the highest order of
interactions that are present in the system \cite{Neidhart2013}. Comparing different 
NK interaction structures at a given $k$ thus amounts to exploring effects that go
beyond the interaction order and involve more subtle aspects of
genetic architecture. In this regard, our analysis shows that two
structural paradigms that can be regarded as extremes in a spectrum of
possible architectures, the perfectly modular BN structure and the
strongly hierarchical SN structure, also represent extremes with
respect to the ruggedness of the resulting fitness landscape: The BN
landscape has the largest number of fitness maxima, whereas the number
of maxima in the SN landscape remains finite for $L \to \infty$. 
The (deterministic) AN structure and the (random) RN
structure are intermediate between these two limits, but AN
landscapes are more rugged than RN landscapes for large $k$. 

We hope that these analyses can serve as a starting point for further
exploration of other, empirically motivated interaction
schemes. Recent high-throughput experiments on protein fitness
landscapes suggest that it is principally feasible to extract the interaction structure by
determining the type of epistasis between pairs of loci
\cite{Bank2016,Pokusaeva2017}. 

\subsection{Universality}

One reason for the broad appeal of the NK models in the description of fitness landscapes lies in their promise of universality, in that quantities like $\ProbMax$ and $\nmax$ depend only on the 
gross parameters $L$ and $k$ (at least when both are large), and are robust against changes in the detailed interaction structure and the underlying base fitness distribution. Our new analyses presented in
Sec.~\ref{Sec:LocalFitnessMaxima} confirm that universality holds,  but it is more restricted than previously appreciated. Specifically, we find evidence for two distinct universality classes characterized by different asymptotic behaviors of $\ln \growthFactor{k}{}=\lim_{L \to \infty} L^{-1} \ln \ProbMax$ for large $k$. It should nevertheless be emphasized that the degree of universality with respect to the base fitness distribution $p_f$ is very strong, as evidenced by the results of \cite{Limic2004} as well as by our computation for the RN model in Sec.~\ref{sec:LocalMaximaRN}.
In this respect the NK landscapes differ markedly from the Rough Mount Fuji (RMF) model, 
another class of tunably rugged fitness landscapes for which an explicit expression for $\ProbMax$ can be derived, and where the asymptotic behavior of this quantity is dominated 
by the tail properties of $p_f$ \cite{Neidhart2014}. 

Whereas the number of fitness maxima remains the most commonly used quantifier of ruggedness, the statistics of accessible pathways and adaptive walks reviewed in 
Sections~\ref{Sec:AccessiblePathways} and \ref{Sec:AdaptiveWalks} address the 
searchability of fitness landscapes in a more direct way. Following the terminology first introduced by Weinreich and collaborators, a pathway is called accessible if it is monotonically
increasing in fitness, and a landscape is accessible if the global fitness maximum can be reached through an accessible pathway starting from its antipodal point 
\cite{Carneiro2010,Franke2011,Weinreich2005,Weinreich2006}. The central result outlined in Sec.~\ref{Sec:LocallyBoundedNK} 
states that the probability for an NK fitness landscape to be accessible decays exponentially in $L$ whenever
the interaction structure is locally bounded, a property that applies to all commonly used structures. Somewhat counterintuitively, this implies that NK landscapes are much less 
accessible than uncorrelated HoC landscapes, for which the decay is only algebraic and moreover accessibility can be boosted simply by choosing a starting point of low fitness 
\cite{Berestycki2016,Hegarty2014}. This shows that local fitness peaks and accessible pathways reflect distinct properties of fitness landscapes that cannot easily be subsumed into a single
notion of ruggedness.  Importantly, the exponential decay of accessibility was not seen in earlier numerical work on the NK model because of the extreme scarcity of the crucial GRSE motifs
for large $k$.

\subsection{Outlook}
\label{Sec:Outlook}

The results described in Sec.~\ref{Sec:LocalFitnessMaxima} suggest  a number of promising avenues for future work on NK fitness landscapes. 
On the side of mathematical analysis, a more rigorous treatment of the joint limit ($L, k \to \infty$ at fixed $\alpha = k/L$) for the $\beta = 2$ universality 
class comprising the MF and RN models would be desirable. Also the intriguing role of the BN model in providing a possibly universal
upper bound on the number of maxima among all interaction structures and base fitness distributions should be elucidated. Finally,
it seems important to direct the attention to the way fitness maxima are organized in sequence space, rather than just focusing on their sheer number. 
A numerical investigation reported in \cite{Nowak2015} found that local maxima are strongly clustered, and the degree of clustering is highly dependent
on the interaction structure. A better understanding of the organization of maxima would also be helpful in strengthening the link between 
static landscape properties and the dynamics of adaptive walks
evolving on the landscape, which is so far quite sketchy (see
Sec.~\ref{Sec:AdaptiveWalks}).  
A useful tool for such an analysis is a network approach where the
vertices are fitness maxima and the links quantify the overlap between their
respective basins of attraction \cite{Tomassini2008}.  

Among the plethora of research problems that present themselves beyond the specific context of NK models, 
we here choose to point the reader to the study of time-dependent fitness landscapes which are sometimes referred to as fitness seascapes \cite{Mustonen2009}.
Natural fitness landscapes are never entirely static, and time-dependent effects are crucial for the explanation of fundamental evolutionary phenomena such as the selective advantage of recombination
\cite{Nowak2014}.  In 1999, Wilke and Martinetz introduced a time-dependent variant of the NK model \cite{Wilke1999}, which subsequently was picked up by the glass physics community  \cite{Isner2006}
and is meanwhile used routinely for the description of periodically stressed disordered solids \cite{Fiocco2014}. 
This example shows that the transfer of concepts across the interface between evolutionary biology and statistical
physics can go both ways, and that further exchanges in this area can be expected to produce surprising and innovative results.

\paragraph{Acknowledgments.} We thank David Dean for useful
discussions, and an anonymous reviewer for constructive remarks on the
manuscript. 
JK acknowledges the kind hospitality of the MPI for Physics of Complex Systems (Dresden) 
and the Kavli Institute for Theoretical Physics (Santa Barbara) during the completion 
of the paper.  This research was supported by DFG within SFB 680 \textit{Molecular basis of evolutionary innovations} and 
SPP1590 \textit{Probabilistic structures in evolution}, and in part 
by the National Science Foundation
Grant No. NSF PHY-1125915, NIH Grant No. R25GM067110, and the Gordon
and Betty Moore Foundation Grant No. 2919.01.

\bibliographystyle{spmpsci}
\bibliography{biblio}

\appendix

\section{Asymptotics of $ \ProbMaxModel{MF} $ in the joint limit $k, L
\to \infty$}
\label{App:JointLimit}
We start from \eqref{ProbMaxMF}. 
Rescaling $y \to \frac{\eta y}{\sqrt{2}}$, we rewrite the equation in
terms of the CDF of a standard Gaussian distribution  $ \Phi(y) $ as
\begin{align}
\ProbMaxModel{MF}
&= \sqrt{\frac{L \eta^2}{4\pi}  } \int dy  e^{-L \eta^2 y^2/4} \left[
\frac{1}{2} \left(\text{erf}\left(\frac{y}{\sqrt{2}}\right)+1\right)
\right]^L  \nonumber \\
& = \sqrt{\frac{\mu}{2\pi}  } \int dy  e^{-\mu y^2/2} 
\Phi(y)^L,
\end{align}
where $\mu \equiv \frac{L \eta^2}{2}$ which converges to $\frac{(2-\alpha)}{\alpha} $ in the joint limit as can be seen from \eqref{MFSingleParameter}.

Interestingly, the only $L$-dependence shown in the above equation
appears as an $L$-th power of the CDF $\Phi(y)$, which converges
monotonically to unity as $ y \to \infty$. 
This implies that the conventional saddle point method cannot be applied here due to the absence of a maximum.
Instead, we can rely on the extreme value theory by interpreting the term $\Phi(y)^L$ as the probability that $L$ randomly sampled standard Gaussian random variables are less than $y$.
This leads immediately to the limit relation \cite{Haan2006} 
\begin{align}
	\Phi\left (\frac{x}{a_L} + b_L\right )^L \to G(x)( 1 + o(1)),
\end{align}
where $G(x)$ is the Gumbel CDF defined by $G(x) = e^{- e^{-x}}$, and
the two scaling factors are given by $a_L = \sqrt{2 \ln L}$ and 
\begin{align}
	b_L = \sqrt{2 \ln L} - \frac{\ln \ln L + \ln 4\pi }{2 \sqrt{2 \ln L}}.
\end{align}
After making the change of variable $y = \frac{x}{a_L} + b_L$, the integral is now of the form
\begin{align}
	\ProbMaxModel{MF}   & = \frac{1}{a_L} \sqrt{\frac{\mu}{2\pi}  } \int dx  e^{-\mu \left(  \frac{x}{a_L} + b_L  \right)^2/2} 
	\Phi\left (\frac{x}{a_L} + b_L\right )^L \nonumber\\
	&= \frac{1}{a_L} \sqrt{\frac{\mu}{2\pi}  } \int dx  e^{- \mu \left(  \frac{x}{a_L} + b_L  \right)^2/2} 
	G(x) \left(1 + o(1)\right).
\end{align}
The evaluation of the integral with respect to $x$  
is greatly simplified once one notices that the term $\frac{x^2}{a_L^2}$ in the exponent is sub-leading in $L$. Ignoring this term gives 
\begin{align}
	\ProbMaxModel{MF}   & =\frac{1}{a_L} \sqrt{\frac{\mu}{2\pi}  } \int dx  e^{- \mu \left( b_L^2 + 2  \frac{b_L x}{a_L}   \right)/2} 
	G(x) \left(1 + o(1)\right)\nonumber \\
	&= \frac{1}{a_L} \sqrt{\frac{\mu}{2\pi}  } e^{-\mu b_L^2/2} \Gamma(
		\mu)\left(1 + o(1)\right),
\end{align} 
where we have used the identity
\begin{align}
	\int_{-\infty }^{\infty } G(x) \exp (-M x) \, dx = \Gamma (M)
\end{align}
for positive $M$.
Next, expanding $a_L$ and $b_L$ and rearranging the terms gives
\begin{align}
	\ProbMaxModel{MF}  
	&= \frac{1}{a_L} \sqrt{\frac{ \mu }{2\pi}  } e^{- \frac{\mu}{2} \left[
		2 \ln L - \left(\ln \ln L + \ln 4 \pi + o(1)\right)
		\right]} \Gamma(\mu)\left(1 + o(1)\right) \nonumber \\
	&= \sqrt{\mu} {\frac{ \left(4\pi \ln L\right)^{\mu/2} }{\left(4\pi \ln L\right)^{1/2}}}   \Gamma(
	\mu) L^{-\mu}  \left(1 + o(1)\right).
\end{align}
As expected from the formal analysis in \secref{sec:LocalMaximaMF},
the leading order behavior is given by a power law with exponent $\mu
= (2-\alpha)/\alpha$. By contrast, the existence of a non-trivial
logarithmic correction is unexpected, in particular since such a
correction does not appear in the exact result $ \ProbMaxModel{HoC} =
(L+1)^{-1} $ for the HoC model ($\alpha = \mu = 1$). 
Remarkably, the logarithmic factors precisely cancel in this
particular case. 

\section{Variational analysis at the maximum of $\growthFactor{k}{AN}$}
\label{appendix:VariationalAnalysis}
In \figref{fig:AdjacencyNKGamma}, we observed that $\growthFactor{2}{AN}$ for the negative
gamma distribution with shape parameter $s$ is maximized at $s=1/2$.
Furthermore, we claimed that this can be naturally generalized to arbitrary values of $k$ if we replace the shape parameter by $1/k$.
As a next question, one might further ask if 
$\growthFactor{k}{AN}$ is an extremum also with respect to arbitrary variations in the space of base fitness distributions $p_f$.
Here, we prove that this is indeed the case for distributions with support limited to
the negative real axis.

Let us first evaluate the $k$-fold convolution of the gamma distribution needed to compute \eqref{ProbMaxAN}.
This is easily achieved using the property that the gamma distribution is closed under the convolution operation, i.e., the $k$-fold convolution of the gamma distribution with shape parameter $s$ is the gamma distribution with shape parameter $s k$.
If we choose as our base distribution the negative gamma distribution with shape parameter $s=1/k$, 
\begin{equation}
\label{defshape}
p_f(x) = p_{1/k}(x) \equiv g_{1/k}(-x),
\end{equation}
the $k$-fold convolution yields the gamma distribution with unit shape parameter a.k.a. a (negative) exponential distribution, 
characterized by the CDF $ \tilde{F}_{1/k}^{(k)}(z) = e^z $ for $z<0$. 
Since $ \tilde{F}_{1/k}^{(k)}(y_1 + y_2 + \cdots) = e^{y_1} e^{y_2} \cdots$, \eqref{ProbMaxAN} is fully factorized as
\begin{align}
\ProbMaxModel{AN} = \left (\int dy \, g_{1/k}(-y)  e^{ky} \right )^L = 	(k+1)^{-L/k},
\label{Probability}
\end{align}
which is exactly the result for the block model obtained in \eqref{ProbMaxBN}.

Next, let us derive a useful general formula for $ \tilde{F}^{(k)}(z) $.
Using the convolution theorem, it satisfies
\begin{align}
\tilde{F}^{(k)}(z) &= \int_{-\infty}^{z} dz' \int_{z'}^{\infty} dy\, p_f(y) \, p^{(k-1)}_f(z'-y) 
\end{align}
where $ p^{(k-1)}_f(z) $ is the PDF of the $k-1$ fold convolution of $ p_f(z)  $.
It will later be convenient to exchange the order of integrals:
\begin{align}
\tilde{F}^{(k)}(z) &= \int_{-\infty}^{z} dy\, p_f(y) \int_{-\infty}^{y} dz' \, p^{(k-1)}_f(z'-y) 
+ \int_{z}^{\infty} dy\, p_f(y) \int_{-\infty}^{z} dz' \, p^{(k-1)}_f(z'-y) \nonumber \\
&= \int_{-\infty}^{z} dy\, p_f(y) 
+ \int_{z}^{\infty} dy\, p_f(y) \tilde{F}_s^{(k-1)}(z-y).
\label{Convolution}
\end{align}
In the first equality, we split the integral into two pieces to accommodate the condition $p^{(k-1)}_f(z) =0$ for positive $z$. In the next equality, we have
used the fact that $\tilde{F}^{(k-1)}(0) =1$.

Now, we want to show that $\ProbMaxModel{AN}$ is maximized 
when the base fitness distribution is given by \eqref{defshape}.
To this end, let us introduce a small perturbation $p_f(y) = p_{1/k}(y) + \epsilon \eta(y)$, with the properties that $\int dy \, \eta(y) = 0$ and 
$\eta(y) = 0$ for $y > 0$.
Since the probability \eqref{ProbMaxAN} is given by the product of $2L$ terms, 
there will be $2L$ linear terms in $O(\epsilon)$, i.e. $\ProbMaxModel{AN}$ changes by 
\begin{align}
\delta \ProbMaxModel{AN}  =& \epsilon L \int dy\, \eta(y) \int \left(\prod_{r=2}^{L} dy_r p_{1/k}(y_r) \right) \prod_{l=0}^{L-1} \tilde{F}_{1/k}^{(k)}\left ( \sum_{m=1 }^{k} y_{(l + m) \Mod L} \right ) \nonumber\\
& + L \int \left(\prod_{r=1}^{L} dy_r p_{1/k}(y_r) \right) 
\delta \tilde{F}^{(k)}\left( \sum_{m=1 }^{k}y_{m} \right )  \prod_{l=1}^{L-1} \tilde{F}_{1/k}^{(k)}\left ( \sum_{m=1 }^{k} y_{(l + m) \Mod L} \right ) \nonumber \\
\equiv & L (J_1 + J_2). 
\label{ProbVariation}
\end{align}
The first term is straightforward to evaluate. Since $ \tilde{F}^{(k)}_{1/k}\left ( \sum_{m=1}^{k} y_{(l + m) \Mod L} \right )$ is factorized, it readily follows that
\begin{align}
J_1 &=  \epsilon  \int dy\, \eta(y) \int \left(\prod_{r=1}^{L-1} dy_r p_{1/k}(y_r) \right) \prod_{l=0}^{L-1} \tilde{F}^{(k)}_{1/k}\left ( \sum_{m=1 }^{k} y_{(l + m) \Mod L} \right ) \nonumber \\
&= \epsilon  \int dy\, \eta(y)  e^{ky} (k+1)^{-(L-1)/k}.
\end{align}

To evaluate $J_2$, let us rewrite it in the following way:
\begin{align}
J_2 = &\int \left(\prod_{r=1}^{L} dy_r p_{1/k}(y_r) \right) 
\delta \tilde{F}^{(k)}\left( \sum_{m=1 }^{k}y_{m} \right )
\prod_{l=1}^{L-1} \tilde{F}^{(k)}_{1/k}\left ( \sum_{m=1 }^{k} y_{(l + m) \Mod L} \right ) \nonumber\\
=&(k+1)^{-(L-k)/k}  \int \left(\prod_{r=1}^{k} dy_r p_{1/k}(y_r)  e^{(k-1) y_r}\right) \delta \tilde{F}^{(k)}\left( \sum_{m=1 }^{k}y_{m} \right ).
\end{align}
The argument of $\delta \tilde{F}^{(k)}$ is the sum of the variables $y_r$ 
that remain to be integrated over.  
To make them independent, let us introduce a delta function through the identity 
\begin{align}
1 = \int dY \delta\left( \sum_{m=1 }^{k}y_{m} -Y\right) \Theta(-Y)
\end{align}
or, in the Fourier representation,
\begin{align}
1 = \int \frac{dY dZ}{2\pi} e^{ -i Z (\sum_{m=1 }^{k}y_{m}  - Y) } \Theta(-Y),
\end{align} 
where we impose the negativity of $Y$ by inserting an additional theta function.
Using the property $\int dx \delta(x-a) f(x) = \int dx \delta(x-a) f(a)$, we may now complete the integrations over the $y_r$ as
\begin{align}
&\int \frac{dY dZ}{2\pi} \Theta(-Y) \int \left(\prod_{r=1}^{k} dy_r g_{1/k}(y_r)  e^{(k-1) y_r}\right)  e^{ iZ (Y - \sum_{m}^{k} y_m )} 
\delta \tilde{F}^{(k)}(Y) \nonumber \\
= & \int \frac{dY dZ}{2\pi} \Theta(-Y) (k-i Z)^{-1}  e^{ iZ Y} \delta \tilde{F}^{(k)}(Y) 
= \int dY \Theta(-Y) e^{ k Y} \delta \tilde{F}^{(k)}(Y),
\end{align}
where we used Jordan's lemma to evaluate the integral with respect to $Z$.
With this result, $J_2$ is of the relatively simple form
\begin{align}
	J_2 = (k+1)^{-(L-k)/k}  \int dY \Theta(-Y) e^{ k Y}  \delta \tilde{F}^{(k)}\left( Y \right ).
\end{align}
Next, let us evaluate $\delta \tilde{F}^{(k)}(z)$.
Using \eqref{Convolution}, we find that 
\begin{align}
\delta \tilde{F}^{(k)}(z) &= \epsilon k \left[
\int_{-\infty}^{z} dy\, \eta(y) 
+ \int_{z}^{\infty} dy\, \eta(y) \tilde{F}^{(k-1)}_{1/k}(z-y)
\right] \nonumber\\
&= \epsilon k \left[
\int_{-\infty}^{\infty} dy\, \eta(y) 
+ \int_{z}^{\infty} dy\, \eta(y) \left(\tilde{F}^{(k-1)}_{1/k}(z-y) -1\right)
\right] \nonumber\\
&= \epsilon k
\int_{z}^{\infty} dy\, \eta(y) \left(\tilde{F}^{(k-1)}_{1/k}(z-y) -1\right)  \nonumber\\
&= \epsilon k
\int_{-\infty}^{\infty} dy\, \eta(y) \left(\tilde{F}^{(k-1)}_{1/k}(z-y) -1\right) \Theta(y-z),
\end{align}
where the factor $k$  comes from the $k$ different choices of $p_f(y)$ in 
the variation of $\tilde{F}^{(k)}$ and the fact that $\int dy \, \eta(y) =0$ is used to eliminate the first term in the second equality.
As expected, this implies that any perturbation made in the range $(-\infty,z)$ does not change the behavior of $ \tilde{F}^{(k)}(z) $.
Inserting this result into $J_2$ gives
\begin{align}
J_2=&  (k+1)^{-(L-k)/k}  \int dY \Theta(-Y) e^{ k Y}  
\int dy\, \eta(y) \nonumber\\
& \times \epsilon k \left(\tilde{F}_{1/k}^{(k-1)}(Y-y) -1\right) \Theta(y-Y). 
\end{align} 
Now, the only technical point left is the integration with respect to $Y$. 
The integral domain is determined by two theta functions $\Theta(-Y)$ and $\Theta(y- Y)$,  
but since $\eta(y)$ is assumed to be supported only on the negative real axis, the condition imposed 
by $\Theta(-Y)$ is irrelevant. Finally, using the identity
\begin{align}
\int_{-\infty}^{0}dY\,	k e^{k Y} \left(1-\frac{\Gamma \left(\frac{k-1}{k},-Y\right)}{\Gamma \left(\frac{k-1}{k}\right)}\right) = (k+1)^{\frac{1}{k}-1},
\end{align}
we find
\begin{align}
J_2 =& - \epsilon \int dy\, \eta(y) e^{ky}(k+1)^{-(L-1)/k}.
\end{align} 
Thus, the two terms in \eqref{ProbVariation} perfectly cancel, which completes the proof that $ \delta \ProbMaxModel{AN} = 0 $.

\section{General bounds on $\beta$ for uniform and regular structures with Gaussian fitness}
\label{appendix:betabounds}

In this appendix we derive some general upper and lower bounds on the
coefficient $\beta$, defined in \eqref{Defbeta}, for NK structures that are both uniform and regular.
For this purpose we write the probability of $\sigma$ being a local optimum as
\begin{equation}
    \pi_\text{max} = \mean{\prod_{l=1}^{L} \Theta\left(-\Delta_lF(\sigma)\right)} = \mean{\prod_{l=1}^L \Theta\left(-\sum_{r=1}^{|\nkstruct|} \left(f_r\left(\proj[B_r]\Delta_l \sigma\right)-f_r\left(\proj[B_r]\sigma\right)\right)\right)}.
\end{equation}
All fitness values of the partial landscapes $f_r$ are i.i.d. random variables.
If $l\in B_r$, then $f_r\left(\proj[B_r]\Delta_l\sigma\right)$ and $f_r\left(\proj[B_r]\sigma\right)$ are independent.
Otherwise they are identical.
Thus effectively only the sum over $r$ with $l\in B_r$ remains.
Due to regularity there are $\tilde k = \frac{Nk}{L}$ such elements for each $l$.
For different $r$, the terms are always independent.
The left-hand terms are also independent for different $l$.
However the right-hand terms are correlated for different $l$ but the same $r$, resulting in a non-trivial problem.
Using these observations we can directly integrate out all terms
$f_r\left(\proj[B_r]\Delta_l\sigma\right)$ and arrive at
\begin{equation}
    \pi_\text{max} = \mean{\prod_{l=1}^{L} \Phi_{\tilde k}\left(\sum_{r\;|\;l\in B_r} f_r\left(\proj[B_r]\sigma\right)\right)},
\end{equation}
where $\Phi_{\tilde k}$ is the cumulative distribution function of the sum of $\tilde k$ i.i.d. fitness values.
Introducing the short-hand notation $x_r = f_r\left(\proj[B_r]\sigma\right)$, we can write the sum as a matrix product
\begin{equation}
    \pi_\text{max} = \mean{\prod_{l=1}^{L} \Phi_{\tilde k}\left((\mathbf{B}x)_l\right)}
\end{equation}
where $\mathbf{B}$ is the incidence matrix of the NK structure, i.e. $\mathbf{B}_{lr} = b_{l,r} = 1$ if $l\in B_r$ and $0$ otherwise.

If the base fitness distribution is a standard normal distribution, then the sum of $\tilde k$ i.i.d. fitness values is also normal distributed with variance $\tilde k$.
Consequently we can simplify as
\begin{equation}
    \pi_\text{max} = \mean{\prod_{l=1}^{L} \Phi\left(\frac{1}{\sqrt{\tilde k}}(\mathbf{B}x)_l\right)}.
\end{equation}
The random vector $y = \frac{1}{\sqrt{\tilde k}}\mathbf{B}x$ is then jointly normal distributed with zero mean and covariance matrix $\mathbf{C} = \frac{1}{\tilde k}\mathbf{B}\mathbf{B}^T$.
This matrix is positive-semidefinite, and therefore
\begin{equation}
    \pi_\text{max} = \int_{\mathbb{R}^L} \frac{\mathrm{d}y}{\sqrt{(2\pi)^L\det \mathbf{C}}}\exp\left(-\frac{1}{2}y^T\mathbf{C}^{-1}y + \sum_{l=1}^L \ln\Phi(y_l)\right).
\end{equation}
We can shift the integrand by a yet to be specified vector $z$, which yields
\begin{align}
    \pi_\text{max} = \int_{\mathbb{R}^L}
  \frac{\mathrm{d}y}{\sqrt{(2\pi)^L\det \mathbf{C}}}  \times  \nonumber \\
\times  \exp\left(-\frac{1}{2}y^T\mathbf{C}^{-1}y -\frac{1}{2}z^T\mathbf{C}^{-1}z -z^T\mathbf{C}^{-1}y + \sum_{l=1}^L\ln\Phi(y_l+z_l)\right).
\end{align}
Absorbing the first term in the exponent into a probability measure, we have again
\begin{equation}
    \pi_\text{max} = e^{-\frac{1}{2}z^T\mathbf{C}^{-1}z}\mean{\exp\left(-z^T\mathbf{C}^{-1}y +\sum_{l=1}^L \ln\Phi(y_l+z_l)\right)}
\end{equation}
where $y$ is still jointly normal distributed with covariance matrix $\mathbf{C}$.

Notice that the all-ones vector $\bar 1$ is an eigenvector of $\mathbf{C}$ with the eigenvalue $k$.
This can be seen through the relations $\mathbf{B}\bar 1 = \tilde k\bar 1$ and $\mathbf{B}^T\bar 1 = k\bar 1$, as there are exactly $\tilde k$ ones in each row of $\mathbf{B}$ and $k$ ones in each column.
Thus let the $z_l = \bar z$ be equal for all $l$.
Then
\begin{equation}
    \pi_\text{max} = e^{-L\frac{\bar z^2}{2k}}\prod_{l=1}^L\mean{\exp\left(\sum_{l=1}^L\left(\ln\Phi(y_l+\bar z) - \frac{\bar z}{k} y_l\right)\right)}.
    \label{appendixPmaxG}
\end{equation}

\subsection{Lower bound}
By Jensen's inequality we have
\begin{equation}
    \pi_\text{max} \geq e^{-L\frac{\bar z^2}{2k}}\prod_{l=1}^L\exp\left(\mean{\ln\Phi(y_l+\bar z) - \frac{\bar z}{k} y_l}\right).
\end{equation}
Because $y_l$ has a symmetric distribution, the mean of $\bar z y_l$ vanishes.
The variance of $y_l$ is always $1$, because by regularity and
uniformity the diagonal elements of $\mathbf{B}\mathbf{B}^T$ are
$\tilde k$, which is canceled to $1$ by the pre-factor in $\mathbf{C}$.
If we then assume $\bar z$ to be increasing in our limit of interest
and noting that the Gaussian has a tail falling much quicker to zero
than the tail of $\ln\Phi$ falls to $-\infty$ at
$x\rightarrow-\infty$, we can establish the bound
\begin{equation}
    \pi_\text{max} \geq e^{-L\frac{\bar
        z^2}{2k}}\prod_{l=1}^L\exp\left(\mean{\Phi(y_l+\bar z)-1}(1+o(1))\right)
\end{equation}
which can be evaluated to
\begin{equation}
\label{lower_bound}
    \pi_\text{max} \geq \exp\left(-L\frac{\bar z^2}{2k} + L\left(\Phi\left(\frac{\bar z}{\sqrt{2}}\right)-1\right)(1+o(1))\right).
\end{equation}
If we choose $\bar z = 2\sqrt{\ln k}$, then asymptotically for large $k$
\begin{equation}
    \pi_\text{max} \geq \exp\left(-L\left(\frac{2\ln k}{k} +
        \mathcal{O}\left(\frac{1}{k \sqrt{\ln k}}\right)\right)\right).
\end{equation}
Note that choosing $\bar z = \tilde z \sqrt{\ln k}$ with $\tilde z <
2$ will not give a better bound, 
as the right-hand term in the exponent in \eqref{lower_bound} would
then dominate and approach zero more slowly than $\frac{\ln k}{k}$.
This shows that $\beta \leq 2$ for uniform and regular
structures. With the MF model, which is uniform and regular, we have
an example of a realization of $\beta = 2$. This shows that the
bound is tight. 

\subsection{Upper bound}

Starting from \eqref{appendixPmaxG} we can find an upper bound by simply optimizing each term in the sum.
The resulting sum is then an upper bound on the integrand, and because
the expectation is taken with respect to a probability measure, it is bounded by the same value as well.
If $0 < \frac{\bar z}{k} < \frac{1}{\sqrt{2\pi}}$, the optimum must be at $y_l^\star +\bar z > 0$.
Then by using the simplification $\ln\Phi(y_l+z_l) \leq \Phi(y_l+z_l) -1$, the optimum is found to be at
\begin{equation}
    y_l^\star = \sqrt{2\ln\left(\frac{k}{\sqrt{2\pi}\bar z}\right)}-\bar z.
\end{equation}
Inserting $y_l^\star$ back into the simplified argument of the
expectation and assuming $\bar z\rightarrow\infty$ in the limit of
interest we find 
\begin{equation}
    \pi_\text{max} \leq \exp\left(-L\frac{\bar z^2}{2k} -L\left(\frac{\bar z}{k\sqrt{2\ln\left(\frac{k}{\sqrt{2\pi}\bar z}\right)}}(1+o(1)) + \frac{\bar z}{k}\sqrt{2\ln\left(\frac{k}{\sqrt{2\pi}\bar z}\right)} - \frac{\bar z^2}{k}\right)\right).
    \label{appendixPmaxGUpper}
\end{equation}
The left-most and right-most terms are of equal order, but the second one from the left is always of less significant order than the second from the right, as long as $\bar z = o(k)$.

The second term from the right becomes equal in order to the other two if $\bar z = \tilde z\sqrt{2\ln k}$ with a positive constant $\tilde z$.
This satisfies the condition $\bar z = o(k)$ while still $\bar z \rightarrow \infty$, as required by previous assumptions (given that $k\rightarrow\infty$ in the limit of interest).
With this we have
\begin{equation}
    \pi_\text{max} \leq \exp\left(-L\left(\frac{\ln k}{k}(2\tilde z - \tilde z^2) + \mathcal{O}\left(\frac{\ln\ln k}{k}\right)\right)\right).
\end{equation}
The bound is best for $\tilde z = 1$, and so:
\begin{equation}
    \pi_\text{max} \leq \exp\left(-L\left(\frac{\ln k}{k} + \mathcal{O}\left(\frac{\ln\ln k}{k}\right)\right)\right)
\end{equation}
showing that $\beta \geq 1$ for regular and uniform NK structures with Gaussian fitness.
This bound is realized by the AN and BN structures, for example,
and thus it is tight.

\section{Simulation of the number of local maxima}
\label{appendix:Algorithm}
As first realized in \cite{Buzas2014}, the choice of a Gaussian base fitness 
distribution greatly simplifies the computation of $\ProbMax$ through the numerical evaluation
of \eqref{ProbMaxFormalism}, as it allows us to take advantage of an efficient algorithm. 
With this choice, the integrals over $\Vecbf{q}$ and $\Vecbf{y}$ can be cast into the form of multi-dimensional Gaussian integrals which may be evaluated for generally defined NK structures.
Once these integrals are evaluated, we may construct a covariance matrix $\Sigma$ that satisfies the relation 
\begin{align}
	\ProbMax = \int \mathcal{D} \Vecbf{u} \exp \left(
		-\frac{1}{2}\sum_{j l} u_j \Sigma^{-1}_{j l} u_l 
	\right),
	\label{ProbMaxGaussian}
\end{align}
where $ \int \mathcal{D} \Vecbf{u} = \frac{1}{\sqrt{(2\pi)^L \det \Sigma}} \int_{0}^{\infty} \prod_j du_j $ and the matrix elements of $\Sigma$ are given by 
\begin{align}
	\Sigma_{j l}  = \begin{cases}
	2 \sum_{r} b_{l,r} & j = l \\
	\sum_{r} b_{j,r} b_{l,r} &  j \ne l.
	\end{cases}
\end{align}
Thus, the problem reduces to determining the probability that all the entries of the Gaussian random vector realized by the covariance matrix $\Sigma$ are positive. 
Since finding the probability for rectangular domains of multivariate Gaussian distribution is a well-known problem, an efficient algorithm has been known for a long time \cite{Genz1992} and its implementation has been provided by the original authors as an R library \cite{Genz2017}.

Roughly speaking, this algorithm consists of two steps: i) transforming to an integral over a unit rectangular domain such that a rejection-free Monte-Carlo simulation is possible and ii) finding an ordering of loci that minimizes the variance of the Monte-Carlo step.
However, since the loci in the NK models we consider in this review are statistically identical, the second step is irrelevant in this particular case.
Thus, here we describe briefly how the transformation can be achieved from \eqref{ProbMaxGaussian}.

Since $\Sigma$ is positive-definite, the Cholesky decomposition
ensures that there exists a triangular matrix $C$ such that $\Sigma =
C C^T$. 
The substitution $\Vecbf{u} = C \Vecbf{x}$ then diagonalizes the integral at the cost of nontrivial integral domain,
\begin{align}
	\ProbMax = \frac{1}{(2\pi)^{L/2}}\int_{ \Vecbf{x} \in \mathcal{R}}\prod_{j=1}^{L} dx_j \exp \left(
	-\frac{1}{2}\sum_{j =1} ^L x_j^2
	\right),
\end{align}
where the domain $\mathcal{R} = (a_1, \infty) \times (a_2, \infty) \times \cdots (a_L, \infty)$ and $a_j = - \sum_{l=1}^{j-1} x_l C_{j l }  / C_{j j } $.
Next, performing the canonical transformation to a standard uniform distribution $   z_i =  \Phi(x_i) $, where $\Phi(x)$ is the CDF of the standard Gaussian distribution, the integral becomes 
\begin{align}
	\ProbMax = \int_{ \Vecbf{z} \in \mathcal{R'}}  \prod_{j=1}^{L} dz_j,
\end{align}
where $  \mathcal{R'} =  (d_1, 1) \times (d_2, 1) \times \cdots (d_L, 1) $ and $d_j = \Phi( - \sum_{l=1}^{j-1} \Phi^{-1}(z_l) C_{j l}  / C_{j j })$.
Finally, another linear transformation $z_j = d_j + w_j (1- d_j)$ brings the integral into the form
\begin{align}
\ProbMax  = \int_{\Vecbf{w} \in \mathcal{R''} }   \prod_{j=1}^{L} (1- d_j) dw_j,
\end{align}
where $\mathcal{R''} = (0,1)^{L}$. 
Now that the integral domain is the $L$-dimensional unit rectangle, 
this integral can be evaluated by sampling $L$ random variables from a uniform distribution on $(0,1)$ and subsequently estimating the weight factors $d_j$.

\end{document}